\documentclass[twocolumn,times]{aastex62}
\received{March 22, 2019}
\revised{July 22,2019}
\accepted{July 29, 2019}
\submitjournal{ApJ, accepted for publication.}

\shorttitle{Reverberation-measured quasars for cosmology}
\shortauthors{Mart\'inez-Aldama et al.}

\usepackage{multirow}
\usepackage{longtable} 
\usepackage{mathrsfs}  
\usepackage{amssymb, amsmath, amsbsy}
\usepackage{textcomp}

\def\kms{\,km\,s$^{-1}$}
\def\ergs{erg\,s$^{-1}$}
\def\hb{{\sc{H}}$\beta$\/}
\def\lbol{$L\mathrm{_{bol}}$}
\def\mbh{$M\mathrm{_{BH}}$}
\def\fblr{$f\mathrm{_{BLR}}$}
\def\mdot{$\dot{\mathscr{M}}$}
\def\DRhb{$\Delta R_{\mathrm{H\beta}}$}
\def\LLEdd{$L\mathrm{_{bol}}/L\mathrm{_{Edd}}$}
\def\fvar{$F\mathrm{_{var}}$}
\def\mdotc{$\dot{\mathscr{M}}\mathrm{^{c}}$}
\def\RL{$R\mathrm{_{H\beta}}-L_{5100}$}
\def\LLEddc{$L\mathrm{_{bol}}/L\mathrm{_{Edd}^{c}}$}
\def\fblrc{$f\mathrm{_{BLR}^{\,c}}$}

\begin{document}
\title{\large Can reverberation-measured quasars be used for cosmology?}

\correspondingauthor{M.L. Mart\'inez-Aldama}
\email{mmary@cft.edu.pl}

\author{Mary Loli Mart\'inez-Aldama}
\affiliation{Center for Theoretical Physics, Polish Academy of Sciences, Al. Lotnik\'ow 32/46, 02-668 Warsaw, Poland}

\author{Bo{\.z}ena Czerny}
\affiliation{Center for Theoretical Physics, Polish Academy of Sciences, Al. Lotnik\'ow 32/46, 02-668 Warsaw, Poland}

\author{Damian Kawka}
\affiliation{Center for Theoretical Physics, Polish Academy of Sciences, Al. Lotnik\'ow 32/46, 02-668 Warsaw, Poland}

\author{Vladimir Karas}
\affiliation{Astronomical Institute, Academy of Sciences, Bocni II, CZ-141 31 Prague, Czech Republic}

\author{Swayamtrupta Panda}
\affiliation{Center for Theoretical Physics, Polish Academy of Sciences, Al. Lotnik\'ow 32/46, 02-668 Warsaw, Poland}

\affiliation{Copernicus Astronomical Center, Polish Academy of Sciences,  ul. Bartycka 18, 00-716 Warsaw, Poland}

  
\author{Michal Zaja\v{c}ek}
\affiliation{Center for Theoretical Physics, Polish Academy of Sciences, Al. Lotnik\'ow 32/46, 02-668 Warsaw, Poland}

\author{Piotr T. \. Zycki}
\affiliation{Copernicus Astronomical Center, Polish Academy of Sciences,  ul. Bartycka 18, 00-716 Warsaw, Poland}



\begin{abstract}

Quasars have been proposed as a new class of standard candles analogous to Supernovae, since their large redshift range and high luminosities make them excellent candidates. Reverberation mapping (RM) method enables to estimate the distance to the source from the time delay measurement of the emission lines with respect to the continuum, since the time delay  depends on the absolute luminosity of the source. The radius-luminosity relation exhibits a low scatter {and offers a potential use in cosmology}. However, in the recent years {the inclusion of new sources, particularly the super-Eddington accreting QSO, has increased} the dispersion in the radius-luminosity relation, with many objects showing time delays shorter than the expected. Using 117 \hb\ RM AGN with  $0.002<z<0.9$ and $41.5<\mathrm{log}~L_{5100}<45.9$, we find a  correction for the time delay based on the dimensionless accretion rate (\mdot) considering a virial factor anti-correlated with the FWHM of \hb. This correction decreases the scattering of the accretion parameters compared with typical values used, which is directly reflected by suppressing the radius-luminosity relation dispersion. We also confirm the anti-correlation between the excess of variability and the accretion parameters. With this correction we are able to build the Hubble diagram {and estimate the cosmological constants $\Omega_m$ and $\Omega_\Lambda$, which are consistent with standard cosmological model at 2$\sigma$ confidence level}. Therefore, RM results can be used to constrain cosmological models in the future.

\end{abstract}




\section{Introduction}
\label{sec:intro}

Understanding dark energy is one of the greatest puzzles of the modern physics. In order to test numerous theories
proposed to explain the phenomenon of accelerated expansion of the Universe, we first need to measure it precisely. There are well-established methods such as the studies of the {Cosmic Microwave Background, Supernovae Ia (SNIa), Baryon Acoustic Oscillations, and weak lensing}. Combination of these methods currently sets the following
cosmological parameters to these values: $H_0 = 67.66 \pm 0.42$ km s$^{-1}$ Mpc$^{-1}$, $\Omega_{\Lambda} = 0.6889 \pm 0.0056$, $\Omega_m = 0.3111 \pm 0.0056$ \citep{planck2018}. These results are consistent with the simplest interpretation of the  cosmological constant in dark matter dominated Universe in the form of $\Lambda$ cold dark matter model \citep{planckXIV}. However, there is some tension now with the local measurements of the Hubble constant \citep{riess2018}, amplitude  of  matter  fluctuations  in  the  late time Universe compared to cosmic shear measurements \citep{joudaki2017,hildebrandt2017}, and the number counts of galaxy clusters (\citet{PCGC}; see \citet{pacaud2018} for most recent results). Therefore, new objects are proposed as tools to constrain better the Universe expansion, and active galactic nuclei (AGN) are among them \citep[e.g.][]{czerny2018SSR}.

AGN cover a broad range of redshift, and they do not show strong evolution with the redshift - even for example, the most distant quasars have metallicities similar to the nearby AGN (close to the solar value, or slightly higher \citep[e.g.][]{groves2006}. {This could be caused by the combination of efficient rotational mixing and powerful stellar winds of early massive stars \citep{meader2000,brott2011,ekstrom2012}, which could transport the heavier nucleosynthesis products to the surface of massive stars already within $\sim 10\,{\rm Myr}$ of the stellar evolution \citep{stanway2019}. Powerful stellar winds would further enrich the ISM of early quasars.}

Reverberation campaigns revealed a very strong and tight correlation between the broad line region (BLR) size {($R\mathrm{_{H\beta}}$)} and the monochromatic luminosity at 5100\AA\ {($L_{5100}$)}, producing the well-know radius-luminosity relation, \RL~\citep{kaspi2000,peterson2004,bentz2013}. {$R\mathrm{_{H\beta}}$ is estimated from the time delay ($\tau\mathrm{_{obs}}$) between continuum and emission line variations and assuming the velocity as the speed of light ($c$), i.e., $R\mathrm{_{H\beta}}=\tau\mathrm{_{obs}} \cdot c$}. The radius-luminosity relation offers prospects of cosmological applications. After proper calibration, the time delay measurement allows us to determine the absolute luminosity, and to use a generalized standard candle approach to obtain the cosmological parameters \citep{watson2011,haas2011,czerny2013,king2014}. 

The problem has started with the detection of some outliers from the radius-luminosity relation \citep{bentz2013}. First, outliers from the \RL\ relation have been found among the highly accreting AGN, which are the subject of the super-Eddington accreting massive black holes (SEAMBHs) campaign  \citep{du2014, wangSEAMBH2014, hu2015, du2015,du2016,du2018}. The interpretation is that the measured delays much shorter than implied by the standard $R\mathrm{_{H\beta}}-L_{5100}$ relation \citep{bentz2013} are caused by  the self-shielding of geometrically thick accretion disk which subsequently modifies the radiation field seen by the surrounding material forming the BLR  \citep{wangshielding2014}. 

Recently, more sources with considerably shorter than expected time delays were found by \citet{grier2017} and \citet{du2018}. A significant fraction of them have low values of the Eddington ratio, and they cannot be simply eliminated from the sample. If the $R\mathrm{_{H\beta}}-L_{5100}$  relation has such a large scatter, application of AGN to cosmology based on this relation is problematic, unless we understand what additional parameter is responsible for the departure from the original $R\mathrm{_{H\beta}}-L_{5100}$ relation, and are able to correct for this trend. It poses also a question about the nature of the standard radius-luminosity relation and the physical reasons for the departures from this law. These shortened lags could be explained for example by retrograde accretion \citep{wangspin2014,du2018}, the inner disk evaporation, or replacing the dust-based model of BLR formation \citep{czerny2011,czerny2015,czerny2017} with the old model based on assumption of ampleness of gaseous material close to the nucleus and formation of the BLR where the ionization parameter has the optimum value \citep{czerny2019}.

In this paper we analyze in detail how the properties of active galaxies correlate with their location with respect to the standard \RL\ relation. Section \ref{sec:method} gives a description of the different \hb\ reverberation--measured sub-samples considered in this work and the relations used to estimate the main physical parameters, such as virial factor, black hole mass, accretion parameters, variability, etc. Section \ref{sec:results} describes the correction for the time delay based on the accretion parameters recovering the low scatter along the \RL\ relation. We confirm the anti-correlation between the variability and the accretion parameters as well. Section \ref{sec:cosmo} presents the Hubble diagram built with the reverberation--measured sample and the possible cosmological implications. {We also discuss some remarks of the presented method which affect the implementation of quasar in cosmology in Section \ref{sec:discussion}.} In Section \ref{sec:conclusions}, we review the main result of this work. Absolute values of the luminosity are given assuming the cosmological parameters: $H_0=67$ km s$^{-1}$ Mpc$^{-1}$, $\Omega_{\Lambda}=0.68$, $\Omega_m=0.32$ \citep{planck2013}.

\section{Method}
\label{sec:method}

\subsection{Observational data}
\label{sec:obs_data}

Our sample of \hb\ reverberation-measured AGN is a compilation of the results published earlier in the literature. It was previously used by \citet{czerny2019}: luminosity ($L_{5100}$), time delay ($\tau\mathrm{_{obs}}$) and FWHM are the same as considered by them. We have collected a total of 117 sources, plus 2 objects which have been discarded from the analysis (see details in Section \ref{sec:dis_obj}). The first sub-sample is composed of 25 highly accreting AGN observed by the SEAMBH (Super-Eddington Accretion in Massive Black Holes) project \citep{du2014, wangSEAMBH2014, hu2015, du2015,du2016, du2018}. SEAMBH project group have been monitoring super-Eddington sources since 2012 obtaining important results for this kind of objects.  The second sub-sample contains 44 objects from the recent SDSS-RM (Sloan Digital Sky Survey Reverberation Measurement) project \citep{grier2017}; two sources of this sample have been discarded. This sample comes from  a larger sample of \citet{shen2015}, recently they have published an update of the catalog with additional information such as the variability properties \citep{shen2018}. The third sub-sample is a collection of 48 sources from a long-term monitoring projects, where the majority  of the sources have been summarized by \citet{bentz2013}. We include in this sample other sources monitored in the recent years \citep{bentz2009, bentz2014, barth2013, pei2014, bentz2016a, bentz2016b,fausnaugh2017}. {This collective sample is henceforth referred to as the Bentz Collection.} The fourth sub-sample includes NGC 5548 and 3C 273 (PG 1226+023) monitored  by \citet{lu2016} and \citet{zhang2018}, respectively. 3C 273 was previously monitored by \citet{kaspi2000} and \citet{peterson2004}, however new results from GRAVITY Collaboration \citep{gravity2018} resolved the BLR with a much better angular resolution of 10$^{-5}\,\arcsec$ indicating a smaller BLR size ($\sim$ 150 light days) than the reverberation mapping results. {Almost at the same time, \citet{zhang2018} report a new value for BLR size from a RM monitoring performed for $\sim$10 years, which is similar to the one given by \citet{gravity2018}. We will use this new estimation for 3C273 henceforth in the paper.}

These sources do not form a uniform sample. Sources from the Bentz Collection were selected to cover broad range of redshift (0.002 $\lesssim z \lesssim $ 0.292), from nearby sources studied earlier for example by \citet{peterson2004} to more distant PG quasar sample from \citet{kaspi2000}. The average luminosity and dimensionless accretion rate are log$L_{5100}$=43.4 \ergs\ and \mdotc$\sim$0.8 (See Section \ref{sec:bhm-accretion}), respectively. Sources from SDSS-RM sample are on average slightly more luminous, log$L_{5100}$=43.9 \ergs, and cover systematically larger redshifts, 0.116 $\lesssim z \lesssim $ 0.89. On the other hand, SEAMBH sample has been selected with the aim to study super-Eddington sources, they are nearby objects (0.017 $\lesssim z \lesssim $ 0.4), but with the largest accretion rates, \mdotc$_{mean}\sim$ 14.6 (See Section \ref{sec:bhm-accretion}). 

The full sample includes 117 sources and covers a large redshift range (0.002 $\lesssim z \lesssim $ 0.89), which is convenient in order to test cosmological models. A detail description of the sample is shown in Table \ref{tab:samples}. 


\subsection{Discarded objects and biases in the sample} \label{sec:dis_obj}

Two objects from SDSS-RM sample have been discarded from the analysis: J141856 and J141314. In the case of J141856 there is no detection of the blue side of \hb\ line, destroying the profile and prohibiting any possibility of measurement. In J141314, the \hb\ line is at the border of the spectrum, where the S/N is poor and the emission line cannot be observed. 

Observational problems can cause an incorrect estimation in the time delay, see Section \ref{sec:candence} and Appendix \ref{sect:simulation}. It is important to stress that $\sim$30\% of the SDSS-RM sample have a contribution of the host galaxy luminosity $>$50\% with respect to AGN. In order to decompose the quasar and host-galaxy contribution, \citet{shen2015} applied a principal component analysis (PCA). This method could present some systematic uncertainties in the decomposition, due to the limited S/N or insufficient host contribution. However, in all the analyzed cases PCA is successful in decomposing both components, even those where S/N and the equivalent width are low, e.g. J141123.

In addition, some objects from the Bentz collection show a large variability {and seem to not follow the \RL\ relation. For example, the \hb\ broad component in NGC5548 appears and disappears over the years \citep[e.g.][]{sergeev2007}  and shows a stepper radius-luminosity relation \citep{peterson2004}}. However, variations in the measurements seem to be included in the intrinsic scatter of the \RL\ relation \citep{kilerci2015}. {We decide to keep these sources in the sample, like a representation of peculiar behavior and the  uncertainties which entails. Its location is clearly marked in all the plots. }


\subsection{Black hole mass and accretion parameters}
\label{sec:bhm-accretion}

Some of the sources from the sample deviate from the classical $R\mathrm{_{H\beta}}-L_{5100}$ relation \citep{grier2017,czerny2019}, and our aim is to find properties which characterize their departure from the scaling law in the best way. \citet{du2016} suggested that accretion rate is the key parameter, from which we calculate the related parameters uniformly for our sample.

In order to have an agreement in the computations of AGN parameters, we recompute the values following the same methods and using the same constant factors. The black hole mass (\mbh) is estimated following the well-known relation:
\begin{equation}
M\mathrm{_{BH}}=f\mathrm{_{BLR}}\frac{R\mathrm{_{H\beta}} \, v^2}{G}
\label{equ:mass}
\end{equation}
where $G$ is the gravitational constant, \fblr\ is the virial factor,  $R\mathrm{_{H\beta}}$ is the broad line region size and $v$ is the velocity field in the BLR, which is typically represented by the full-width at half maximum (FWHM) {or the line dispersion ($\sigma_\mathrm{line,rms}$) of the emission line measured in the rms spectrum.}

{The virial factor takes into account geometry, kinematics, and inclination angle of the BLR.  Many formalisms have been proposed for its description, for example \citet{peterson2004}, \citet{onken2004}, \citet{collin2006},  \citet{mejia-restrepo2018} and \citet{yu2019}. The best method to calibrate the virial factor is through a comparison with an independent measurement of the black hole mass, like for example the one obtained from the relation between the \mbh\ and the bulge or spheroid stellar velocity dispersion ($\sigma_*$), the relation \mbh--$\sigma_*$  \citep[e.g.][]{woo2015}. However, in some cases it is hard to get a proper measurement of the stellar absorption features, particularly in high-redshift sources. Also, it has not been tested considering super-Eddington sources. The large uncertainties associated with the virial factor introduces an error  in the \mbh\ determination by a factor 2--3.}
 


 In this work, we are going to consider the FWHM as the velocity field in the BLR. Recently, \citet{mejia-restrepo2018} proposed that the virial factor is anti-correlated with the FWHM of broad emission lines. For the \hb\ line the relation is given by,
\begin{equation} \label{equ:fblrc}
f\mathrm{_{BLR}^{\,c}}=\left(\frac{{\mathrm{FWHM_{obs}}}}{4550 {\pm 1000}}\right)^{ -1.17},
\end{equation}
with FWHM$\mathrm{_{obs}}$ in units of \kms. {A similar relation has been recently found by \citet{yu2019}, but with an exponent of -1.11}. This representation could indicate a disk-like geometry for the BLR and/or cloud motions or winds induced by the radiation force. In order to explore the effects of a different virial factor expression over the black hole mass and accretion parameters (Eddington ratio and dimensionless accretion rate), we have computed the \mbh\ using the typical value \fblr=1 and $f\mathrm{_{BLR}^{\,c}}\propto$ FWHM$^{-1.17}$ (See Section \ref{sec:desviation}). In Table \ref{tab:measurements}, we report mass \mbh\ considering  \fblr=1. 

In order to estimate the accretion rate, we use the dimensionless accretion rate introduced by \citet{du2016}: 
\begin{equation} \label{equ:mdot}
\dot{\mathscr{M}}=20.1\left(\frac{\it{l}\mathrm{_{44}}}{\mathrm{cos}\,\it{\theta}}\right)^{3/2} m_{7}^{-2},
\end{equation}
where $\it{l}\mathrm{_{44}}$ is the luminosity at 5100 \AA\ in units of 10$^{44}$ \ergs, $\theta$ is inclination angle of disk to the line of sight, and $m_{7}$ is the black hole mass in units of 10$^7$ $M_\odot$. We considered cos $\it{\theta}$ = 0.75, which is the mean disk inclination for type 1 AGN. It is estimated considering a torus axis co-aligned with the disk axis and a torus covering factor of 0.5  \citep{du2016}. {Justification of this assumption is discussed in Sect. \ref{sec:inclination_angle}.} Sources with \mdot\ $\gtrsim$ 3 are highly accreting AGN and host a slim accretion disk \citep{wangshielding2014}.  \mdot\ with \fblr=1 is reported in the column 3 of Table \ref{tab:measurements}. 


SEAMBH sample has on average the largest dimensionless accretion rate, \mdot\ $=139.5^{+96.8}_{-25.8}$, which is expected due to the selection criteria of the sample. SDSS-RM sample has the mean \mdot$=11.9^{+5.3}_{-2.7}$, where one--third of the sources are  classified as high accretors. The Bentz collection has the smallest mean value, \mdot$=7.9^{+4.5}_{-4.3}$. As for the two remaining objects, 3C 273 is classified as a high accretor (\mdot$=27.4^{+3.9}_{-5.1}$) and NGC5548 is one the source with lowest accretion rate in the sample, \mdot$=0.01^{+0.006}_{-0.005}$.

We also consider the Eddington ratio, \LLEdd, as an estimation of the accretion rate, where $L\mathrm{_{bol}}$ is the bolometric luminosity and $L\mathrm{_{Edd}}$ is the Eddington luminosity defined by $L\mathrm{_{Edd}}=1.5\times10^{38}\left(\frac{M_{BH}}{M_\odot}\right)$. In order to determine \lbol, we use the bolometric correction factor at 5100\AA\ proposed by \citet{richards2006},  BC$_{5100}=10.33$. Although \LLEdd\  depends upon the bolometric correction factor used, it has given good results in the identification of high accretion sources, which show  \LLEdd$>$0.2 \citep{sulentic2017}. The limit considered for \mdot\ and \LLEdd\ in order to identify  highly accreting sources is analogous, since both parameters are well correlated \citep[e.g.][]{ capellupo2016}.  Average \LLEdd\ values are as follow: {1.6$^{+0.5}_{-0.2}$, 0.3$^{+0.04}_{-0.03}$ and 0.2$\pm0.04$} for SEAMBH, SDSS-RM and Bentz collection respectively. \LLEdd\ with a virial factor equal to 1 is shown in the fourth column of Table \ref{tab:measurements}.

{Considering the virial factor anti-correlated with the FWHM of \hb\ ($f\mathrm{_{BLR}^{\,c}}$),  dimensionless accretion rate and Eddington ratio change by a factor  $(f\mathrm{_{BLR}^{\,c}})^{-2}$ and $(f\mathrm{_{BLR}^{\,c}})^{-1}$ respectively, i.e.,}
\begin{equation}\label{equ:ac_fAD}
\begin{gathered}
\dot{\mathscr{M}}\mathrm{^{c}}=(f\mathrm{_{BLR}^{\,c})^{-2}}\,\dot{\mathscr{M}}, \\
\frac{L\mathrm{_{bol}}}{L\mathrm{{_{Edd}^{\,c}}}}=(f\mathrm{_{BLR}^{\,c})^{-1}} \left(\frac{L\mathrm{_{bol}}}{L\mathrm{{_{Edd}}}}\right).  \\
\end{gathered}
\end{equation}
{Therefore, new values for the accretion parameters can be estimated from the ones reported in Table~\ref{tab:measurements}. The virial factor selection changes considerably the accretion parameters and the dispersion associated with other physical parameters (see Section \ref{sec:desviation}). We include a discussion of the uncertainties associated with the assumed virial factor and the implications for the presented analysis in Section \ref{sec:virial_factor}.}

\subsection{Variability characteristics}
\label{sec:fvar}
In order to have an estimation of the {optical} continuum variability amplitude, we will consider the parameter \fvar\ \citep{rodriguez-pascual1997}. It estimates the rms of the intrinsic variability relative to the mean flux,
\begin{equation}
F\mathrm{_{var}}=\frac{(\sigma^2-\Delta^2)^{1/2}}{\langle f \rangle},
\label{equ:fvar}
\end{equation}
where $\sigma^2$ is the variance of the flux, $\Delta$ is the mean square value of the uncertainties ($\Delta_{i}$) associated with each flux measurement ($f_{i}$), and $\langle f \rangle$ is the mean flux. Their definitions are as follows, 
\begin{equation}\label{equ:fvar_def}
\begin{gathered}
\sigma^2=\frac{1}{1-N} \sum_{i=1}^{N} (f_{i}-\langle f \rangle)^{2}, \\
\Delta^2=\frac{1}{N} \sum_{i=1}^{N} \Delta{^{2}_{i}}, \\
\langle f \rangle=\frac{1}{N} \sum_{i=1}^{N} f_{i}. 
\end{gathered}
\end{equation}
\fvar\ parameter has been reported for all the objects of  Bentz collection  \citep{peterson2004, bentz2009, denney2010, bentz2014,barth2013,pei2014,bentz2016a,bentz2016b,fausnaugh2017} and some SEAMBH objects \citep{hu2015}. For the remaining SEAMBH sources, we estimate \fvar\ from the light curves available in the literature \citep{du2015,du2016,du2018} following the Equations~(\ref{equ:fvar}) and (\ref{equ:fvar_def}). In the case of the SDSS-RM sample, we use the \textit{fractional RMS variability} provided by \citet{shen2018} (see their Table 2). Using the luminosities reported in Table \ref{tab:samples}, we can convert this quantity to \fvar. The object J142103 shows  $\sigma^2-\Delta^2<0$, indicating that it does not present a significant variability, such as the one that has been reported in other objects \citep[e.g.][]{sanchez2017}. \fvar\ values are reported in the last column of Table \ref{tab:measurements}.

\section{Accretion rate dependence along the \RL\ relation}
\label{sec:results}

\subsection{\RL\ relation}
\begin{figure*} 
\centering
\includegraphics[width=0.4\textwidth]{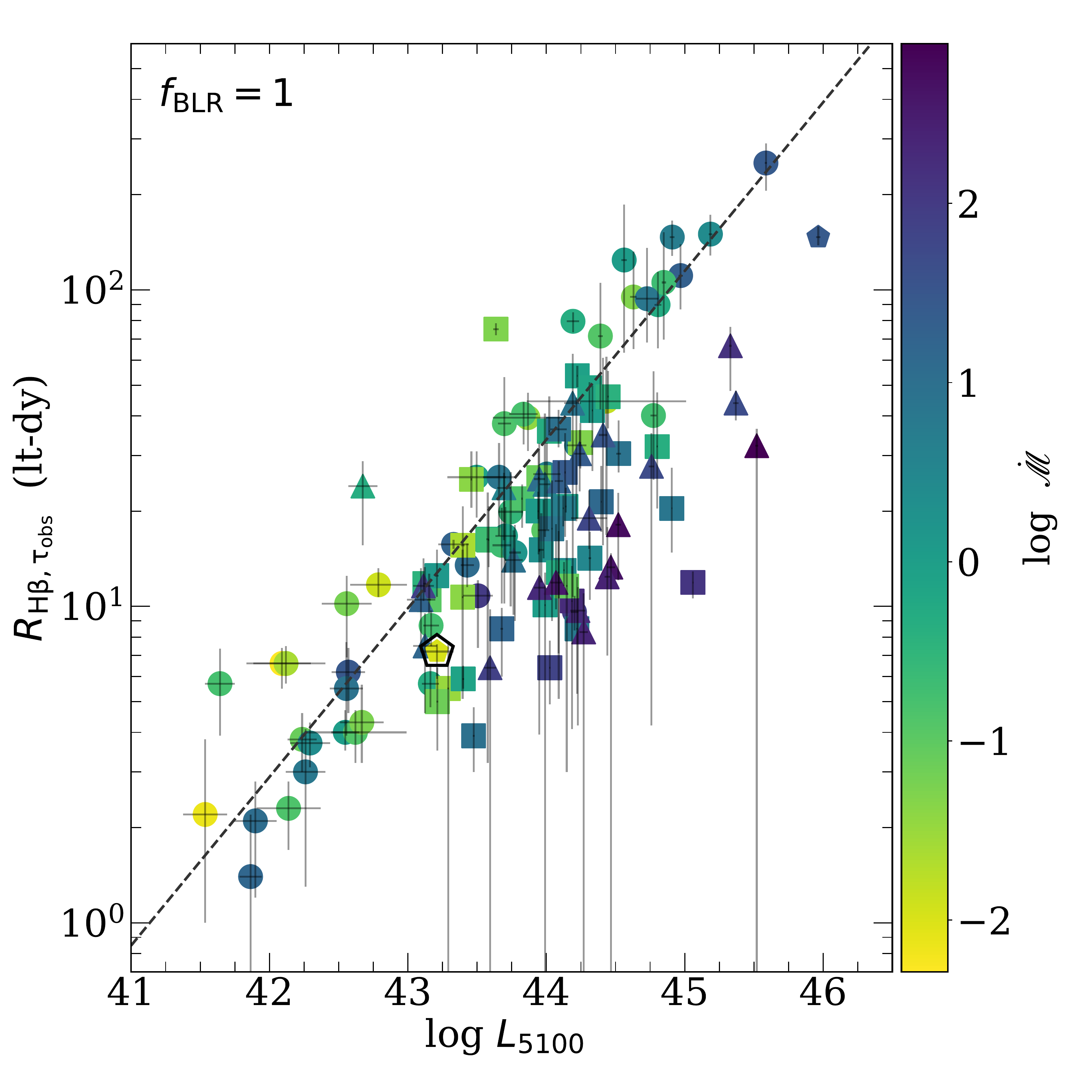}
\includegraphics[width=0.4\textwidth]{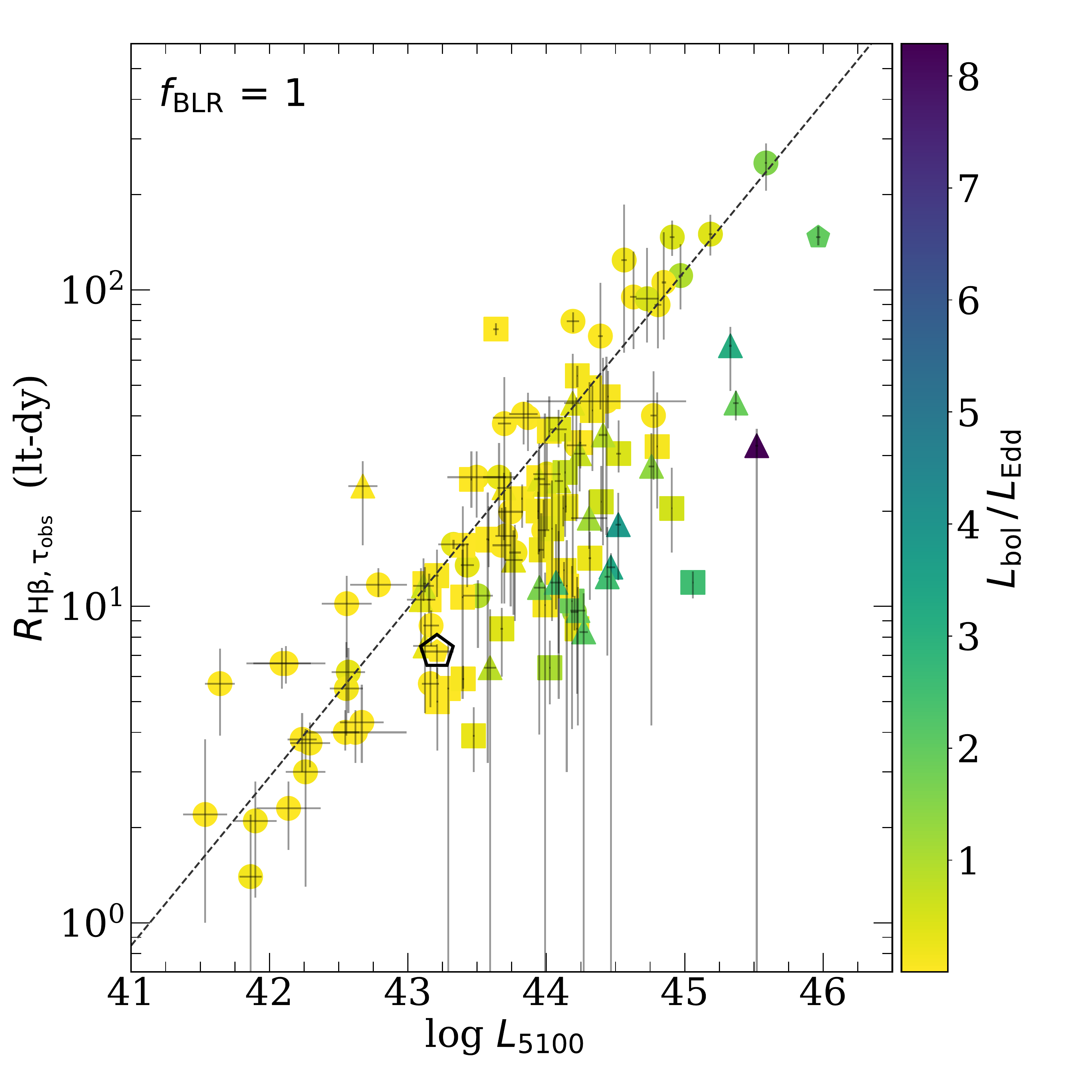}
\caption{$R_{H\beta}-L_{5100}$ relation for SEAMBH (triangles), SDSS-RM (squares), Bentz Collection (circles), NGC5548 and 3C 273 (pentagons). Colors indicate the variation in dimensionless accretion rate (\mdot in log-space, left) and Eddington ratio (\LLEdd, right). Dashed black line corresponds to the expected $R_{H\beta}-L_{5100}$ relation from \citet{bentz2013}. The dimensionless accretion parameter and the Eddington ratio have been computed considering \fblr=1. {Open black pentagon corresponds to NGC 5548 (see Sections \ref{sec:dis_obj} and \ref{sec:inclination_angle}).} \label{fig:RL_mdot} }
\end{figure*}

\begin{figure*} 
\centering
\includegraphics[width=0.4\textwidth]{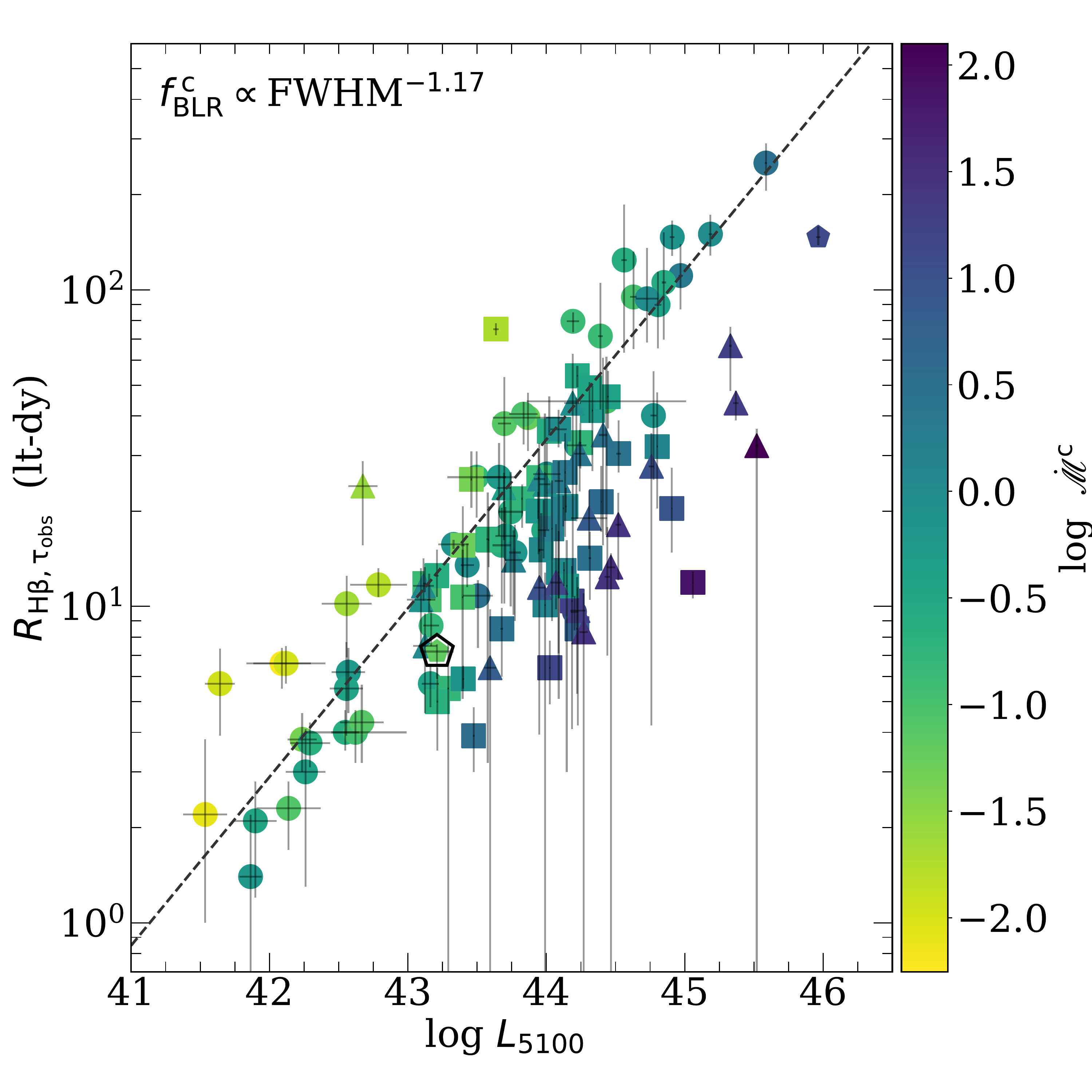}
\includegraphics[width=0.4\textwidth]{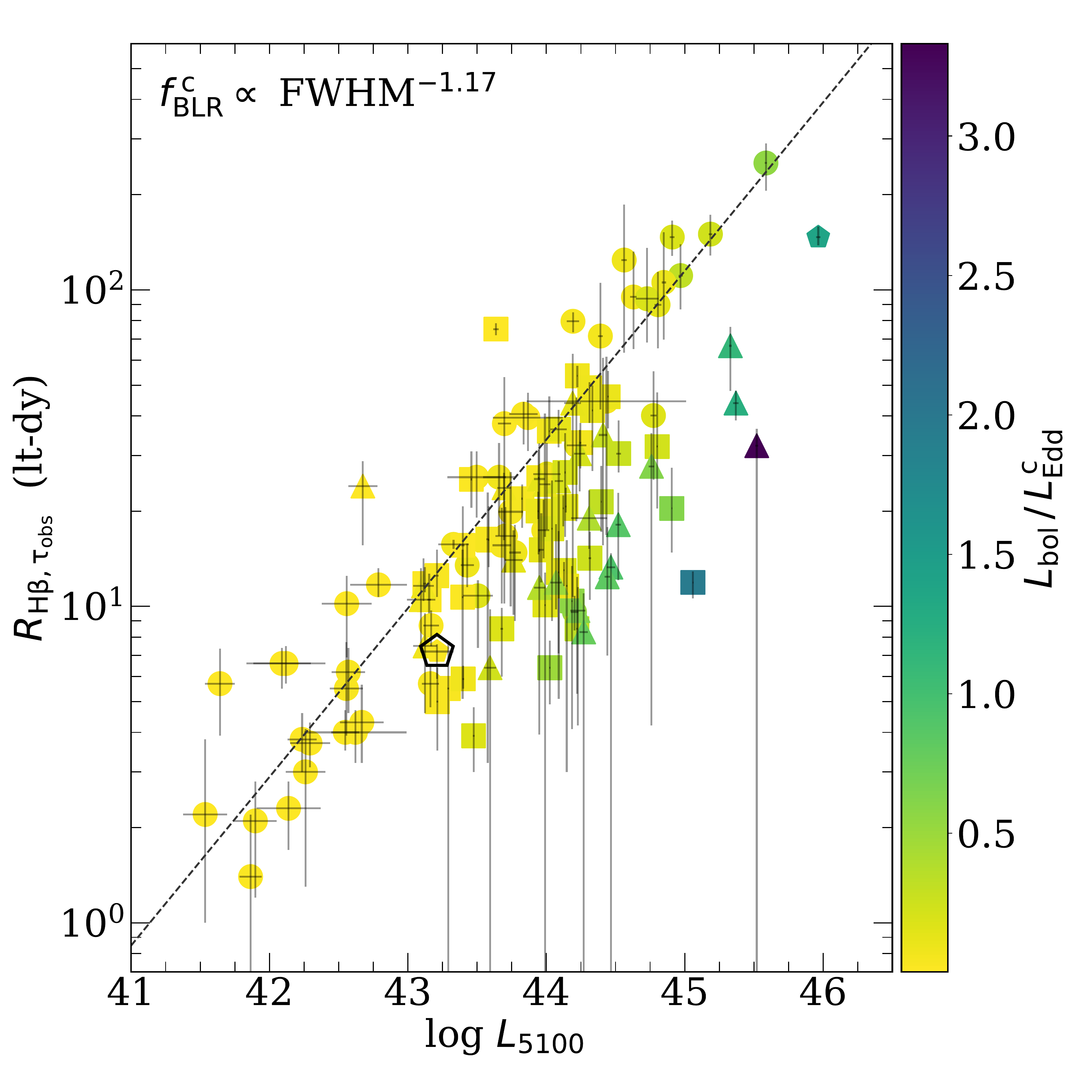}
\caption{The same figure as in Figure~\ref{fig:RL_mdot}, but considering the virial factor dependence $f\mathrm{_{BLR}^{\,c}}\propto$ FWHM$^{-1.17}$. \label{fig:RL_mdot2}}
\end{figure*}

{The radius-luminosity relation used in this paper is given by  \citet{bentz2013}, 
 \begin{align}
\mathrm{log}\left(\frac{R\mathrm{_{H\beta}}}{\mathrm{1 lt-day}}\right)=&(1.527\,\pm\,0.31)\, + \nonumber \\ &\,0.533{^{+0.035}_{-0.033}}\, \mathrm{log}\left(\frac{ L_{5100}}{10^{44}L_\odot}\right).
\label{equ:bentz}
\end{align}
With the information reported in Table \ref{tab:samples}, we are able to build a  $R\mathrm{_{H\beta}}-L_{5100}$ diagram, which is shown in Figures~\ref{fig:RL_mdot} and \ref{fig:RL_mdot2}. In both figures, the variations of the dimensionless accretion rate (left) and Eddington ratio (right) along the diagram, considering \fblr=1 (Figure~\ref{fig:RL_mdot}) and \fblrc$\propto$FWHM$^{-1.17}$ (Figure~\ref{fig:RL_mdot2}) are shown. Independently of the selected virial factor, there is a clear  trend with the accretion parameters. Sources with high accretion rate values show the largest departures from the \RL\ relation \citep{du2015, du2018}.}

{On the other hand, we also explored if FWHM and equivalent width of \hb\ line show a similar trend along the \RL\ diagram, but we can not find any clear pattern.}

\subsection{Testing cadence in SDSS-RM sample}
\label{sec:candence}
The monitoring performed for the SDSS-RM sample is relatively short (only 180 days) taking into account that their sources are at relatively large redshift and some of them are rather bright. Expected delay could be close to the duration of the campaign. In addition, the number of spectroscopic measurements is not very high (only 32). This may cast some doubts whether the estimated time delays are measured reliably. Thus we performed simulations of the expected time delays assuming that the sources follow the standard \RL\ relation and using the observational cadence. The results are presented in the Appendix \ref{sect:simulation} and they show that both  the duration of the observation and the cadence should not strongly affect the measured delays.

\subsection{Correction for the time delay }
\label{sec:desviation}
\begin{figure*} 
\centering
\includegraphics[width=0.4\textwidth]{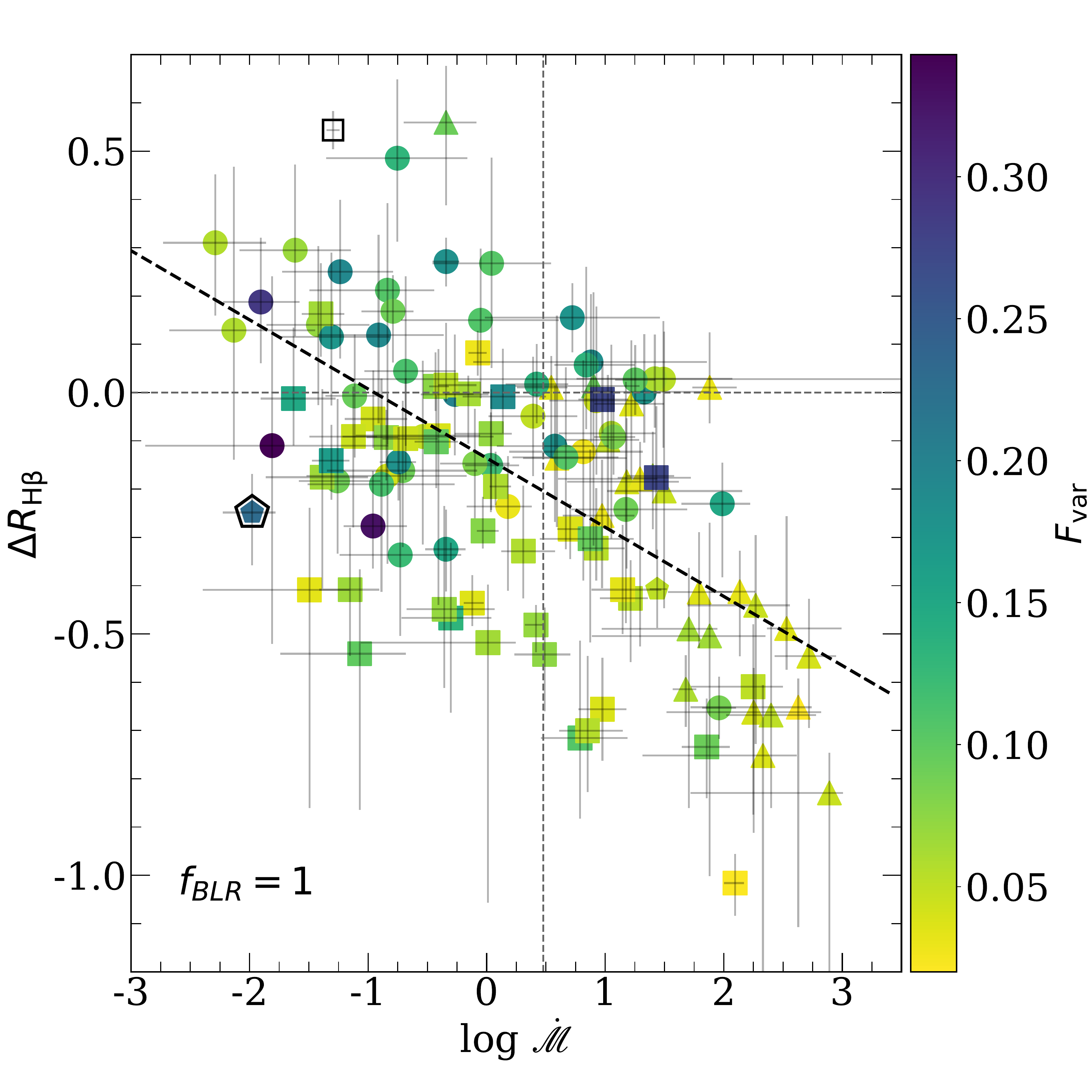}
\includegraphics[width=0.4\textwidth]{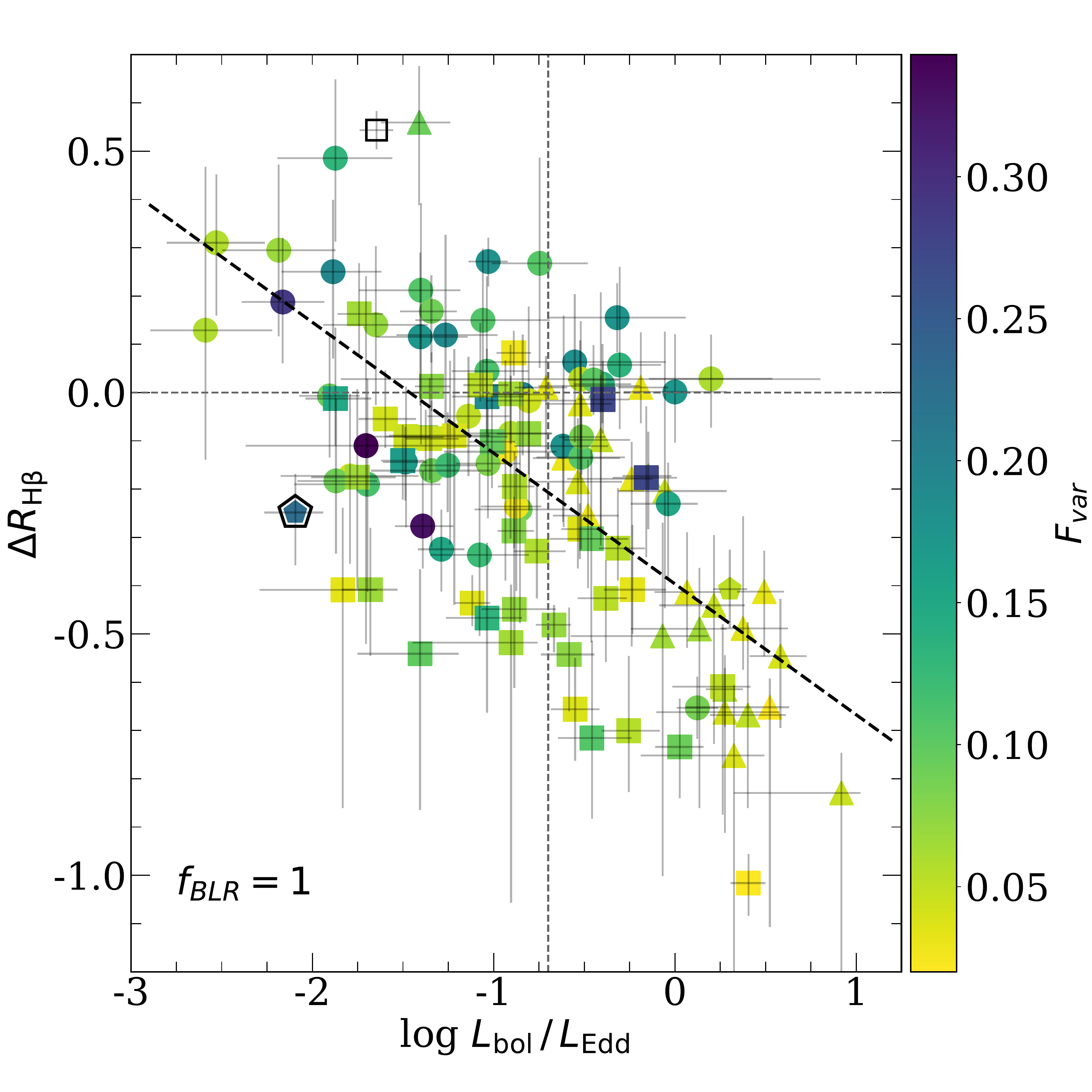}
\caption{Relation between \DRhb\ and \mdot\ (left, in log-space) and \LLEdd\ (right, in log-space) using \fblr=1. Symbols are the same as Figure~\ref{fig:RL_mdot}. Marker colors indicate the variation in \fvar\ {at 5100\AA}. {Open black square marker} corresponds to the quasar J142103 (See Section \ref{sec:fvar}). In both panels dashed horizontal correspond to \DRhb=0  and thick dashed black line show the best orthogonal linear fit. Vertical lines corresponds to \mdot=3 and \LLEdd=0.2.  \label{fig:DRhb_f1}}
\end{figure*}

\begin{figure*} 
\centering
\includegraphics[width=0.4\textwidth]{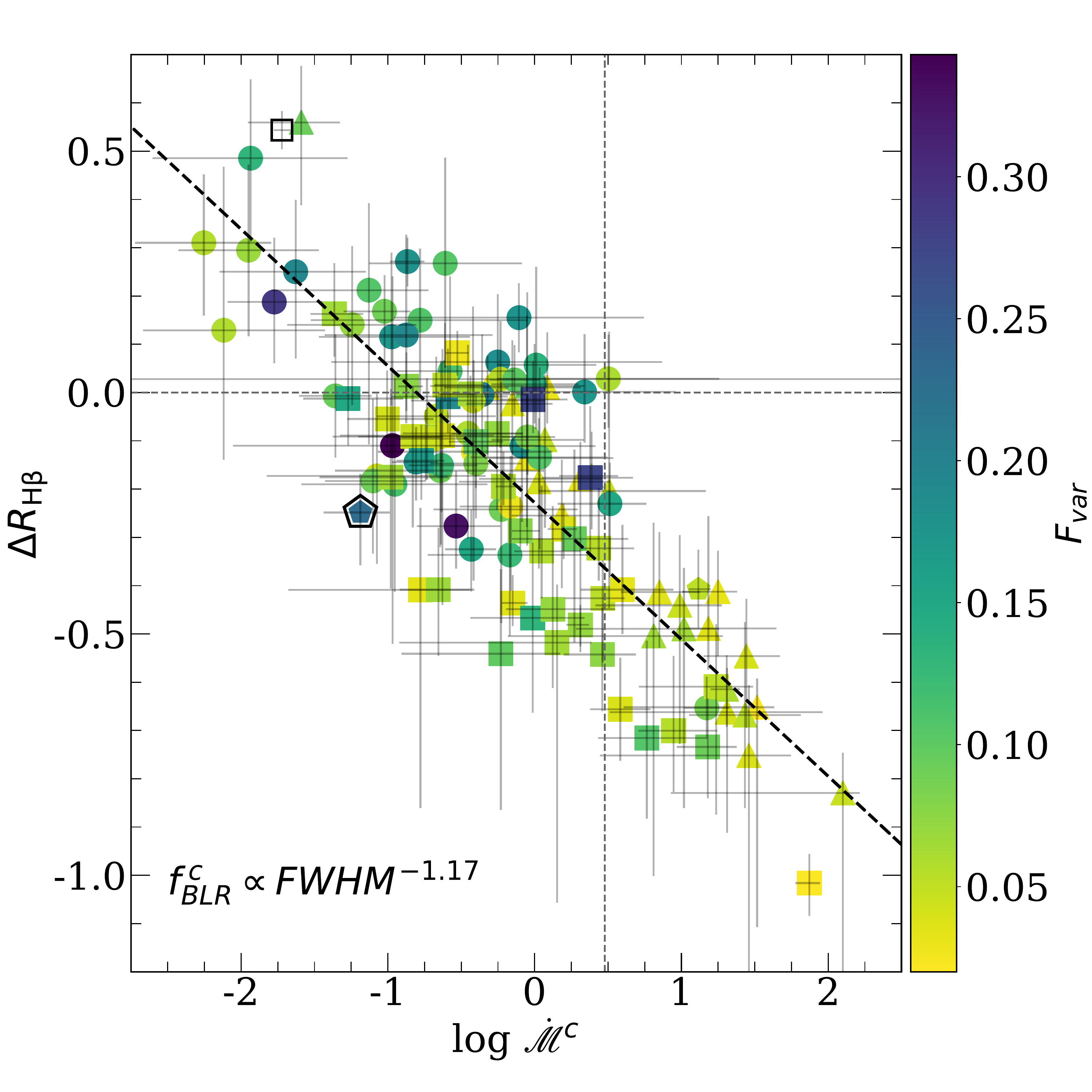}
\includegraphics[width=0.4\textwidth]{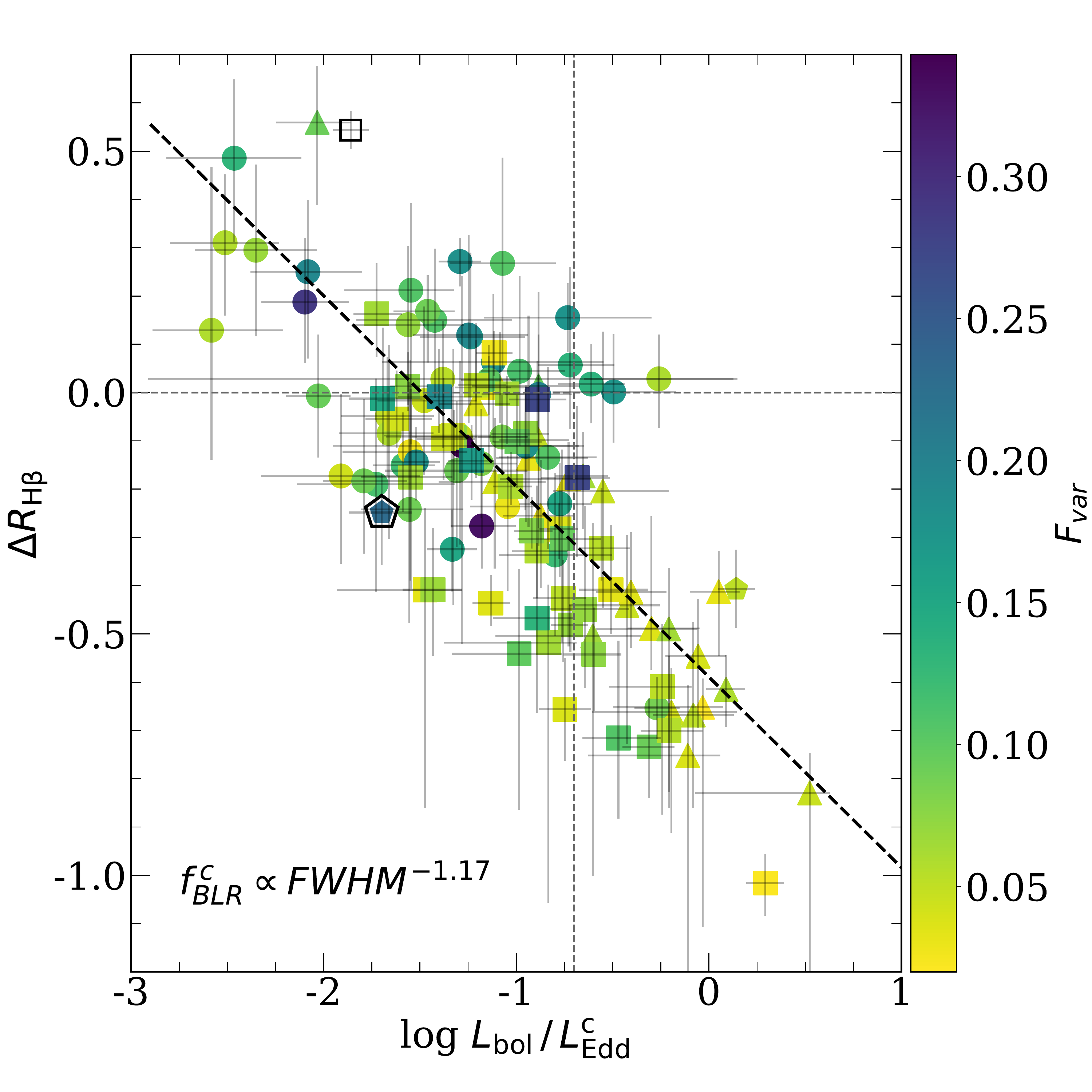}
\caption{Relation between \DRhb\ and \mdotc\ (left) and \LLEddc\ (right) using $f\mathrm{_{BLR}^{\,c}}\propto$ FWHM$^{-1.17}$. Description of markers, colors and lines are the same as in Figure~\ref{fig:DRhb_f1}. \label{fig:DRhb_ffwhm}}
\end{figure*}

SEAMBH team have done an important progress in understanding the reverberation mapping in highly accreting sources \citep{wangSEAMBH2014,du2014, hu2015, du2015,du2016,du2018}, {which have been scarcely included in previous RM samples.} They found that AGN with \mdot$\gtrsim3$ have time delay ($\tau\mathrm{_{obs}}$) shorter than expected from the $R\mathrm{_{H\beta}}-L_{5100}$ relation \citep{bentz2013}. The deviation can be estimated by the parameter, 
\begin{equation}
\Delta R_{\mathrm{H\beta}}=\mathrm{log}\,\left( \frac{R\mathrm{_{H\beta}}}{R\mathrm{_{H\beta_{\,R-L}}}}\right)=\mathrm{log}\,\left( \frac{\tau\mathrm{_{obs}}}{\tau\mathrm{_{{H\beta_{\,R-L}}}}}\right),  \label{equ:DR}
\end{equation}
where $\tau\mathrm{_{{H\beta_{\,R-L}}}}$ is the time delay corresponding to that  expected from the \RL\ relation for the given $L_{5100}$, see Equation~(\ref{equ:bentz}). Figure~\ref{fig:DRhb_f1} and \ref{fig:DRhb_ffwhm} show \DRhb\ as a function of \mdot\ and \LLEdd\ considering \fblr\ and $f\mathrm{_{BLR}^{\,c}}$. In all four cases, the largest \DRhb\ are associated with the highest \mdot\ and \LLEdd, the difference is the scatter along the relations. We perform an orthogonal linear fit in order to get the linear trend, and estimate the Pearson coefficient ($P$) and the root-mean-squared error ($rms$) to measure the correlation and dispersion along the trend line. {The general linear relation is given by:}
\begin{equation}
\Delta R\mathrm{_{H\beta, X}} = \alpha \, \mathrm{log}{\mathrm{X}} +  \beta,    \label{equ:linearfit}
\end{equation}
{where X corresponds to the accretion parameter (dimensionless accretion rate or Eddington ratio) using a virial factor equal 1 or the one anti-correlated with the FWHM. Coefficients of the fit ($\alpha$ and $\beta$), Pearson  and $rms$ values are given in the Table \ref{tab:linearfit}.}

The information given by the Pearson coefficients and $rms$ values indicates that the relations between \DRhb\ and accretion parameters are stronger when the virial factor anti-correlated with the FWHM is used, see Equation~(\ref{equ:linearfit}). We are able to propose a correction for the time delay based on the accretion parameters which recovers the expected value  from the $R_{H\beta}-L_{5100}$ relation. Since $P$ and $rms$ favor \mdotc, the correction used will be based on it and not on \LLEddc. The scattering of Eddington ratio could be higher due to the large uncertainties associated with the bolometric correction factor. 

\begin{figure} 
\centering
\includegraphics[width=0.5\textwidth]{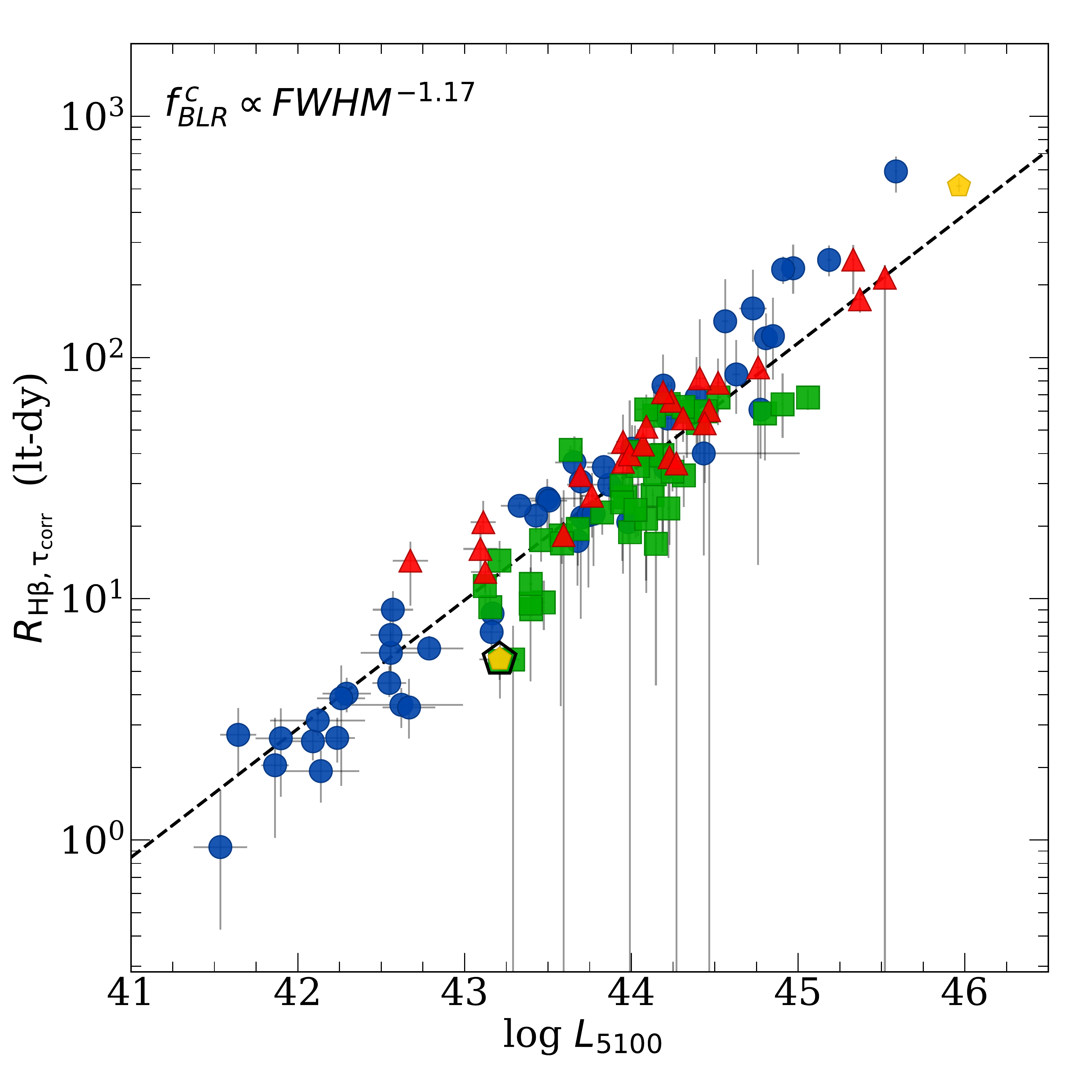}
\caption{\RL\ relation with time delay corrected by dimensionless accretion rate effect. Red triangles correspond to the SEAMBH sample, green squares correspond to the SSDS-RM sample, blue circles correspond to Bentz collection, and yellow pentagons mark the position of NGC 5548 and 3C 273. {Open black pentagon corresponds to NGC 5548.} Dashed black line corresponds to the expected \RL\ relation from \citet{bentz2013}. \label{fig:RL_corr}}
\end{figure}

Time delay corrected for the effect of the dimensionless accretion rate can be estimated by the relation,
\begin{align} \label{equ:taucorr}
\tau_{\mathrm{corr}}(\dot{\mathscr{M}}^{c})\,=\,10^{-\Delta R\mathrm{_{H\beta}( \dot{\mathscr{M}}\mathrm{^{c}}})} \, \cdot \tau_{\mathrm{obs}}.
\end{align}
The quantities \mdotc, $\Delta R\mathrm{_{H\beta} (\dot{\mathscr{M}}\mathrm{^{c})}}$ and $\tau_{\mathrm{corr}}(\dot{\mathscr{M}}\mathrm{^{c}})$ are listed in Table \ref{tab:mea_corr}. The \RL\ relation with the correction for the dimensionless accretion rate is shown in the Figure~\ref{fig:RL_corr}. If we compare this new diagram with the one shown in the left panel of Figure~\ref{fig:RL_mdot2}, it is clear that the scatter decreases significantly along the \RL\ relation, $\sigma\mathrm{_{obs}}=0.684$ vs. $\sigma\mathrm{_{corr}}=0.396$ in log space. With this correction, we are able to build a better luminosity distance--redshift relation or Hubble diagram and compare them with the standard cosmological models (see Section \ref{sec:cosmo}). 

\subsection{Is there a break in \mdot=3 for the \DRhb\ behavior?} \label{sec:break}

As we show in the previous Section, the relation between \DRhb\ and the dimensionless accretion rate (or Eddington ratio) can be represented by a linear fit, independently of the virial factor. However, \citet{du2015, du2018} argue that around \mdot=3 there is a break in the behavior of \DRhb, where highly accreting sources would be characterized by optically thick and geometrically slim accretion \citep{wangshielding2014}. Considering the deviation of the \RL\ relation, sources with \mdot$<$3 would be associated with a \DRhb$\sim$0, while sources with \mdot$>$3 would show a \DRhb, with a decreasing trend as a function of \mdot. This interpretation  should correspond to the relation \DRhb--\mdot\ assuming \fblr=1 (left panel of Figure~\ref{fig:DRhb_f1}), which was analyzed by \citet{du2015, du2018}. However, even in this case, there is a significant difference with our analysis which comes from the location of the SDSS-RM sample in the diagram. \citet{du2018} include this sample in their analysis, but they use the \mbh\ estimated from \citet{grier2017}, who adopt a \fblr=4.47 adequate for a \mbh\ estimation based on $\sigma _{\mathrm{line,rms}}$ \citep{woo2015}. In our estimation, some SDSS-RM objects are located around \mdot=3, covering the empty region shown in their Figure~3. In order to get a possible difference between the sources with \mdot$<3$ and $>3$, we perform a two-sample Kolmogorov-Smirnov (K-S) test. We find a value of 0.344 with a probability of $p\mathrm{_{KS}}\sim0.003$. A similar  $p\mathrm{_{KS}}$ is reported by \citet{du2015}, that according to them is enough to demonstrate that \DRhb$\sim 0$ for \mdot$<3$. If we apply the K-S test for \DRhb--\LLEdd\ with \fblr=1, we get a value 0.392 and  $p\mathrm{_{KS}}=0.0004$, which favors the difference at \mdot=3. On the other hand, if we consider a virial factor anti-correlated with the FWHM of \hb, \DRhb\ decreases progressively as function of \mdotc\ or \LLEddc\ (Figure~\ref{fig:DRhb_ffwhm}). It is confirmed by the K-S test, we get 0.771 with a $p\mathrm{_{KS}}=2.9\times10^{-11}$ and  0.663  with a $p\mathrm{_{KS}}=1.8\times10^{-9}$ for \mdotc\ and \LLEddc, respectively.

In all four cases, there is a different behavior around \mdot=3. For all cases, \DRhb\ median is 0.09 for \mdot$<3$ (or \LLEdd$<0.2$), however the scatter is higher with a virial factor equal to 1. For the high accretion rate, \DRhb\  median is 0.26 and 0.31 for \mdot\ and \LLEdd, and 0.61 and 0.50 for \mdotc\ and \LLEddc, respectively. However, from Figure~\ref{fig:RL_mdot2} we see that very low accretion objects have a \DRhb$>0$ for a virial factor anti-correlated with the FWHM. As we have shown in the previous Section, we can represent the relation between \DRhb\ and accretion parameters by a linear fit in the whole log space, without any break, which is supported by the Pearson coefficient and $rms$ value. We do not see any evidence of a break around \mdot=3, specially in the cases with a virial factor anti-correlated with the FWHM of \hb. 

The virial factor is one of the most important uncertainties in the black hole mass determination and accretion parameters {as it is emphasized in the Section \ref{sec:virial_factor} and \ref{sec:inclination_angle}}. According to \citet{mejia-restrepo2018}, the virial factor \fblrc\ includes the correction for the orientation effect, then \mdotc\ would show an accurate estimation which is reflected in the scattering of the measurement, particularly when it is compared with the other parameters. However, their results are based on Shakura--Sunyaev (SS) disk \citep{shakura1973}, which is appropriate for their sample, where $\sim90\%$ of the objects show a low accretion rate and they can be perfectly presented by a SS model disk \citep{capellupo2016}. One-third of our sample shows a high accretion rate (\mdot $\gtrsim$3) and according to \citet{wangshielding2014}, in high accretors the slim disk produces an anisotropic radiation field, which divides the BLR into two regions with distinct incident ionizing photon fluxes. Using the code BRAINS, \citet{li2018} demonstrated that two region model is better than a simple one for Mrk 142, a typical high accretor source. Therefore, the incident radiation flux and geometry are more complex in highly accreting AGN. New spectral energy distributions (SED) models for slim disk are required, in order to get a better estimation of the virial factor for this kind of objects.

\subsection{Relation between \fvar\ and accretion parameters}

Studies of large quasar samples showed that continuum amplitude of the variability is anti-correlated with Eddington ratio \citep{wilhite2008, macleod2010, simm2016, rakshit2017,sanchez-saenz2018, li2018}. \fvar\ measures the excess of variability above the noise level and it can be used as an estimator of this effect. As we showed in the previous Section, \DRhb\  is anti-correlated with the dimensionless accretion rate and Eddington ratio. In Figure~\ref{fig:DRhb_f1} and \ref{fig:DRhb_ffwhm} we show the change of \fvar\ along the relation \DRhb--\mdot\ and \DRhb-- \LLEdd\ using different virial factors. In all the cases, the minimum \fvar\ at 5100 \AA\ values tend to be associated with the highest \mdot\ (and \LLEdd) and the smallest \DRhb\ values.

\citet{sanchez-saenz2018} report that the amplitude of variability $A$ is strongly related with $\sigma_{rms}$ \citep{sanchez2017}, which is the square of  \fvar. Estimating $\sigma_{rms}$, we compute the Spearman coefficient ($\rho {_{s}}$) in order to confirm the relation between the accretion parameters and variability. The stronger relation is given by \mdot\ and \mdotc, with a $\rho {_{s}}=-0.397$ ($p=7.9\times10^{-6}$) and $\rho {_{s}}=-0.374$ ($p=1.6\times10^{-5}$), respectively. Spearman coefficients for \LLEdd\ and \LLEddc\ are -0.341 ($p=1.8\times10^{-4}$) and -0.259 ($p=1.8\times10^{-3}$), respectively. \citet{sanchez-saenz2018} report a Spearman coefficient for \LLEdd\ of $\rho {_{s}}=-0.22$ ($p=1\times10^{-8}$), which is comparable to the value found by us. Dimensionless accretion rate seems to be  more strongly related with the other physical parameters (e.g. \fvar\ and \DRhb) than the Eddington ratio, which is probably linked with the bolometric luminosity uncertainties. 

According to \citet{allevato2013}, \fvar\ is strongly affected by biases, for example in the structure of variability or length of the light curve, therefore it has to be treated with caution. In our sample, some of the 
SDSS-RM object were observed in the red--edge of the spectrum, where the telluric lines are difficult to remove, and considering the short cadence of the light curve, it could also affect the relation. However, under the proper observational conditions, \fvar\ could be another of the possible variability parameters, giving information about the physical properties of the AGN. Surveys such as the Large Synoptic Survey Telescope \citep{lsst2019} will be able to estimate this parameter for a large quasar sample, and following the relations like those presented in this work, this can provide information about the accretion disk structure, accretion process and the size of the BLR. A multivariate analysis is needed to get a correct relation between the \fvar\ (or $\sigma_{rms}$) and the dimensionless accretion rate,  which is out of the scope of this work. 


\section{Hubble diagram}
\label{sec:cosmo}
\begin{figure*} 
\centering
\includegraphics[width=0.4\textwidth]{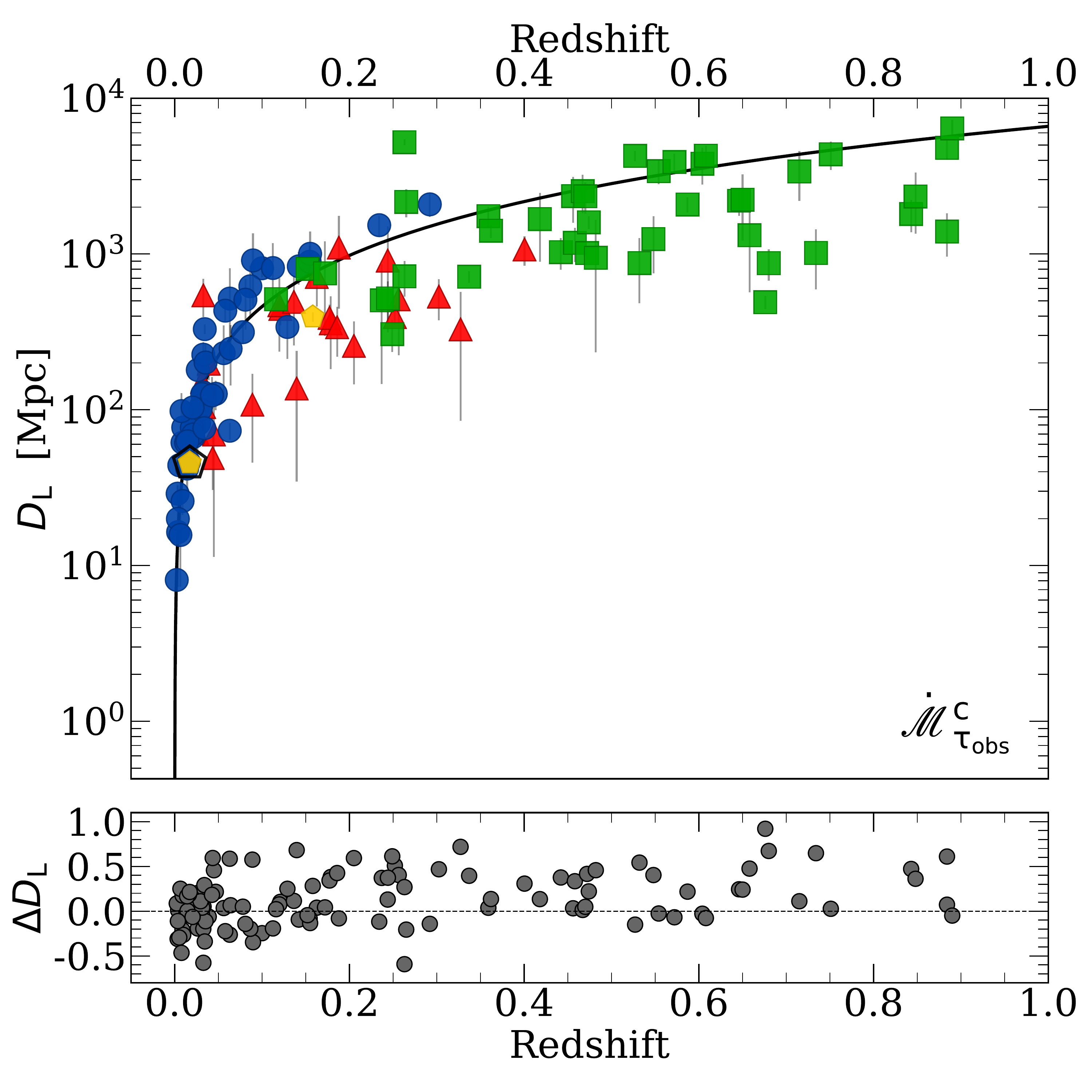}
\includegraphics[width=0.4\textwidth]{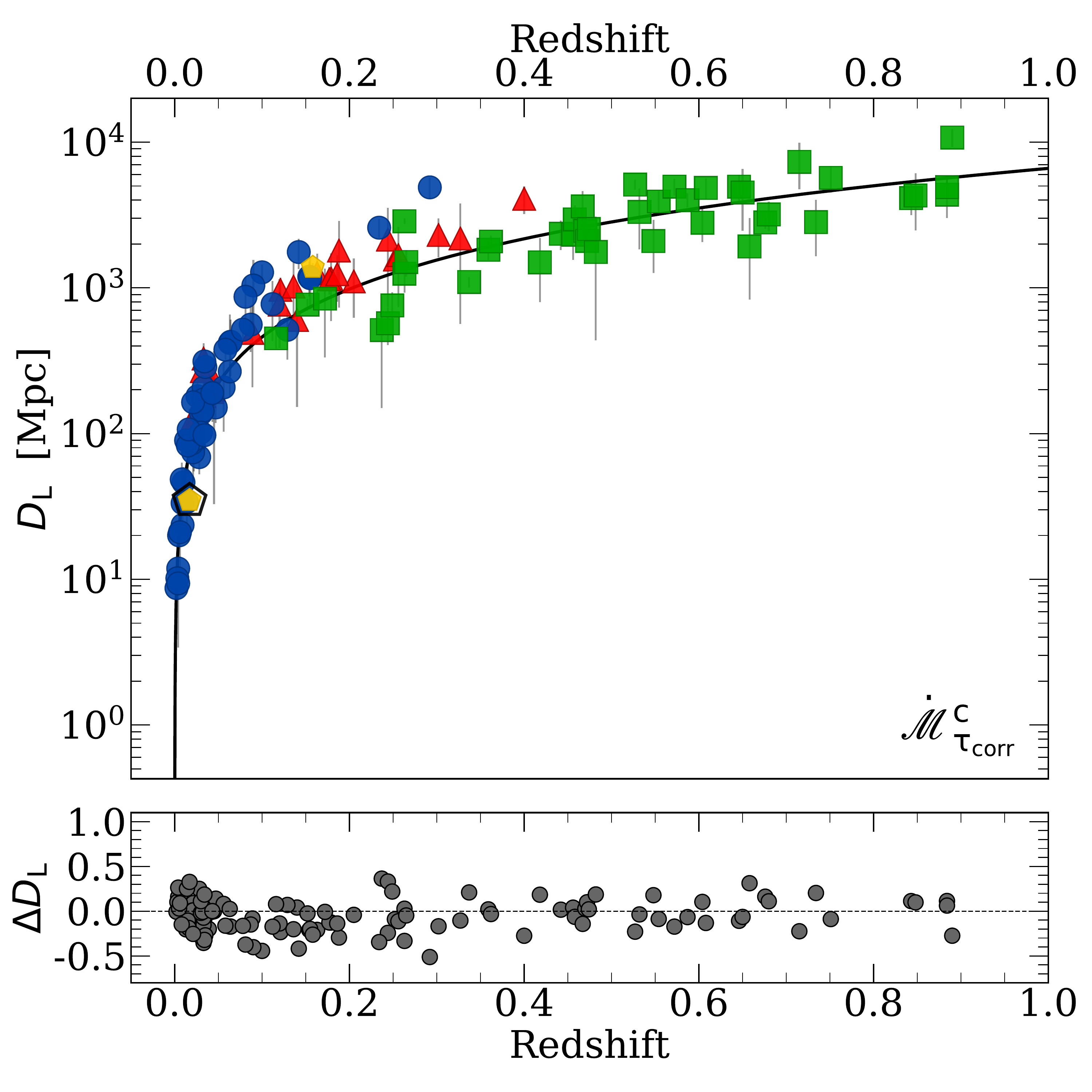}
\caption{Hubble diagram before (left) and after (right) the correction by dimensionless accretion rate. Markers and colors are the same as in Figure~\ref{fig:RL_corr}. The black lines indicates the expected luminosity distance based on the standard $\Lambda$CDM model. In both cases, the bottom panel shows the difference between the expected luminosity distance and the observed one. \label{fig:hubble} }
\end{figure*}

We now perform a simple test of the prospects for quasar application for cosmology by locating the sources on the Hubble diagram. Now we only use their measured time delay ($\tau\mathrm{_{obs}}$ and $\tau\mathrm{_{corr}}$) and the observed flux at 5100 \AA\ rest--frame ($F_{5100}$). We use Equation~(\ref{equ:bentz}) to {\it determine} $L_{5100}$, and finally to measure the luminosity distance, $D\mathrm{_L}$,
\begin{equation}
D\mathrm{_L}= \left(\frac{ L_{5100}}{ 4\,\pi\,F_{5100} }\right)^{1/2}.
\end{equation}
Left panel of Figure~\ref{fig:hubble} shows the luminosity distance estimated with $\tau\mathrm{_{obs}}$, which exhibits large scatter.  We now take all objects from our sample plotted in Figure~\ref{fig:RL_corr} and assume that they follow precisely the \RL\ relation while we relax the assumption that their absolute luminosity $L_{5100}$ is known. Repeating the same exercise, we find that the scatter is strongly reduced (see right panel of Figure~\ref{fig:hubble}). This is better shown in the corresponding bottom panels where we plot the residuals between the logarithm of the expected ($D{\mathrm{_{L,{mod}}}}$) and estimated luminosity distance ($D\mathrm{_{L}}$). 

It is clear that the dispersion decreases after the correction, which is supported by the $rms$ value (0.287 vs. 0.182). {Here plotted uncertainties come only from the uncertainty in the time delay measurement }. As a guide, in both panels we plot the standard $\Lambda$CDM model with the parameters \citep{planck2013}: $H_0=67$ km s$^{-1}$ Mpc$^{-1}$, $\Omega_{\Lambda}=0.68$, $\Omega_m=0.32$  (continuous black line Figure~\ref{fig:hubble}). We see that while in the left panel a significant fraction of points implied too small distances, after the accretion-rate dependent correction the points are distributed close to the line.



\begin{figure} 
\centering
\includegraphics[width=0.45\textwidth]{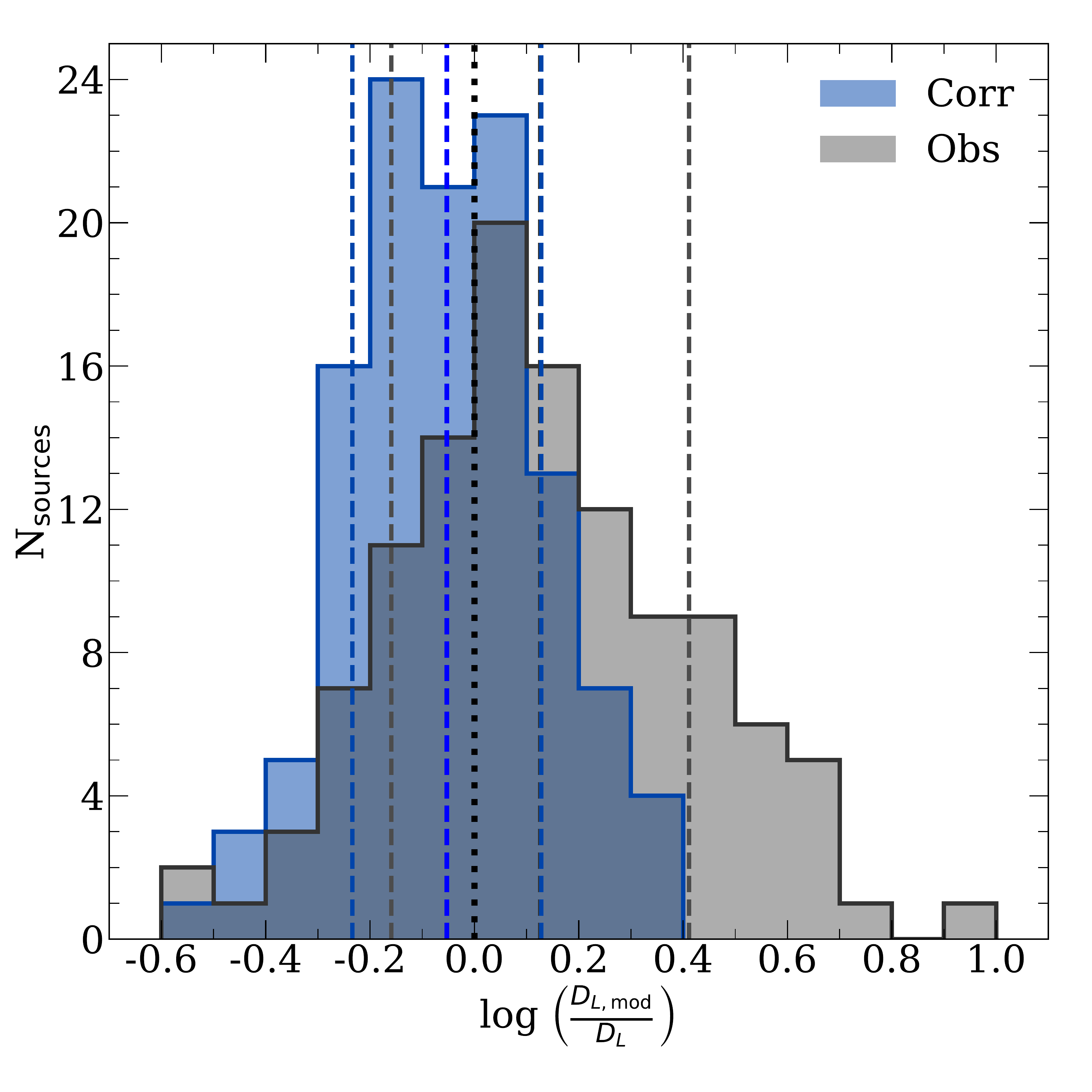}
\caption{Distribution of the difference between the logarithm of the expected luminosity distance (log $\frac{{D_\mathrm{L,mod}}}{D\mathrm{_L}}$) respect to that obtained from the observations before (gray) and after correction (blue). Vertical dashed gray and blue lines correspond to the average (thick), and $\pm 1 \sigma$ values (thin) before and after correction, respectively. Vertical dotted black line marks  log$\frac{D\mathrm{_{L,mod}}}{D\mathrm{_L}}$=0. \label{fig:scatt_LD}}
\end{figure}

To demonstrate this quantitatively, in Figure~\ref{fig:scatt_LD} we give the distribution of the log  $\frac{D\mathrm{_{L,mod}}}{D\mathrm{_{L}}}$. The distribution before the correction shows clear asymmetry to the right, and it is centered at 0.123. After applying the correction, the distribution is centered at 0.055. The standard deviation also shows an improvement (0.19 vs. 0.31). With this information we can think of constraining  the cosmological parameters.

\begin{figure*} 
\centering
\includegraphics[width=0.49\textwidth]{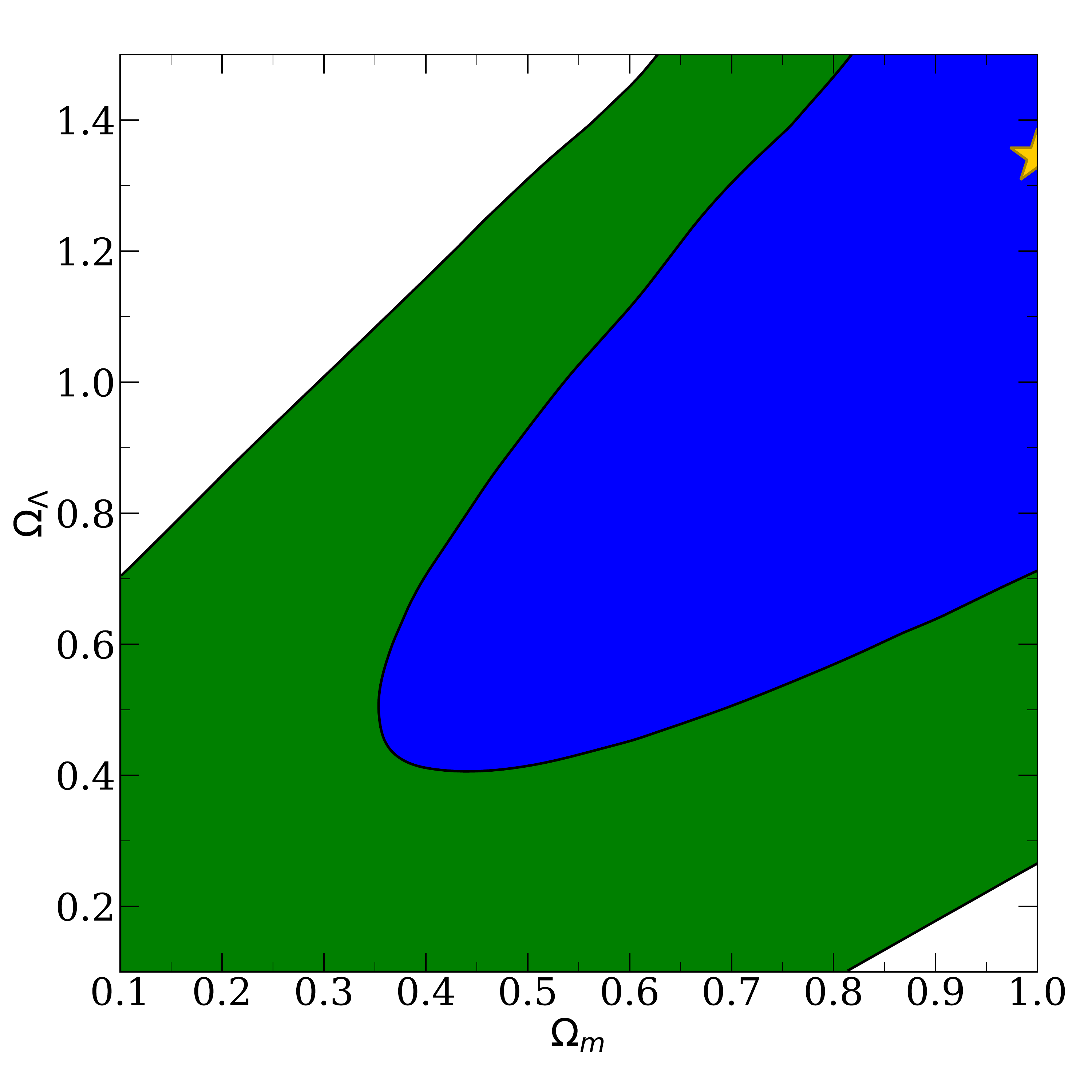}
\includegraphics[width=0.49\textwidth]{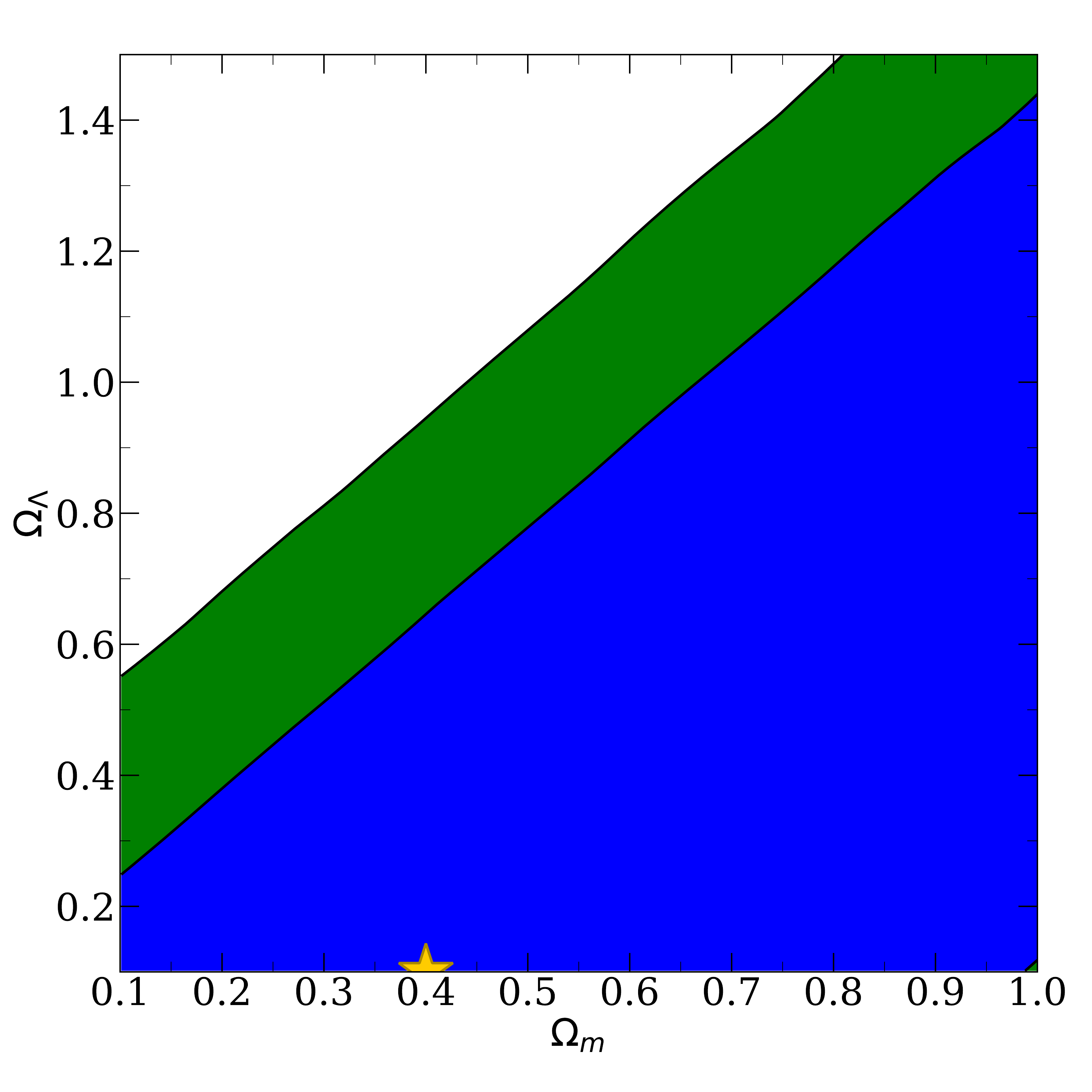}
\caption{$\chi^2$ behavior in the $\Omega_m-\Omega_\Lambda$ space for the full sample (left panel) and for the selected sources with only $z>0.4$ (right panel). In both panels blue and green contours correspond to 1$\sigma$ and 2$\sigma$ confidence levels, respectively. {Yellow symbol} marks the minimum $\chi^2$ value, in both cases at the edge of the domain. \label{fig:cosmo_par}}
\end{figure*}

In order to illustrate better the accuracy of determination of the cosmological parameters, we performed formal computations of the best cosmological model. We assumed a standard $\Lambda$CDM model, and the value of the Hubble constant, $H_0=67$ km s$^{-1}$ Mpc$^{-1}$, is used in the computations of corrections related to the accretion rate (see Section~\ref{sec:desviation}). We did not modify the parameters in Equation~(\ref{equ:bentz}) relating the time delay and the absolute luminosity. Then we searched for the minimum of the function
\begin{equation}
    \chi^2 = \sum^{N}_{i=1} \frac{ (\mathrm{log}(D_\mathrm{L,mod}^{\,i}) - \mathrm{log}(D_\mathrm{L,obs}^{\,i}))^2} {(\mathrm{log}(1 + b_ i)^2 + \sigma^2)},
\end{equation}
where $N$ is the total number of sources in the sample, and $b_i$ is the relative error in the luminosity distance determination implied by uncertainty in the measured time delay. We introduced a quantity $\sigma$ following the approach by \citet{risaliti2015}. This quantity describes the dispersion in the sample, which is larger than the claimed measurement errors. The result is shown in Figure~\ref{fig:cosmo_par} (left panel). The best fit is on the edge of the domain (minimum $\chi^2$), however it is  consistent with the standard cosmology, and the accepted parameter values are consistent with our results within 2$\sigma$. The errors are still large, due to the limited number of objects (117 in the full sample), and possibly also to heterogenic way of data reduction--time delays were measured by various authors, using different methods. We checked if we can improve the constraints by using only higher--redshift sources $z>$0.4 (30 sources, right panel Figure~\ref{fig:cosmo_par}), but this caused the large shift of the best fit (from $\Omega_m = 1.0, \Omega_{\Lambda} = 1.34$ to $\Omega_m = 0.4, \Omega_{\Lambda} = 0.1$) and the errors contour were then even larger due to the reduced number of objects.

The effective use of the method requires an increase of the number of reverberation-measured objects by a factor of at least 5. There are some prospects for new time delay measurements, particularly at larger redshifts. First, the SDSS-RM campaign continues. Second, the project of monitoring 500 quasars with Oz-DES survey \citep{king2015} is likely to finish soon. Thus, the method itself looks quite promising.

\section{Discussion}
\label{sec:discussion}

{One of the most critical points of reverberation mapping technique is the uncertainty around the black hole mass determination, which depends on parameter like  virial factor or  inclination angle. The method proposed in this paper is strongly dependent on these two parameters and a change in them affects the approach to correct the time delay by the accretion rate effect. All these problems are reflected in the large uncertainties associated with the cosmological parameters determination (see Section  \ref{sec:cosmo}). }

{Over the years, type 1a supernovae have been also corrected by uncertainties in some physical properties like mass, peculiar velocity, redshift, etc. Sophisticated statistical methods \citep[e.g.][]{scolnic2018} has been applied in order to decrease the uncertainty of these parameters, resulting in accurate cosmological estimations. In future, these kind of methods must be implemented for quasars as well.} 

{We discuss the most critical points of virial factor and inclination angle, which would be resolved with the arrival of new data, instruments and improvements in the methods in the following.}

\subsection{{Virial factor remarks}}
\label{sec:virial_factor}

{It is clear that the selection of the virial factor anti-correlated with the FWHM of the line has an implication over the correction proposed for the observed time delay. As we previously argued, among both virial factors analyzed in this work, the best one is the one proposed by \citet{mejia-restrepo2018}. It is not the first time that a virial factor anti-correlated with the FWHM is proposed \citep[e.g.][]{collin2006, strochi-bergmann2017}. However, there are some caveats to consider under this assumption.}

{\citet{mejia-restrepo2018} sample includes 37 AGN at $z\sim1.5$, with  1,600$<$FWHM$<$10,100 \kms\ and a median value of $\sim$4700 \kms. Only 16$\%$ of the sample shows  FWHM$<2,000$ \kms. On the other hand, our sample includes sources with  780$<$FWHM$<$10,400 \kms, with a median value of 3,000 \kms, where $\sim31\%$ of the sample show FWHM$<2,000$ \kms. It indicates that our sample has an over representation of narrow profiles compared with the ones shown by  \citet{mejia-restrepo2018} sample. The narrow FWHM regime is populated mainly by the super-Eddington sources (Du sample), which are relative new in the AGN analysis. The scarce represented narrow profiles in \citet{mejia-restrepo2018} analysis has a direct effect over the exponent in the anti-correlation with the FWHM.  Recently,  \citet{yu2019} perform a new analysis following the formalism proposed by \citet{collin2006} in order to estimate the virial factor and \mbh. Their sample includes  $26\%$ of narrow profiles and find an anti-correlation given by \fblr$\propto$ FWHM$^{-1.11}$, very close to the one proposed by Mej\'ia-Restrepo. This confirms the anti-correlation between the virial factor and the FWHM of the line, and suggests that variation in the exponent is not so large when the number of narrow profiles increases.  However, it still has to be tested. }

{\citet{woo2015} calibrate the virial factor using the relation \mbh--$\sigma^*$ for 93 Narrow-Line Seyfert 1 galaxies (NLS1) and 29 RM AGN (where only one-third of the sample is NLS1), obtaining a value of  \fblr$\sim$1.12. This virial factor  is basically the same as the one used to estimate \mbh\ and \mdot\ in Figure~\ref{fig:RL_mdot}
and \ref{fig:DRhb_f1} respectively, which show a large scatter with respect to the one estimated from the \fblrc. The large scatter around them could be related with an effect of the inclination angle, that according to \citet{peterson2004} affects strongly the FWHM measurements.} 


{We performed a test to estimate the average \fblr\ in our sample. We independently estimate the \mbh\ from the \mbh -- $\sigma_{*}$ relation \citep{gultekin2009} for 15 sources from the Bentz Collection and for 25 sources from \citet{grier2017}, which have stellar velocity dispersion measurements. We consider the centroid of the $\sigma_{*}$ distribution obtained for this sub-sample i.e., $\log\sigma = 2.142 \pm 0.173$. Then, using the Equation~3 in \citet{gultekin2009}, we derive the average \mbh\ . We use the mean of the $\tau\mathrm{_{obs}}$ measurements for this sub-sample and the \mbh\ as obtained from the previous step, and substitute in the Equation~(\ref{equ:mass}) to get an average \fblr -- FWHM relation i.e.,  $\log \overline{f_\mathrm{BLR}} \approx 6.667 - 2$ log FWHM. The $\tau\mathrm{_{obs}}$ measurements are taken from Table \ref{tab:samples}.} 

{Considering the centroid of the FWHM distribution for this sub-sample i.e. $\sim$ 4000 \kms\ , we get an \fblr\ that is substantially smaller i.e. $\sim 0.29$. We estimate the \fblr\ for a representative FWHM = 2000 \kms\ , which is $\sim 1.16$, while it is relatively higher for cases with lower FWHM (e.g. Mrk493 has a FWHM $\approx$ 778 \kms\ which would give an \fblr $\sim 7.67$). This exercise already indicates the importance of a variable \fblr\ and how it can affect the \mbh\ estimates that are derived using the virial relation.  The \fblr\ depends on the inclination angle as well as on the gas distribution. A case-by-case study is needed to estimate the $f_\mathrm{BLR}$ since this is quite evident from this study as well as many previous works \citep{collin2006,panda2019b} that a constant, singular value of \fblr\ can not explain and should not to be assumed for all sources.}

\subsection{{Inclination angle remarks}}
\label{sec:inclination_angle}

{One the most important parameters included in the virial factor is the inclination angle. In this work, we are assuming an angle of $\theta\sim40^\circ$ for the dimensionless accretion rate estimation (Equation~\ref{equ:mdot}), which is a typical value for the type 1 AGN. However, this generalization can overestimate \mdot\ value in some objects. }

{\citet{strochi-bergmann2017} estimate and collect the inclination angle for a sample of double-peak H$\alpha$ AGN, where seven sources of our sample are included. These objects are marked in the last column of Table \ref{tab:samples}. They estimate a range for the inclination angle of $17<\theta<38 \, ^\circ$, with an average value of $\theta \sim 27^\circ$. This implies a variation in \mbh\ and accretion rate estimations by an order of 3 and
2 respectively, in seven sources of our sample.}

{According to the model assumed by \citet{strochi-bergmann2017}, the inclination angle is correlated with FWHM and anti-correlated with the virial factor.  It has been suggested by \citet{collin2006} as well. It means that narrow profiles correspond to low inclination angles and high virial factors. Since one-third of our sample shows FWHM$<2000$ \kms, this suggests that $30\%$  of the sample should have angles  $\theta<40^\circ$. }

{Recently, \citet{negrete2018} constrain the spectroscopic behavior of the super-Eddington sources in the optical regime, which show FWHM$<4000$ \kms\ and inclination angles of $\theta\sim20^\circ$.  Following the formalism proposed by \citet{marzianisulentic2014}, Negrete et al. argue that these sources can be used as standard candles. In order to decrease the scatter in the Hubble diagram, they proposed a correction for the inclination affects based on the virial luminosity (estimated from the FWHM) and cosmological distance (estimated by the redshift of the source). With this correction, they obtain a correction in the luminosity distance of 0.02-0.08 mag, which is promising. However, this formalism is only valid for the sources radiating close to the Eddington limit and cannot apply to the rest of the AGN populations. }

{One clear example of the uncertainties in the virial factor and inclination angle is coming from the novel results provided by \citet{gravity2018}. They estimate an angle for 3C 273 of $\theta \sim 12^\circ$. It is a high-Eddington source, therefore the angle estimated is in concordance with the one estimated for this kind of populations by \citet{negrete2018}. On the other hand, using the Equation~2 of \citet{mejia-restrepo2018} the virial factor is 4.6 considering a ratio H/R=0.1, which is almost a factor 3 larger than that used for the estimation reported in this paper (\fblrc$\sim$1.45). Additionally, Gravity results favor very thick BLR (opening angle of the torus is 45$^\circ$, implying H/R = 1.0), well outside the range favored statistically by \citet{mejia-restrepo2018}.}

{Other example is NGC 5548, which has been monitored by for almost 40 years, showing a change from Seyfert 1 to Seyfert 1.8. The FWHM of \hb\ varied from 4,000\kms\ to 10,000\kms. Assuming that black hole mass and inclination angle have a slow variation, the big changes in the source can  be attributed to the accretion rate \citep{bon2018}. However, they are not so significant to change from sub to super-Eddington regime. In general, NGC 5548 is not following the \RL\ relation, it shows a steeper slope than 0.5 \citep{peterson2004}, but within the uncertainties of the relation. \citet{pancoast2014} performed a dynamical modelling of the BLR in NGC 5548 which does not require a virial factor. They assume an inclination angle $\theta=38.8^\circ$ and find \mbh=$3.39^{+2.87}_{-1.49}\times10^7$ M$_{\odot}$. This result is in agreement with the one obtained from reverberation mapping using the dispersion of the line in rms spectrum and a virial factor of 5.5. It supports the analysis which indicate that  $\sigma_\mathrm{{line,rms}}$ is less affected by inclination than FWHM \citep{peterson2004, collin2006}. }


{Determination the best angle for the presented sample is complicated due to the diversity of properties observed in the sample. The variable FWHM seems to be the best option, as it apparently includes the variation of the inclination angle. However more analysis with homogeneous samples are still needed in order to clarify the convolution between virial factor and inclination angle and the related uncertainties.}


\subsection{{Luminosity distance remarks}}

{The estimation of the distance to the astronomical sources is fundamental for cosmology. One of the most important results of the reverberation mapping technique is an independent estimation of the luminosity distance. As we show, the \RL\ relation is not followed by the super-Eddington sources and a correction is required (Equation~\ref{equ:taucorr}). However, as it is  emphasized in Sections \ref{sec:virial_factor} and \ref{sec:inclination_angle}, this correction is strongly affected by  uncertainties in the virial factor and  inclination angle. All these uncertainties are reflected in the estimation of the luminosity distance and in the determination of the cosmological constants. Water masers or torus diameter estimation could provide accurate measurements of the luminosity distance \citep{humphreys2013, honing2014}, however the scarcity of water masers and the required long time for the monitoring limit the feasibility of these methods. }

{The luminosity distance has been estimated by remarkable methods in two of our sources: NGC4151 and 3C  273.  Since the dusty torus is larger by a factor 4 than the BLR, it is relatively easily resolved it by the optical long-baseline interferometers. On other hand, due to  dust response to the continuum variations, the distance from the central continuum to the dusty torus can be estimated by the reverberation mapping method. Combining these two techniques,  
 \citet{honing2014} estimated a distance to NGC4151, $D_{\mathrm{L}}$ = $19.0^{+2.4}_{-2.6}$ Mpc, which is in agreement with the one reported by \citet{tsvetkov2018} using SNe. They also report a virial factor value of $5.2<f\mathrm{^{dust}_{BLR}}<6.5$. Before and after application of the correction related to accretion rate, we obtain a luminosity distance of 12.9$\pm$1.5 Mpc and 10.2$\pm$1.5 Mpc, respectively. It indicates an underestimation in the luminosity distance by 0.27 dex compared to the dust-parallax method. This difference is related with the large uncertainty associated with the black hole mass. Combining reverberation-mapped results from optical and ultraviolet emission lines, \citet{bentz2006} found a weighted mean \mbh\ for NGC4151  of $4.57^{+0.57}_{-0.47}\times10^{7} \, M_\odot$, which compared to the results of the \mbh--$\sigma^*$ relation is underestimated by a factor of 7, indicating a big uncertainty in the reverberation-based masses determination. }
 
{3C 273 is the first quasar where the linear and angular size of the broad line region have been measured \citep{gravity2018, zhang2018}. Both results provide a size of the BLR of $\sim$145 ltd. Recently, \citet{wang2019} joined both techniques in order to determine the cosmological distance. The novelty of the method is the distance determination without invoking otherwise calibrations through known cosmic ladders. They determine a distance of 551.5$^{+97.3}_{-78.7}$ Mpc within $15\%$ of the average accuracy. Before and after application of the correction related to accretion rate, we obtain a distance of 394 Mpc and 1381 Mpc respectively, an order of 0.4 dex.}

{The information provided by NGC4151 and 3C 273 is an indication of the uncertainties associated with the present method and the limitation for their use in cosmology. More observations are require in order to test and improve the correction based on the accretion parameters. The arrival of novel and  sophisticated results, like \citet{gravity2018}, will  provide information that help us to improve the use of quasar in cosmology and also determine the origin of the uncertainties associated with the classical methods like reverberation mapping.}

\section{Conclusions}
\label{sec:conclusions}

In this work, we confirmed that the time delay measured during the reverberation mapping campaigns is affected by the accretion rate of individual sources. Considering the deviation from the \RL\ relation and the corresponding accretion rate, we propose a correction based on \DRhb, which is a power--law function of the dimensionless accretion rate parameter.  This correction recovers the expected time delay, decreasing the scatter and providing a proper estimation of the BLR size. Using the corrected values, we built the Hubble diagram, obtaining consistent results with the $\Lambda$CDM model within 2$\sigma$ confidence level. However, uncertainties are still large, which could be mitigated by significantly increasing the number of sources, especially towards larger redshifts.  

We used the dimensionless accretion rate and the Eddington ratio to estimate the effect of the accretion rate. The former one appears to show a better correlation with other physical quantities, which could be related by large uncertainties associated with the bolometric luminosity. We also explore the use of a virial factor anti-correlated with the FWHM of \hb\ line { \citep{mejia-restrepo2018}, which is in agreement with the one found by \citet{yu2019}}. The black hole mass and the accretion parameters give a lower dispersion in the comparison with the case \fblr=1. Since the virial factor anti-correlated with the FWHM includes the correction for the source orientation, it can explained the lower scatter. However, the virial factor used needs a more detailed analysis to confirm the use of this value, especially for the highly accreting objects.

In addition, we also confirm the anti-correlation between the continuum variability and the accretion parameters, using the parameter \fvar. Although \fvar\ is sensible to the monitoring conditions and the quality of observations, it shows the same behavior as similar parameters, such as the variability amplitude. Large surveys such as LSST will observe this kind of properties and it is important to establish the relation with other physical parameters. This result supports that the accretion parameter (the Eddington ratio or the dimensionless accretion rate) is the main driver of many of the quasar properties \citep{marziani2001, shenandho2014} such as variability or outflows. It could be linked to a different type of the  accretion disk structure \citep{wangshielding2014}. 

{There are large uncertainties associated with the proposed method in this work. The determination of the virial factor and inclination angle is essential for the black hole mass determination and accretion rate and it is still an open problem in the AGN field. The necessity to include the broad range of AGN properties (like FWHM, accretion rate, orientation angle, redshift) in the modeling of these key parameters is still is a pending task. All these uncertainties are also reflected in the luminosity distance. A comparison with independent methods for two sources show a difference by 0.27 and 0.4 dex. These uncertainties can not be immediately diminished, but probably with the arrival of new data and novel and sophisticated techniques can permit us to calibrate the results from reverberation mapping, and we can improve their use for cosmology purposes. }

\section*{Acknowledgements}
{We thank the referee for valuable suggestions that helped to improve the paper.} The project was partially supported by National Science Centre, Poland, grant No. 2017/26/A/ST9/00756 (Maestro 9), and by the MNiSW grant DIR/WK/2018/12. VK acknowledges Czech Science Foundation No. 17-16287S. We acknowledges P. Du, Y.-R. Li and J.-M. Wang for the provided data and comments that improved the paper. We acknowledge  M. Bentz, C. J. Grier, J. Kuraszkiewicz,  A. del Olmo, M. Fausnaugh, J. Mej\'ia--Restrepo and Y. Shen for the extra information required for the development of this work.

\appendix
\section{Monte Carlo simulations of the SDSS-RM setup}
\label{sect:simulation}
Monitoring presented in \citet{grier2017} is relatively short, taking into account rather high redshifts and luminosities
of the sources. Therefore, we performed Monte Carlo simulations with the aim to check independently whether the observational setup forces the measured  time delays to be shorter than otherwise expected for a given source redshift and luminosity.

For this purpose, we simulated each source independently, taking into account its monochromatic luminosity, $L_{5100}$, and its redshift. For each source we first created 100 artificial dense light curves using \citet{timmer1995} algorithm.
We model the overall power density spectrum (PDS) shape assuming a power law shape with two breaks and three slopes. The high frequency slope was taken as $2.5$ following Kepler results for quasars from \citet{mushotzky2011}. The high frequency break was set at 200 days for all sources, as a mean value taken from \citet{simm2016}. The slope of the middle part of PDS was taken as $s_1 = 1.2$ \citep{czerny1999,simm2016}.  The low frequency break was fixed somewhat arbitrarily at the value ten times lower than the high frequency break, and the low frequency slope was assumed to be zero. 
We then exponentiated the resulting light curve, following \citet{uttley2005}, since this reproduces the log-normal distribution characteristic for the light curves of accreting sources.

The light curve describing the emission line was generated from the dense light curve above describing the continuum by shifting it by the delay expected from Equation~(\ref{equ:bentz}) and smearing it with a Gaussian of the width equal to 0.1 of the expected time delay. Next a continuum and line light curve were created
 out of the dense light curve adopting the cadence in the SDSS-RM monitoring (32 observations, separated as indicated by Figure~2 in \citet{grier2017}, {and sent to us by the authors).}   After correcting the cadence for the redshift of a given source we perform simulations in the rest frame of each source.
For each of the 100 random realizations of the process we now calculate the time delay using the Interpolated Cross Correlation Function (ICCF). Finally, from these 100 realizations we calculate the mean time delay and the dispersion.
\begin{figure}
\centering
\includegraphics[width=0.5\textwidth]{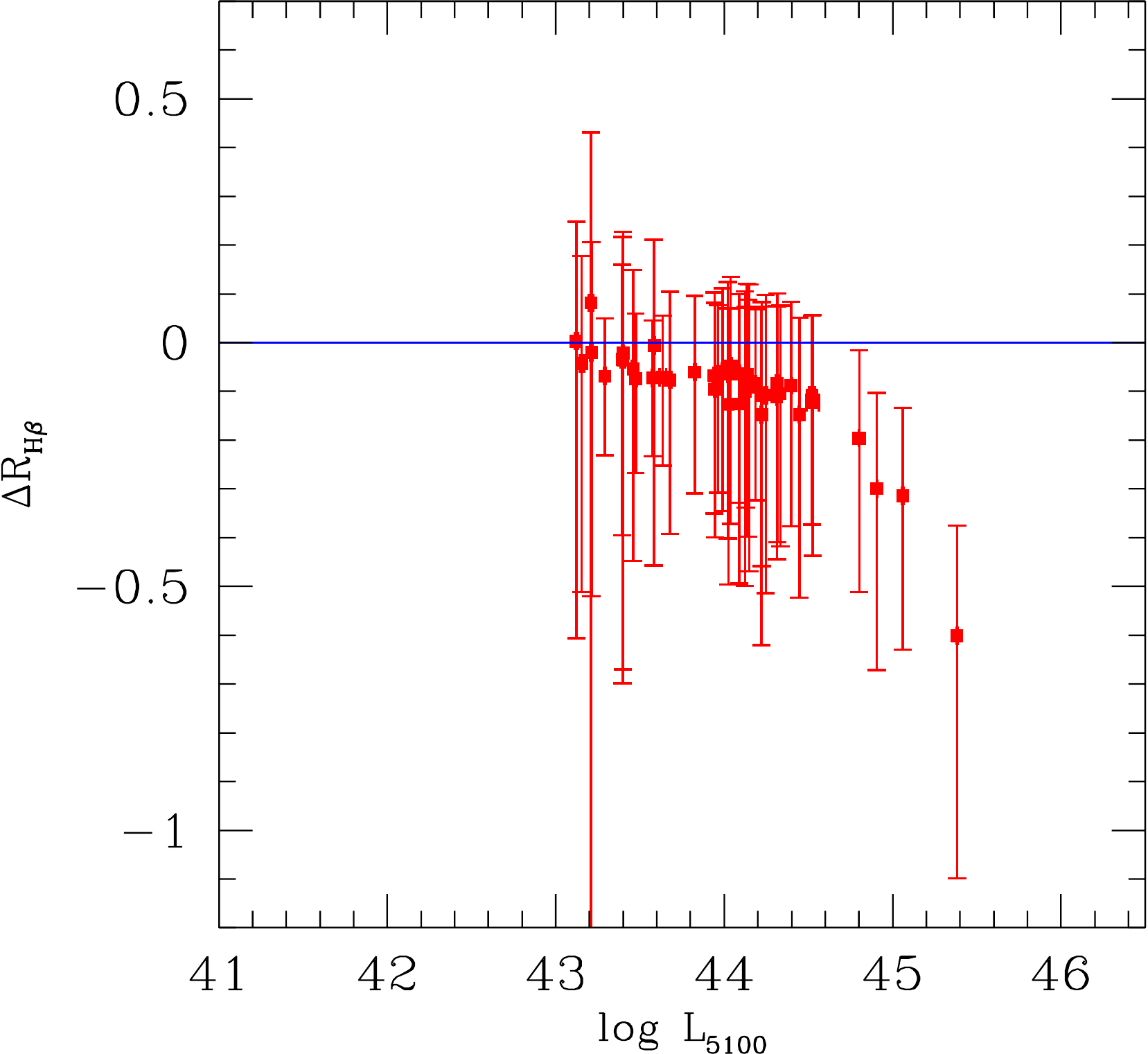}
\caption{The estimate of the systematic offset in the measured time delay in the sub-sample of \citet{grier2017} due to the cadence. In the case of four brightest objects the measured delay is underestimated, but the effect seems very strong only in the case of one object, J141856. }
\label{fig:simul}
\end{figure}
In Fig.~\ref{fig:simul} we compare the time delay from our Monte Carlo simulations with the time delay expected from Equation~(\ref{equ:bentz}) and used in the simulations. We see that the dispersion in the time delay obtained from simulations is, in general, higher than the errors quoted by \citet{grier2017}. We see some trend to obtain somewhat shorter time delay than assumed, due to the specific cadence used in the observations. However, for the majority of the sources the implied underestimation of the BLR radius (\DRhb) is below 0.1, much smaller that the actual departure from the standard \RL\ law, and well within the error. Only one source, J141856, is strongly affected by the cadence of the SDSS-RM, and its measured delay is likely much higher than 15.8 days reported by \citet{grier2017}. This source was not considered in the presented analysis, due to the low quality of its spectra, where the \hb\ profile is completely destroyed. In our simulations, assuming a delay of 203 days in the quasar rest-frame, we obtained the probability of 0.32 to  obtain the delay within the upper limit of the delay measured for this source (21.8 days), and a probability of 0.29 to get a delay shorter than the measured value of 15.8 days.



\bibliographystyle{aasjournal}
\bibliography{delays}

\begin{thebibliography}{}
\expandafter\ifx\csname natexlab\endcsname\relax\def\natexlab#1{#1}\fi
\providecommand{\url}[1]{\href{#1}{#1}}
\providecommand{\dodoi}[1]{doi:~\href{http://doi.org/#1}{\nolinkurl{#1}}}
\providecommand{\doeprint}[1]{\href{http://ascl.net/#1}{\nolinkurl{http://ascl.net/#1}}}
\providecommand{\doarXiv}[1]{\href{https://arxiv.org/abs/#1}{\nolinkurl{https://arxiv.org/abs/#1}}}

\bibitem[{{Allevato} {et~al.}(2013){Allevato}, {Paolillo}, {Papadakis}, \&
  {Pinto}}]{allevato2013}
{Allevato}, V., {Paolillo}, M., {Papadakis}, I., \& {Pinto}, C. 2013, \apj,
  771, 9, \dodoi{10.1088/0004-637X/771/1/9}

\bibitem[{{Barth} {et~al.}(2013){Barth}, {Pancoast}, {Bennert}, {Brewer},
  {Canalizo}, {Filippenko}, {Gates}, {Greene}, {Li}, {Malkan}, {Sand}, {Stern},
  {Treu}, {Woo}, {Assef}, {Bae}, {Buehler}, {Cenko}, {Clubb}, {Cooper},
  {Diamond-Stanic}, {H{\"o}nig}, {Joner}, {Laney}, {Lazarova}, {Nierenberg},
  {Silverman}, {Tollerud}, \& {Walsh}}]{barth2013}
{Barth}, A.~J., {Pancoast}, A., {Bennert}, V.~N., {et~al.} 2013, \apj, 769,
  128, \dodoi{10.1088/0004-637X/769/2/128}

\bibitem[{{Bentz} {et~al.}(2016{\natexlab{a}}){Bentz}, {Cackett}, {Crenshaw},
  {Horne}, {Street}, \& {Ou-Yang}}]{bentz2016a}
{Bentz}, M.~C., {Cackett}, E.~M., {Crenshaw}, D.~M., {et~al.}
  2016{\natexlab{a}}, \apj, 830, 136, \dodoi{10.3847/0004-637X/830/2/136}

\bibitem[{{Bentz} {et~al.}(2016{\natexlab{b}}){Bentz}, {Cackett}, {Crenshaw},
  {Horne}, {Street}, \& {Ou-Yang}}]{bentz2016b}
---. 2016{\natexlab{b}}, \apj, 830, 136, \dodoi{10.3847/0004-637X/830/2/136}

\bibitem[{{Bentz} {et~al.}(2009){Bentz}, {Peterson}, {Netzer}, {Pogge}, \&
  {Vestergaard}}]{bentz2009}
{Bentz}, M.~C., {Peterson}, B.~M., {Netzer}, H., {Pogge}, R.~W., \&
  {Vestergaard}, M. 2009, \apj, 697, 160, \dodoi{10.1088/0004-637X/697/1/160}

\bibitem[{{Bentz} {et~al.}(2006){Bentz}, {Denney}, {Cackett}, {Dietrich},
  {Fogel}, {Ghosh}, {Horne}, {Kuehn}, {Minezaki}, \& {Onken}}]{bentz2006}
{Bentz}, M.~C., {Denney}, K.~D., {Cackett}, E.~M., {et~al.} 2006, \apj, 651,
  775, \dodoi{10.1086/507417}

\bibitem[{{Bentz} {et~al.}(2013){Bentz}, {Denney}, {Grier}, {Barth},
  {Peterson}, {Vestergaard}, {Bennert}, {Canalizo}, {De Rosa}, {Filippenko},
  {Gates}, {Greene}, {Li}, {Malkan}, {Pogge}, {Stern}, {Treu}, \&
  {Woo}}]{bentz2013}
{Bentz}, M.~C., {Denney}, K.~D., {Grier}, C.~J., {et~al.} 2013, \apj, 767, 149,
  \dodoi{10.1088/0004-637X/767/2/149}

\bibitem[{{Bentz} {et~al.}(2014){Bentz}, {Horenstein}, {Bazhaw},
  {Manne-Nicholas}, {Ou-Yang}, {Anderson}, {Jones}, {Norris}, {Parks},
  {Saylor}, {Teems}, \& {Turner}}]{bentz2014}
{Bentz}, M.~C., {Horenstein}, D., {Bazhaw}, C., {et~al.} 2014, \apj, 796, 8,
  \dodoi{10.1088/0004-637X/796/1/8}

\bibitem[{{Bon} {et~al.}(2018){Bon}, {Bon}, \& {Marziani}}]{bon2018}
{Bon}, N., {Bon}, E., \& {Marziani}, P. 2018, Frontiers in Astronomy and Space
  Sciences, 5, 3, \dodoi{10.3389/fspas.2018.00003}

\bibitem[{{Brott} {et~al.}(2011){Brott}, {de Mink}, {Cantiello}, {Langer}, {de
  Koter}, {Evans}, {Hunter}, {Trundle}, \& {Vink}}]{brott2011}
{Brott}, I., {de Mink}, S.~E., {Cantiello}, M., {et~al.} 2011, \aap, 530, A115,
  \dodoi{10.1051/0004-6361/201016113}

\bibitem[{{Capellupo} {et~al.}(2016){Capellupo}, {Netzer}, {Lira},
  {Trakhtenbrot}, \& {Mej{\'{\i}}a-Restrepo}}]{capellupo2016}
{Capellupo}, D.~M., {Netzer}, H., {Lira}, P., {Trakhtenbrot}, B., \&
  {Mej{\'{\i}}a-Restrepo}, J. 2016, \mnras, 460, 212,
  \dodoi{10.1093/mnras/stw937}

\bibitem[{{Collin} {et~al.}(2006){Collin}, {Kawaguchi}, {Peterson}, \&
  {Vestergaard}}]{collin2006}
{Collin}, S., {Kawaguchi}, T., {Peterson}, B.~M., \& {Vestergaard}, M. 2006,
  \aap, 456, 75, \dodoi{10.1051/0004-6361:20064878}

\bibitem[{{Czerny} \& {Hryniewicz}(2011)}]{czerny2011}
{Czerny}, B., \& {Hryniewicz}, K. 2011, \aap, 525, L8,
  \dodoi{10.1051/0004-6361/201016025}

\bibitem[{{Czerny} {et~al.}(2013){Czerny}, {Hryniewicz}, {Maity},
  {Schwarzenberg-Czerny}, {{\.Z}ycki}, \& {Bilicki}}]{czerny2013}
{Czerny}, B., {Hryniewicz}, K., {Maity}, I., {et~al.} 2013, \aap, 556, A97,
  \dodoi{10.1051/0004-6361/201220832}

\bibitem[{{Czerny} {et~al.}(1999){Czerny}, {Schwarzenberg-Czerny}, \&
  {Loska}}]{czerny1999}
{Czerny}, B., {Schwarzenberg-Czerny}, A., \& {Loska}, Z. 1999, \mnras, 303,
  148, \dodoi{10.1046/j.1365-8711.1999.02196.x}

\bibitem[{{Czerny} {et~al.}(2015){Czerny}, {Modzelewska}, {Petrogalli}, {Pych},
  {Adhikari}, {{\.Z}ycki}, {Hryniewicz}, {Krupa}, {{\'S}wie{\c t}o{\'n}}, \&
  {Niko{\l}ajuk}}]{czerny2015}
{Czerny}, B., {Modzelewska}, J., {Petrogalli}, F., {et~al.} 2015, Advances in
  Space Research, 55, 1806, \dodoi{10.1016/j.asr.2015.01.004}

\bibitem[{{Czerny} {et~al.}(2017){Czerny}, {Li}, {Hryniewicz}, {Panda},
  {Wildy}, {Sniegowska}, {Wang}, {Sredzinska}, \& {Karas}}]{czerny2017}
{Czerny}, B., {Li}, Y.-R., {Hryniewicz}, K., {et~al.} 2017, \apj, 846, 154,
  \dodoi{10.3847/1538-4357/aa8810}

\bibitem[{{Czerny} {et~al.}(2018){Czerny}, {Beaton}, {Bejger}, {Cackett},
  {Dall'Ora}, {Holanda}, {Jensen}, {Jha}, {Lusso}, {Minezaki}, {Risaliti},
  {Salaris}, {Toonen}, \& {Yoshii}}]{czerny2018SSR}
{Czerny}, B., {Beaton}, R., {Bejger}, M., {et~al.} 2018, \ssr, 214, 32,
  \dodoi{10.1007/s11214-018-0466-9}

\bibitem[{{Czerny} {et~al.}(2019){Czerny}, {Wang}, {Du}, {Hryniewicz}, {Karas},
  {Li}, {Panda}, {Sniegowska}, {Wildy}, \& {Yuan}}]{czerny2019}
{Czerny}, B., {Wang}, J.-M., {Du}, P., {et~al.} 2019, \apj, 870, 84,
  \dodoi{10.3847/1538-4357/aaf396}

\bibitem[{{Denney} {et~al.}(2010){Denney}, {Peterson}, {Pogge}, {Adair},
  {Atlee}, {Au-Yong}, {Bentz}, {Bird}, {Brokofsky}, {Chisholm}, {Comins},
  {Dietrich}, {Doroshenko}, {Eastman}, {Efimov}, {Ewald}, {Ferbey}, {Gaskell},
  {Hedrick}, {Jackson}, {Klimanov}, {Klimek}, {Kruse}, {Lad{\'e}route}, {Lamb},
  {Leighly}, {Minezaki}, {Nazarov}, {Onken}, {Petersen}, {Peterson},
  {Poindexter}, {Sakata}, {Schlesinger}, {Sergeev}, {Skolski}, {Stieglitz},
  {Tobin}, {Unterborn}, {Vestergaard}, {Watkins}, {Watson}, \&
  {Yoshii}}]{denney2010}
{Denney}, K.~D., {Peterson}, B.~M., {Pogge}, R.~W., {et~al.} 2010, \apj, 721,
  715, \dodoi{10.1088/0004-637X/721/1/715}

\bibitem[{{Du} {et~al.}(2014){Du}, {Hu}, {Lu}, {Wang}, {Qiu}, {Li}, {Bai},
  {Kaspi}, {Netzer}, {Wang}, \& {SEAMBH Collaboration}}]{du2014}
{Du}, P., {Hu}, C., {Lu}, K.-X., {et~al.} 2014, \apj, 782, 45,
  \dodoi{10.1088/0004-637X/782/1/45}

\bibitem[{{Du} {et~al.}(2015){Du}, {Hu}, {Lu}, {Huang}, {Cheng}, {Qiu}, {Li},
  {Zhang}, {Fan}, {Bai}, {Bian}, {Yuan}, {Kaspi}, {Ho}, {Netzer}, {Wang}, \&
  {SEAMBH Collaboration}}]{du2015}
---. 2015, \apj, 806, 22, \dodoi{10.1088/0004-637X/806/1/22}

\bibitem[{{Du} {et~al.}(2016){Du}, {Lu}, {Zhang}, {Huang}, {Wang}, {Hu}, {Qiu},
  {Li}, {Fan}, {Fang}, {Bai}, {Bian}, {Yuan}, {Ho}, {Wang}, \& {SEAMBH
  Collaboration}}]{du2016}
{Du}, P., {Lu}, K.-X., {Zhang}, Z.-X., {et~al.} 2016, \apj, 825, 126,
  \dodoi{10.3847/0004-637X/825/2/126}

\bibitem[{{Du} {et~al.}(2018){Du}, {Zhang}, {Wang}, {Huang}, {Zhang}, {Lu},
  {Hu}, {Li}, {Bai}, {Bian}, {Yuan}, {Ho}, {Wang}, \& {SEAMBH
  Collaboration}}]{du2018}
{Du}, P., {Zhang}, Z.-X., {Wang}, K., {et~al.} 2018, \apj, 856, 6,
  \dodoi{10.3847/1538-4357/aaae6b}

\bibitem[{{Ekstr{\"o}m} {et~al.}(2012){Ekstr{\"o}m}, {Georgy}, {Eggenberger},
  {Meynet}, {Mowlavi}, {Wyttenbach}, {Granada}, {Decressin}, {Hirschi},
  {Frischknecht}, {Charbonnel}, \& {Maeder}}]{ekstrom2012}
{Ekstr{\"o}m}, S., {Georgy}, C., {Eggenberger}, P., {et~al.} 2012, \aap, 537,
  A146, \dodoi{10.1051/0004-6361/201117751}

\bibitem[{{Fausnaugh} {et~al.}(2017){Fausnaugh}, {Grier}, {Bentz}, {Denney},
  {De Rosa}, {Peterson}, {Kochanek}, {Pogge}, {Adams}, {Barth}, {Beatty},
  {Bhattacharjee}, {Borman}, {Boroson}, {Bottorff}, {Brown}, {Brown},
  {Brotherton}, {Coker}, {Crawford}, {Croxall}, {Eftekharzadeh}, {Eracleous},
  {Joner}, {Henderson}, {Holoien}, {Horne}, {Hutchison}, {Kaspi}, {Kim},
  {King}, {Li}, {Lochhaas}, {Ma}, {MacInnis}, {Manne-Nicholas}, {Mason},
  {Montuori}, {Mosquera}, {Mudd}, {Musso}, {Nazarov}, {Nguyen}, {Okhmat},
  {Onken}, {Ou-Yang}, {Pancoast}, {Pei}, {Penny}, {Poleski}, {Rafter},
  {Romero-Colmenero}, {Runnoe}, {Sand}, {Schimoia}, {Sergeev}, {Shappee},
  {Simonian}, {Somers}, {Spencer}, {Starkey}, {Stevens}, {Tayar}, {Treu},
  {Valenti}, {Van Saders}, {Villanueva}, {Villforth}, {Weiss}, {Winkler}, \&
  {Zhu}}]{fausnaugh2017}
{Fausnaugh}, M.~M., {Grier}, C.~J., {Bentz}, M.~C., {et~al.} 2017, \apj, 840,
  97, \dodoi{10.3847/1538-4357/aa6d52}

\bibitem[{{Gravity Collaboration} {et~al.}(2018){Gravity Collaboration},
  {Sturm}, {Dexter}, {Pfuhl}, {Stock}, {Davies}, {Lutz}, {Cl{\'e}net},
  {Eckart}, {Eisenhauer}, {Genzel}, {Gratadour}, {H{\"o}nig}, {Kishimoto},
  {Lacour}, {Millour}, {Netzer}, {Perrin}, {Peterson}, {Petrucci}, {Rouan},
  {Waisberg}, {Woillez}, {Amorim}, {Brandner}, {F{\"o}rster Schreiber},
  {Garcia}, {Gillessen}, {Ott}, {Paumard}, {Perraut}, {Scheithauer},
  {Straubmeier}, {Tacconi}, \& {Widmann}}]{gravity2018}
{Gravity Collaboration}, {Sturm}, E., {Dexter}, J., {et~al.} 2018, \nat, 563,
  657, \dodoi{10.1038/s41586-018-0731-9}

\bibitem[{{Grier} {et~al.}(2017){Grier}, {Trump}, {Shen}, {Horne}, {Kinemuchi},
  {McGreer}, {Starkey}, {Brandt}, {Hall}, {Kochanek}, {Chen}, {Denney},
  {Greene}, {Ho}, {Homayouni}, {I-Hsiu Li}, {Pei}, {Peterson}, {Petitjean},
  {Schneider}, {Sun}, {AlSayyad}, {Bizyaev}, {Brinkmann}, {Brownstein},
  {Bundy}, {Dawson}, {Eftekharzadeh}, {Fernandez-Trincado}, {Gao},
  {Hutchinson}, {Jia}, {Jiang}, {Oravetz}, {Pan}, {Paris}, {Ponder}, {Peters},
  {Rogerson}, {Simmons}, {Smith}, \& {Wang}}]{grier2017}
{Grier}, C.~J., {Trump}, J.~R., {Shen}, Y., {et~al.} 2017, \apj, 851, 21,
  \dodoi{10.3847/1538-4357/aa98dc}

\bibitem[{{Groves} {et~al.}(2006){Groves}, {Heckman}, \&
  {Kauffmann}}]{groves2006}
{Groves}, B.~A., {Heckman}, T.~M., \& {Kauffmann}, G. 2006, \mnras, 371, 1559,
  \dodoi{10.1111/j.1365-2966.2006.10812.x}

\bibitem[{{G{\"u}ltekin} {et~al.}(2009){G{\"u}ltekin}, {Richstone}, {Gebhardt},
  {Lauer}, {Tremaine}, {Aller}, {Bender}, {Dressler}, {Faber}, {Filippenko},
  {Green}, {Ho}, {Kormendy}, {Magorrian}, {Pinkney}, \&
  {Siopis}}]{gultekin2009}
{G{\"u}ltekin}, K., {Richstone}, D.~O., {Gebhardt}, K., {et~al.} 2009, \apj,
  698, 198, \dodoi{10.1088/0004-637X/698/1/198}

\bibitem[{{Haas} {et~al.}(2011){Haas}, {Chini}, {Ramolla}, {Pozo Nu{\~n}ez},
  {Westhues}, {Watermann}, {Hoffmeister}, \& {Murphy}}]{haas2011}
{Haas}, M., {Chini}, R., {Ramolla}, M., {et~al.} 2011, \aap, 535, A73,
  \dodoi{10.1051/0004-6361/201117325}

\bibitem[{{Hildebrandt} {et~al.}(2017){Hildebrandt}, {Viola}, {Heymans},
  {Joudaki}, {Kuijken}, {Blake}, {Erben}, {Joachimi}, {Klaes}, {Miller},
  {Morrison}, {Nakajima}, {Verdoes Kleijn}, {Amon}, {Choi}, {Covone}, {de
  Jong}, {Dvornik}, {Fenech Conti}, {Grado}, {Harnois-D{\'e}raps}, {Herbonnet},
  {Hoekstra}, {K{\"o}hlinger}, {McFarland}, {Mead}, {Merten}, {Napolitano},
  {Peacock}, {Radovich}, {Schneider}, {Simon}, {Valentijn}, {van den Busch},
  {van Uitert}, \& {Van Waerbeke}}]{hildebrandt2017}
{Hildebrandt}, H., {Viola}, M., {Heymans}, C., {et~al.} 2017, \mnras, 465,
  1454, \dodoi{10.1093/mnras/stw2805}

\bibitem[{{H{\"o}nig} {et~al.}(2014){H{\"o}nig}, {Watson}, {Kishimoto}, \&
  {Hjorth}}]{honing2014}
{H{\"o}nig}, S.~F., {Watson}, D., {Kishimoto}, M., \& {Hjorth}, J. 2014, \nat,
  515, 528, \dodoi{10.1038/nature13914}

\bibitem[{{Hu} {et~al.}(2015){Hu}, {Du}, {Lu}, {Li}, {Wang}, {Qiu}, {Bai},
  {Kaspi}, {Ho}, {Netzer}, {Wang}, \& {SEAMBH Collaboration}}]{hu2015}
{Hu}, C., {Du}, P., {Lu}, K.-X., {et~al.} 2015, \apj, 804, 138,
  \dodoi{10.1088/0004-637X/804/2/138}

\bibitem[{{Humphreys} {et~al.}(2013){Humphreys}, {Reid}, {Moran}, {Greenhill},
  \& {Argon}}]{humphreys2013}
{Humphreys}, E.~M.~L., {Reid}, M.~J., {Moran}, J.~M., {Greenhill}, L.~J., \&
  {Argon}, A.~L. 2013, \apj, 775, 13, \dodoi{10.1088/0004-637X/775/1/13}

\bibitem[{{Ivezi{\'c}} {et~al.}(2019){Ivezi{\'c}}, {Kahn}, {Tyson}, {Abel},
  {Acosta}, {Allsman}, {Alonso}, {AlSayyad}, {Anderson}, {Andrew}, \&
  et~al.}]{lsst2019}
{Ivezi{\'c}}, {\v Z}., {Kahn}, S.~M., {Tyson}, J.~A., {et~al.} 2019, \apj, 873,
  111, \dodoi{10.3847/1538-4357/ab042c}

\bibitem[{{Joudaki} {et~al.}(2017){Joudaki}, {Mead}, {Blake}, {Choi}, {de
  Jong}, {Erben}, {Fenech Conti}, {Herbonnet}, {Heymans}, {Hildebrandt},
  {Hoekstra}, {Joachimi}, {Klaes}, {K{\"o}hlinger}, {Kuijken}, {McFarland},
  {Miller}, {Schneider}, \& {Viola}}]{joudaki2017}
{Joudaki}, S., {Mead}, A., {Blake}, C., {et~al.} 2017, \mnras, 471, 1259,
  \dodoi{10.1093/mnras/stx998}

\bibitem[{{Kaspi} {et~al.}(2000){Kaspi}, {Smith}, {Netzer}, {Maoz}, {Jannuzi},
  \& {Giveon}}]{kaspi2000}
{Kaspi}, S., {Smith}, P.~S., {Netzer}, H., {et~al.} 2000, \apj, 533, 631,
  \dodoi{10.1086/308704}

\bibitem[{{Kilerci Eser} {et~al.}(2015){Kilerci Eser}, {Vestergaard},
  {Peterson}, {Denney}, \& {Bentz}}]{kilerci2015}
{Kilerci Eser}, E., {Vestergaard}, M., {Peterson}, B.~M., {Denney}, K.~D., \&
  {Bentz}, M.~C. 2015, \apj, 801, 8, \dodoi{10.1088/0004-637X/801/1/8}

\bibitem[{{King} \& {Lasota}(2014)}]{king2014}
{King}, A., \& {Lasota}, J.-P. 2014, \mnras, 444, L30,
  \dodoi{10.1093/mnrasl/slu105}

\bibitem[{{King} {et~al.}(2015){King}, {Martini}, {Davis}, {Denney},
  {Kochanek}, {Peterson}, {Skielboe}, {Vestergaard}, {Huff}, {Watson},
  {Banerji}, {McMahon}, {Sharp}, \& {Lidman}}]{king2015}
{King}, A.~L., {Martini}, P., {Davis}, T.~M., {et~al.} 2015, \mnras, 453, 1701,
  \dodoi{10.1093/mnras/stv1718}

\bibitem[{{Li} {et~al.}(2018){Li}, {Songsheng}, {Qiu}, {Hu}, {Du}, {Lu},
  {Huang}, {Bai}, {Bian}, {Yuan}, {Ho}, \& {Wang}}]{li2018}
{Li}, Y.-R., {Songsheng}, Y.-Y., {Qiu}, J., {et~al.} 2018, \apj, 869, 137,
  \dodoi{10.3847/1538-4357/aaee6b}

\bibitem[{{Lu} {et~al.}(2016){Lu}, {Du}, {Hu}, {Li}, {Zhang}, {Wang}, {Huang},
  {Bi}, {Bai}, {Ho}, \& {Wang}}]{lu2016}
{Lu}, K.-X., {Du}, P., {Hu}, C., {et~al.} 2016, \apj, 827, 118,
  \dodoi{10.3847/0004-637X/827/2/118}

\bibitem[{{MacLeod} {et~al.}(2010){MacLeod}, {Ivezi{\'c}}, {Kochanek},
  {Koz{\l}owski}, {Kelly}, {Bullock}, {Kimball}, {Sesar}, {Westman}, {Brooks},
  {Gibson}, {Becker}, \& {de Vries}}]{macleod2010}
{MacLeod}, C.~L., {Ivezi{\'c}}, {\v Z}., {Kochanek}, C.~S., {et~al.} 2010,
  \apj, 721, 1014, \dodoi{10.1088/0004-637X/721/2/1014}

\bibitem[{{Maeder} \& {Meynet}(2000)}]{meader2000}
{Maeder}, A., \& {Meynet}, G. 2000, \aap, 361, 159

\bibitem[{{Marziani} \& {Sulentic}(2014)}]{marzianisulentic2014}
{Marziani}, P., \& {Sulentic}, J.~W. 2014, \mnras, 442, 1211,
  \dodoi{10.1093/mnras/stu951}

\bibitem[{{Marziani} {et~al.}(2001){Marziani}, {Sulentic}, {Zwitter},
  {Dultzin-Hacyan}, \& {Calvani}}]{marziani2001}
{Marziani}, P., {Sulentic}, J.~W., {Zwitter}, T., {Dultzin-Hacyan}, D., \&
  {Calvani}, M. 2001, \apj, 558, 553, \dodoi{10.1086/322286}

\bibitem[{{Mej{\'{\i}}a-Restrepo} {et~al.}(2018){Mej{\'{\i}}a-Restrepo},
  {Lira}, {Netzer}, {Trakhtenbrot}, \& {Capellupo}}]{mejia-restrepo2018}
{Mej{\'{\i}}a-Restrepo}, J.~E., {Lira}, P., {Netzer}, H., {Trakhtenbrot}, B.,
  \& {Capellupo}, D.~M. 2018, Nature Astronomy, 2, 63,
  \dodoi{10.1038/s41550-017-0305-z}

\bibitem[{{Mushotzky} {et~al.}(2011){Mushotzky}, {Edelson}, {Baumgartner}, \&
  {Gandhi}}]{mushotzky2011}
{Mushotzky}, R.~F., {Edelson}, R., {Baumgartner}, W., \& {Gandhi}, P. 2011,
  \apjl, 743, L12, \dodoi{10.1088/2041-8205/743/1/L12}

\bibitem[{{Negrete} {et~al.}(2018){Negrete}, {Dultzin}, {Marziani}, {Esparza},
  {Sulentic}, {del Olmo}, {Mart{\'\i}nez-Aldama}, {Garc{\'\i}a L{\'o}pez},
  {D'Onofrio}, \& {Bon}}]{negrete2018}
{Negrete}, C.~A., {Dultzin}, D., {Marziani}, P., {et~al.} 2018, \aap, 620,
  A118, \dodoi{10.1051/0004-6361/201833285}

\bibitem[{{Onken}(2004)}]{onken2004}
{Onken}, C.~A. 2004, in American Astronomical Society Meeting Abstracts, Vol.
  205, 118.06

\bibitem[{{Pacaud} {et~al.}(2018){Pacaud}, {Pierre}, {Melin}, {Adami},
  {Evrard}, {Galli}, {Gastaldello}, {Maughan}, {Sereno}, {Alis}, {Altieri},
  {Birkinshaw}, {Chiappetti}, {Faccioli}, {Giles}, {Horellou}, {Iovino},
  {Koulouridis}, {Le F{\`e}vre}, {Lidman}, {Lieu}, {Maurogordato},
  {Moscardini}, {Plionis}, {Poggianti}, {Pompei}, {Sadibekova}, {Valtchanov},
  \& {Willis}}]{pacaud2018}
{Pacaud}, F., {Pierre}, M., {Melin}, J.-B., {et~al.} 2018, \aap, 620, A10,
  \dodoi{10.1051/0004-6361/201834022}

\bibitem[{{Pancoast} {et~al.}(2014){Pancoast}, {Brewer}, {Treu}, {Park},
  {Barth}, {Bentz}, \& {Woo}}]{pancoast2014}
{Pancoast}, A., {Brewer}, B.~J., {Treu}, T., {et~al.} 2014, \mnras, 445, 3073,
  \dodoi{10.1093/mnras/stu1419}

\bibitem[{{Panda} {et~al.}(2019){Panda}, {Marziani}, \& {Czerny}}]{panda2019b}
{Panda}, S., {Marziani}, P., \& {Czerny}, B. 2019, arXiv e-prints,
  arXiv:1905.01729.
\newblock \doarXiv{1905.01729}

\bibitem[{{Pei} {et~al.}(2014){Pei}, {Barth}, {Aldering}, {Briley}, {Carroll},
  {Carson}, {Cenko}, {Clubb}, {Cohen}, {Cucchiara}, {Desjardins}, {Edelson},
  {Fang}, {Fedrow}, {Filippenko}, {Fox}, {Furniss}, {Gates}, {Gregg},
  {Gustafson}, {Horst}, {Joner}, {Kelly}, {Lacy}, {Laney}, {Leonard}, {Li},
  {Malkan}, {Margon}, {Neeleman}, {Nguyen}, {Prochaska}, {Ross}, {Sand},
  {Searcy}, {Shivvers}, {Silverman}, {Smith}, {Suzuki}, {Smith}, {Tytler},
  {Werk}, \& {Worseck}}]{pei2014}
{Pei}, L., {Barth}, A.~J., {Aldering}, G.~S., {et~al.} 2014, \apj, 795, 38,
  \dodoi{10.1088/0004-637X/795/1/38}

\bibitem[{{Peterson} {et~al.}(2004){Peterson}, {Ferrarese}, {Gilbert}, {Kaspi},
  {Malkan}, {Maoz}, {Merritt}, {Netzer}, {Onken}, {Pogge}, {Vestergaard}, \&
  {Wandel}}]{peterson2004}
{Peterson}, B.~M., {Ferrarese}, L., {Gilbert}, K.~M., {et~al.} 2004, ApJ, 613,
  682, \dodoi{10.1086/423269}

\bibitem[{{Planck Collaboration}(2013)}]{planck2013}
{Planck Collaboration}. 2013, ArXiv e-prints.
\newblock \doarXiv{1303.5062}

\bibitem[{{Planck Collaboration}(2016{\natexlab{a}})}]{planckXIV}
---. 2016{\natexlab{a}}, \aap, 594, A14, \dodoi{10.1051/0004-6361/201525814}

\bibitem[{{Planck Collaboration}(2016{\natexlab{b}})}]{PCGC}
---. 2016{\natexlab{b}}, \aap, 594, A24, \dodoi{10.1051/0004-6361/201525833}

\bibitem[{{Planck Collaboration}(2018)}]{planck2018}
---. 2018, arXiv e-prints.
\newblock \doarXiv{1807.06209}

\bibitem[{{Rakshit} \& {Stalin}(2017)}]{rakshit2017}
{Rakshit}, S., \& {Stalin}, C.~S. 2017, \apj, 842, 96,
  \dodoi{10.3847/1538-4357/aa72f4}

\bibitem[{{Richards} {et~al.}(2006){Richards}, {Lacy}, {Storrie-Lombardi},
  {Hall}, {Gallagher}, {Hines}, {Fan}, {Papovich}, {Vanden Berk}, {Trammell},
  {Schneider}, {Vestergaard}, {York}, {Jester}, {Anderson}, {Budav{\'a}ri}, \&
  {Szalay}}]{richards2006}
{Richards}, G.~T., {Lacy}, M., {Storrie-Lombardi}, L.~J., {et~al.} 2006, \apjs,
  166, 470, \dodoi{10.1086/506525}

\bibitem[{{Riess} {et~al.}(2018){Riess}, {Casertano}, {Yuan}, {Macri},
  {Bucciarelli}, {Lattanzi}, {MacKenty}, {Bowers}, {Zheng}, {Filippenko},
  {Huang}, \& {Anderson}}]{riess2018}
{Riess}, A.~G., {Casertano}, S., {Yuan}, W., {et~al.} 2018, \apj, 861, 126,
  \dodoi{10.3847/1538-4357/aac82e}

\bibitem[{{Risaliti} \& {Lusso}(2015)}]{risaliti2015}
{Risaliti}, G., \& {Lusso}, E. 2015, \apj, 815, 33,
  \dodoi{10.1088/0004-637X/815/1/33}

\bibitem[{{Rodr{\'{\i}}guez-Pascual} {et~al.}(1997){Rodr{\'{\i}}guez-Pascual},
  {Alloin}, {Clavel}, {Crenshaw}, {Horne}, {Kriss}, {Krolik}, {Malkan},
  {Netzer}, {O'Brien}, {Peterson}, {Reichert}, {Wamsteker}, {Alexander},
  {Barr}, {Blandford}, {Bregman}, {Carone}, {Clements}, {Courvoisier}, {De
  Robertis}, {Dietrich}, {Dottori}, {Edelson}, {Filippenko}, {Gaskell},
  {Huchra}, {Hutchings}, {Kollatschny}, {Koratkar}, {Korista}, {Laor},
  {MacAlpine}, {Martin}, {Maoz}, {McCollum}, {Morris}, {Perola}, {Pogge},
  {Ptak}, {Recondo-Gonz{\'a}lez}, {Rodr{\'{\i}}guez-Espinoza}, {Rokaki},
  {Santos-Lle{\'o}}, {Sekiguchi}, {Shull}, {Snijders}, {Sparke}, {Stirpe},
  {Stoner}, {Sun}, {Wagner}, {Wanders}, {Wilkes}, {Winge}, \&
  {Zheng}}]{rodriguez-pascual1997}
{Rodr{\'{\i}}guez-Pascual}, P.~M., {Alloin}, D., {Clavel}, J., {et~al.} 1997,
  \apjs, 110, 9, \dodoi{10.1086/312996}

\bibitem[{{S{\'a}nchez} {et~al.}(2017){S{\'a}nchez}, {Lira}, {Cartier},
  {P{\'e}rez}, {Miranda}, {Yovaniniz}, {Ar{\'e}valo}, {Milvang-Jensen},
  {Fynbo}, {Dunlop}, {Coppi}, \& {Marchesi}}]{sanchez2017}
{S{\'a}nchez}, P., {Lira}, P., {Cartier}, R., {et~al.} 2017, \apj, 849, 110,
  \dodoi{10.3847/1538-4357/aa9188}

\bibitem[{{S{\'a}nchez-S{\'a}ez} {et~al.}(2018){S{\'a}nchez-S{\'a}ez}, {Lira},
  {Mej{\'{\i}}a-Restrepo}, {Ho}, {Ar{\'e}valo}, {Kim}, {Cartier}, \&
  {Coppi}}]{sanchez-saenz2018}
{S{\'a}nchez-S{\'a}ez}, P., {Lira}, P., {Mej{\'{\i}}a-Restrepo}, J., {et~al.}
  2018, \apj, 864, 87, \dodoi{10.3847/1538-4357/aad7f9}

\bibitem[{{Scolnic} {et~al.}(2018){Scolnic}, {Jones}, {Rest}, {Pan},
  {Chornock}, {Foley}, {Huber}, {Kessler}, {Narayan}, \& {Riess}}]{scolnic2018}
{Scolnic}, D.~M., {Jones}, D.~O., {Rest}, A., {et~al.} 2018, \apj, 859, 101,
  \dodoi{10.3847/1538-4357/aab9bb}

\bibitem[{{Sergeev} {et~al.}(2007){Sergeev}, {Doroshenko}, {Dzyuba},
  {Peterson}, {Pogge}, \& {Pronik}}]{sergeev2007}
{Sergeev}, S.~G., {Doroshenko}, V.~T., {Dzyuba}, S.~A., {et~al.} 2007, \apj,
  668, 708, \dodoi{10.1086/520697}

\bibitem[{{Shakura} \& {Sunyaev}(1973)}]{shakura1973}
{Shakura}, N.~I., \& {Sunyaev}, R.~A. 1973, \aap, 24, 337

\bibitem[{{Shen} \& {Ho}(2014)}]{shenandho2014}
{Shen}, Y., \& {Ho}, L.~C. 2014, \nat, 513, 210, \dodoi{10.1038/nature13712}

\bibitem[{{Shen} {et~al.}(2015){Shen}, {Greene}, {Ho}, {Brandt}, {Denney},
  {Horne}, {Jiang}, {Kochanek}, {McGreer}, {Merloni}, {Peterson}, {Petitjean},
  {Schneider}, {Schulze}, {Strauss}, {Tao}, {Trump}, {Pan}, \&
  {Bizyaev}}]{shen2015}
{Shen}, Y., {Greene}, J.~E., {Ho}, L.~C., {et~al.} 2015, \apj, 805, 96,
  \dodoi{10.1088/0004-637X/805/2/96}

\bibitem[{{Shen} {et~al.}(2018){Shen}, {Hall}, {Horne}, {Zhu}, {McGreer},
  {Simm}, {Trump}, {Kinemuchi}, {Brandt}, {Green}, {Grier}, {Guo}, {Ho},
  {Homayouni}, {Jiang}, {I-Hsiu Li}, {Morganson}, {Petitjean}, {Richards},
  {Schneider}, {Starkey}, {Wang}, {Chambers}, {Kaiser}, {Kudritzki}, {Magnier},
  \& {Waters}}]{shen2018}
{Shen}, Y., {Hall}, P.~B., {Horne}, K., {et~al.} 2018, arXiv e-prints.
\newblock \doarXiv{1810.01447}

\bibitem[{{Simm} {et~al.}(2016){Simm}, {Salvato}, {Saglia}, {Ponti},
  {Lanzuisi}, {Trakhtenbrot}, {Nandra}, \& {Bender}}]{simm2016}
{Simm}, T., {Salvato}, M., {Saglia}, R., {et~al.} 2016, \aap, 585, A129,
  \dodoi{10.1051/0004-6361/201527353}

\bibitem[{{Stanway} \& {Eldridge}(2019)}]{stanway2019}
{Stanway}, E.~R., \& {Eldridge}, J.~J. 2019, \aap, 621, A105,
  \dodoi{10.1051/0004-6361/201834359}

\bibitem[{{Storchi-Bergmann} {et~al.}(2017){Storchi-Bergmann}, {Schimoia},
  {Peterson}, {Elvis}, {Denney}, {Eracleous}, \&
  {Nemmen}}]{strochi-bergmann2017}
{Storchi-Bergmann}, T., {Schimoia}, J.~S., {Peterson}, B.~M., {et~al.} 2017,
  \apj, 835, 236, \dodoi{10.3847/1538-4357/835/2/236}

\bibitem[{{Sulentic} {et~al.}(2017){Sulentic}, {del Olmo}, {Marziani},
  {Mart{\'{\i}}nez-Carballo}, {D'Onofrio}, {Dultzin}, {Perea},
  {Mart{\'{\i}}nez-Aldama}, {Negrete}, {Stirpe}, \& {Zamfir}}]{sulentic2017}
{Sulentic}, J.~W., {del Olmo}, A., {Marziani}, P., {et~al.} 2017, \aap, 608,
  A122, \dodoi{10.1051/0004-6361/201630309}

\bibitem[{{Timmer} \& {Koenig}(1995)}]{timmer1995}
{Timmer}, J., \& {Koenig}, M. 1995, \aap, 300, 707

\bibitem[{{Tsvetkov} {et~al.}(2019){Tsvetkov}, {Baklanov}, {Potashov},
  {Oknyansky}, {Mikailov}, {Huseynov}, {Alekberov}, {Khalilov}, {Pavlyuk}, \&
  {Metlov}}]{tsvetkov2018}
{Tsvetkov}, D.~Y., {Baklanov}, P.~V., {Potashov}, M.~S., {et~al.} 2019, \mnras,
  487, 3001, \dodoi{10.1093/mnras/stz1474}

\bibitem[{{Uttley} {et~al.}(2005){Uttley}, {McHardy}, \&
  {Vaughan}}]{uttley2005}
{Uttley}, P., {McHardy}, I.~M., \& {Vaughan}, S. 2005, \mnras, 359, 345,
  \dodoi{10.1111/j.1365-2966.2005.08886.x}

\bibitem[{{Wang} {et~al.}(2014{\natexlab{a}}){Wang}, {Du}, {Li}, {Ho}, {Hu}, \&
  {Bai}}]{wangspin2014}
{Wang}, J.-M., {Du}, P., {Li}, Y.-R., {et~al.} 2014{\natexlab{a}}, \apjl, 792,
  L13, \dodoi{10.1088/2041-8205/792/1/L13}

\bibitem[{{Wang} {et~al.}(2014{\natexlab{b}}){Wang}, {Qiu}, {Du}, \&
  {Ho}}]{wangshielding2014}
{Wang}, J.-M., {Qiu}, J., {Du}, P., \& {Ho}, L.~C. 2014{\natexlab{b}}, \apj,
  797, 65, \dodoi{10.1088/0004-637X/797/1/65}

\bibitem[{{Wang} {et~al.}(2019){Wang}, {Songsheng}, {Li}, {Du}, \&
  {Zhang}}]{wang2019}
{Wang}, J.-M., {Songsheng}, Y.-Y., {Li}, Y.-R., {Du}, P., \& {Zhang}, Z.-X.
  2019, arXiv e-prints, arXiv:1906.08417.
\newblock \doarXiv{1906.08417}

\bibitem[{{Wang} {et~al.}(2014{\natexlab{c}}){Wang}, {Du}, {Hu}, {Netzer},
  {Bai}, {Lu}, {Kaspi}, {Qiu}, {Li}, {Wang}, \& {SEAMBH
  Collaboration}}]{wangSEAMBH2014}
{Wang}, J.-M., {Du}, P., {Hu}, C., {et~al.} 2014{\natexlab{c}}, \apj, 793, 108,
  \dodoi{10.1088/0004-637X/793/2/108}

\bibitem[{{Watson} {et~al.}(2011){Watson}, {Denney}, {Vestergaard}, \&
  {Davis}}]{watson2011}
{Watson}, D., {Denney}, K.~D., {Vestergaard}, M., \& {Davis}, T.~M. 2011,
  \apjl, 740, L49, \dodoi{10.1088/2041-8205/740/2/L49}

\bibitem[{{Wilhite} {et~al.}(2008){Wilhite}, {Brunner}, {Grier}, {Schneider},
  \& {vanden Berk}}]{wilhite2008}
{Wilhite}, B.~C., {Brunner}, R.~J., {Grier}, C.~J., {Schneider}, D.~P., \&
  {vanden Berk}, D.~E. 2008, \mnras, 383, 1232,
  \dodoi{10.1111/j.1365-2966.2007.12655.x}

\bibitem[{{Woo} {et~al.}(2015){Woo}, {Yoon}, {Park}, {Park}, \&
  {Kim}}]{woo2015}
{Woo}, J.-H., {Yoon}, Y., {Park}, S., {Park}, D., \& {Kim}, S.~C. 2015, \apj,
  801, 38, \dodoi{10.1088/0004-637X/801/1/38}

\bibitem[{{Yu} {et~al.}(2019){Yu}, {Bian}, {Wang}, {Zhao}, \& {Ge}}]{yu2019}
{Yu}, L.-M., {Bian}, W.-H., {Wang}, C., {Zhao}, B.-X., \& {Ge}, X. 2019, arXiv
  e-prints, arXiv:1907.00315.
\newblock \doarXiv{1907.00315}

\bibitem[{{Zhang} {et~al.}(2018){Zhang}, {Du}, {Smith}, {Zhao}, {Hu}, {Xiao},
  {Li}, {Huang}, {Wang}, {Bai}, {Ho}, \& {Wang}}]{zhang2018}
{Zhang}, Z.-X., {Du}, P., {Smith}, P.~S., {et~al.} 2018, arXiv e-prints.
\newblock \doarXiv{1811.03812}

\end{thebibliography}

\newpage

\begin{center}
\scriptsize
\small
\begin{longtable}{ccccccc}
\caption{Sample description} \label{tab:samples} \\ 
 \hline\hline\noalign{\vskip 0.1cm}          
\multirow{2}{*}{Object} & \multirow{2}{*}{$z$} & \multirow{1}{*}{log $L_{5100}$} & \multirow{1}{*}{$\tau_{\mathrm{obs}}$} & \multirow{1}{*}{FWHM}  & \multirow{2}{*}{Reference} \\
 & & [\ergs] & [days] & [\kms] \\
(1) & (2) & (3) & (4) & (5) & (6)  \\

 \hline \noalign{\vskip 0.1cm} 																					
\multicolumn{7}{c}{SEAMBH sample}	\\																			
\hline \noalign{\vskip 0.1cm}																					
Mrk335	&	0.0258	&	43.764	$\pm$	0.067	&	14.0	$^{+	4.6	}	_{-	3.4	}$	&	2096	$\pm$	170	&	1	\\
Mrk142	&	0.0449	&	43.594	$\pm$	0.044	&	6.4	$^{+	7.3	}	_{-	3.4	}$	&	1588	$\pm$	58	&	1	\\
IRASF12397	&	0.0435	&	44.229	$\pm$	0.054	&	9.7	$^{+	5.5	}	_{-	1.8	}$	&	1802	$\pm$	560	&	1	\\
Mrk486	&	0.0389	&	43.694	$\pm$	0.050	&	23.7	$^{+	7.5	}	_{-	2.7	}$	&	1942	$\pm$	67	&	2	\\
Mrk382	&	0.0337	&	43.124	$\pm$	0.085	&	7.5	$^{+	2.9	}	_{-	2.0	}$	&	1462	$\pm$	296	&	2	\\
IRAS04416	&	0.0889	&	44.467	$\pm$	0.030	&	13.3	$^{+	13.9	}	_{-	1.4	}$	&	1522	$\pm$	44	&	2	\\
MCG+06-26-012	&	0.0328	&	42.675	$\pm$	0.106	&	24.0	$^{+	8.4	}	_{-	4.8	}$	&	1334	$\pm$	80	&	2	\\
Mrk493	&	0.0313	&	43.112	$\pm$	0.075	&	11.6	$^{+	1.2	}	_{-	2.6	}$	&	778	$\pm$	12	&	2	\\
Mrk1044	&	0.0165	&	43.095	$\pm$	0.102	&	10.5	$^{+	3.3	}	_{-	2.7	}$	&	1178	$\pm$	22	&	2	\\
J080101	&	0.1396	&	44.270	$\pm$	0.030	&	8.3	$^{+	9.7	}	_{-	2.7	}$	&	1930	$\pm$	18	&	3	\\
J081456	&	0.1197	&	43.990	$\pm$	0.040	&	24.3	$^{+	7.7	}	_{-	16.4	}$	&	2409	$\pm$	61	&	3	\\
J093922	&	0.1859	&	44.070	$\pm$	0.040	&	11.9	$^{+	2.1	}	_{-	6.3	}$	&	1209	$\pm$	16	&	3	\\
J080131	&	0.1786	&	43.950	$\pm$	0.040	&	11.5	$^{+	7.5	}	_{-	3.7	}$	&	1290	$\pm$	13	&	4	\\
J085946	&	0.2438	&	44.410	$\pm$	0.030	&	34.8	$^{+	19.2	}	_{-	26.3	}$	&	1718	$\pm$	16	&	4	\\
J102339	&	0.1364	&	44.090	$\pm$	0.030	&	24.9	$^{+	19.8	}	_{-	3.9	}$	&	1733	$\pm$	29	&	4	\\
J074352	&	0.2520	&	45.370	$\pm$	0.020	&	43.9	$^{+	5.2	}	_{-	4.2	}$	&	3156	$\pm$	36	&	5	\\
J075051	&	0.4004	&	45.330	$\pm$	0.010	&	66.6	$^{+	18.7	}	_{-	9.9	}$	&	1904	$\pm$	9	&	5	\\
J075101	&	0.1209	&	44.240	$\pm$	0.040	&	30.4	$^{+	7.3	}	_{-	5.8	}$	&	1679	$\pm$	35	&	5	\\
J075949	&	0.1879	&	44.190	$\pm$	0.060	&	43.9	$^{+	33.1	}	_{-	19.0	}$	&	1783	$\pm$	17	&	5	\\
J081441	&	0.1626	&	43.950	$\pm$	0.040	&	25.3	$^{+	10.4	}	_{-	7.5	}$	&	1782	$\pm$	16	&	5	\\
J083553	&	0.2051	&	44.440	$\pm$	0.020	&	12.4	$^{+	5.4	}	_{-	5.4	}$	&	1758	$\pm$	16	&	5	\\
J084533	&	0.3024	&	44.520	$\pm$	0.020	&	18.1	$^{+	6.0	}	_{-	4.7	}$	&	1297	$\pm$	12	&	5	\\
J093302	&	0.1772	&	44.310	$\pm$	0.130	&	19.0	$^{+	3.8	}	_{-	4.3	}$	&	1800	$\pm$	25	&	5	\\
J100402	&	0.3272	&	45.520	$\pm$	0.010	&	32.2	$^{+	43.5	}	_{-	4.2	}$	&	2088	$\pm$	1	&	5	\\
J101000	&	0.2564	&	44.760	$\pm$	0.020	&	27.7	$^{+	23.5	}	_{-	7.6	}$	&	2311	$\pm$	11	&	5	\\
\hline \noalign{\vskip 0.1cm} 																					
\multicolumn{7}{c}{SDSS-RM sample}\\																			
\hline \noalign{\vskip 0.1cm} 																					
J140812	&	0.1160	&	43.154	$\pm$	0.013	&	10.5	$^{+	1.0	}	_{-	2.2	}$	&	4345	$\pm$	558	&	6	\\
J141923	&	0.1520	&	43.122	$\pm$	0.010	&	11.8	$^{+	0.7	}	_{-	1.5	}$	&	2945	$\pm$	20	&	6	\\
J140759	&	0.1720	&	43.577	$\pm$	0.009	&	16.3	$^{+	13.1	}	_{-	6.6	}$	&	3662	$\pm$	27	&	6	\\
J141729	&	0.2370	&	43.291	$\pm$	0.007	&	5.5	$^{+	5.7	}	_{-	2.1	}$	&	9208	$\pm$	269	&	6	\\
J141645.15	&	0.2440	&	43.213	$\pm$	0.007	&	5.0	$^{+	1.5	}	_{-	1.4	}$	&	7409	$\pm$	113	&	6	\\
J142135	&	0.2490	&	43.475	$\pm$	0.007	&	3.9	$^{+	0.9	}	_{-	0.9	}$	&	3090	$\pm$	66	&	6	\\
J141625	&	0.2630	&	43.964	$\pm$	0.019	&	15.1	$^{+	3.2	}	_{-	4.6	}$	&	3515	$\pm$	17	&	6	\\
J142103	&	0.2630	&	43.636	$\pm$	0.019	&	75.2	$^{+	3.2	}	_{-	3.3	}$	&	2990	$\pm$	48	&	6	\\
J142038	&	0.2650	&	43.458	$\pm$	0.006	&	25.2	$^{+	4.7	}	_{-	5.7	}$	&	4700	$\pm$	55	&	6	\\
J142043	&	0.3370	&	43.400	$\pm$	0.005	&	5.9	$^{+	0.4	}	_{-	0.6	}$	&	4429	$\pm$	105	&	6	\\
J141041	&	0.3590	&	43.824	$\pm$	0.005	&	21.9	$^{+	4.2	}	_{-	2.4	}$	&	5034	$\pm$	35	&	6	\\
J141318	&	0.3620	&	43.941	$\pm$	0.005	&	20.0	$^{+	1.1	}	_{-	3.0	}$	&	3428	$\pm$	37	&	6	\\
J141955	&	0.4180	&	43.395	$\pm$	0.005	&	10.7	$^{+	5.6	}	_{-	4.4	}$	&	6789	$\pm$	580	&	6	\\
J141645.58	&	0.4420	&	43.679	$\pm$	0.009	&	8.5	$^{+	2.5	}	_{-	1.4	}$	&	2178	$\pm$	44	&	6	\\
J141324	&	0.4560	&	43.945	$\pm$	0.004	&	25.5	$^{+	10.9	}	_{-	5.8	}$	&	6076	$\pm$	121	&	6	\\
J141214	&	0.4580	&	44.397	$\pm$	0.004	&	21.4	$^{+	4.2	}	_{-	6.4	}$	&	2652	$\pm$	302	&	6	\\
J140518	&	0.4670	&	44.333	$\pm$	0.004	&	41.6	$^{+	14.8	}	_{-	8.3	}$	&	3406	$\pm$	22	&	6	\\
J141018	&	0.4700	&	43.584	$\pm$	0.005	&	16.2	$^{+	2.9	}	_{-	4.5	}$	&	4329	$\pm$	298	&	6	\\
J141123	&	0.4720	&	44.128	$\pm$	0.004	&	13.0	$^{+	1.4	}	_{-	0.8	}$	&	4106	$\pm$	38	&	6	\\
J142039	&	0.4740	&	44.141	$\pm$	0.004	&	20.7	$^{+	0.9	}	_{-	3.0	}$	&	4259	$\pm$	90	&	6	\\
J141724	&	0.4820	&	43.992	$\pm$	0.004	&	10.1	$^{+	12.5	}	_{-	2.7	}$	&	5230	$\pm$	76	&	6	\\
J141004	&	0.5270	&	44.224	$\pm$	0.003	&	53.5	$^{+	4.2	}	_{-	4.0	}$	&	2918	$\pm$	62	&	6	\\
J141706	&	0.5320	&	44.186	$\pm$	0.003	&	10.4	$^{+	6.3	}	_{-	3.0	}$	&	1682	$\pm$	14	&	6	\\
J142010	&	0.5480	&	44.088	$\pm$	0.003	&	12.8	$^{+	5.7	}	_{-	4.5	}$	&	6050	$\pm$	541	&	6	\\
J141712	&	0.5540	&	43.209	$\pm$	0.012	&	12.5	$^{+	1.8	}	_{-	2.6	}$	&	2226	$\pm$	405	&	6	\\
J141115	&	0.5720	&	44.313	$\pm$	0.003	&	49.1	$^{+	11.1	}	_{-	2.0	}$	&	3442	$\pm$	51	&	6	\\
J141112	&	0.5870	&	44.123	$\pm$	0.003	&	20.4	$^{+	2.5	}	_{-	2.0	}$	&	2765	$\pm$	36	&	6	\\
J141417	&	0.6040	&	43.397	$\pm$	0.013	&	15.6	$^{+	3.2	}	_{-	5.1	}$	&	6476	$\pm$	793	&	6	\\
J141031	&	0.6080	&	44.022	$\pm$	0.003	&	35.8	$^{+	1.1	}	_{-	10.3	}$	&	3495	$\pm$	118	&	6	\\
J141941	&	0.6460	&	44.522	$\pm$	0.017	&	30.4	$^{+	3.9	}	_{-	8.3	}$	&	2818	$\pm$	48	&	6	\\
J141135	&	0.6500	&	44.040	$\pm$	0.004	&	17.6	$^{+	8.6	}	_{-	7.4	}$	&	2515	$\pm$	61	&	6	\\
J140904	&	0.6580	&	44.147	$\pm$	0.004	&	11.6	$^{+	8.6	}	_{-	4.6	}$	&	10405	$\pm$	1094	&	6	\\
J142052	&	0.6760	&	45.059	$\pm$	0.003	&	11.9	$^{+	1.3	}	_{-	1.0	}$	&	3646	$\pm$	14	&	6	\\
J141147	&	0.6800	&	44.025	$\pm$	0.004	&	6.4	$^{+	1.5	}	_{-	1.4	}$	&	2338	$\pm$	65	&	6	\\
J141532	&	0.7150	&	44.136	$\pm$	0.004	&	26.5	$^{+	9.9	}	_{-	8.8	}$	&	1615	$\pm$	38	&	6	\\
J142023	&	0.7340	&	44.222	$\pm$	0.006	&	8.5	$^{+	3.2	}	_{-	3.9	}$	&	4446	$\pm$	135	&	6	\\
J142049	&	0.7510	&	44.446	$\pm$	0.003	&	46.0	$^{+	9.5	}	_{-	9.5	}$	&	4665	$\pm$	97	&	6	\\
J142112	&	0.8430	&	44.315	$\pm$	0.008	&	14.2	$^{+	3.7	}	_{-	3.0	}$	&	4428	$\pm$	295	&	6	\\
J141606	&	0.8480	&	44.801	$\pm$	0.003	&	32.0	$^{+	11.6	}	_{-	15.5	}$	&	7307	$\pm$	213	&	6	\\
J141859	&	0.8840	&	44.907	$\pm$	0.003	&	20.4	$^{+	5.6	}	_{-	7.0	}$	&	4999	$\pm$	53	&	6	\\
J141952	&	0.8840	&	44.246	$\pm$	0.006	&	32.9	$^{+	5.6	}	_{-	5.1	}$	&	7726	$\pm$	319	&	6	\\
J142417	&	0.8900	&	44.089	$\pm$	0.060	&	36.3	$^{+	4.5	}	_{-	5.5	}$	&	1721	$\pm$	147	&	6	\\
J141856$^\dagger$	&	0.9760	&	45.382	$\pm$	0.002	&	15.8	$^{+	6.0	}	_{-	1.9	}$	&	3120	$\pm$	58	&	6$^\dagger$	\\
J141314$^\dagger$	&	1.0260	&	44.524	$\pm$	0.038	&	43.9	$^{+	4.9	}	_{-	4.3	}$	&	1412	$\pm$	183	&	6$^\dagger$	\\
\hline \noalign{\vskip 0.1cm} 																					
\multicolumn{7}{c}{Bentz collection} \\																					
\hline \noalign{\vskip 0.1cm} 																					
PG0026+129	&	0.1420	&	44.970	$\pm$	0.016	&	111.0	$^{+	24.1	}	_{-	28.3	}$	&	1719	$\pm$	495	&	7	\\
PG0052+251	&	0.1545	&	44.807	$\pm$	0.025	&	89.8	$^{+	24.5	}	_{-	24.1	}$	&	4165	$\pm$	381	&	7	\\
Fairall9	&	0.0470	&	43.981	$\pm$	0.041	&	17.4	$^{+	3.2	}	_{-	4.3	}$	&	6901	$\pm$	707	&	7	\\
Mrk590	&	0.0264	&	43.496	$\pm$	0.212	&	25.6	$^{+	6.5	}	_{-	5.3	}$	&	2220	$\pm$	701	&	7	\\
3C120	&	0.0330	&	44.004	$\pm$	0.100	&	26.2	$^{+	8.7	}	_{-	6.6	}$	&	2372	$\pm$	501	&	7	\\
Ark120	&	0.0327	&	43.867	$\pm$	0.253	&	39.5	$^{+	8.5	}	_{-	7.8	}$	&	5410	$\pm$	360	&	7	\\
Mrk79	&	0.0222	&	43.677	$\pm$	0.067	&	15.6	$^{+	5.4	}	_{-	4.9	}$	&	4852	$\pm$	1554	&	7	\\
PG0804+761	&	0.1000	&	44.910	$\pm$	0.017	&	146.9	$^{+	18.8	}	_{-	18.9	}$	&	2012	$\pm$	845	&	7	\\
Mrk110	&	0.0353	&	43.658	$\pm$	0.115	&	25.6	$^{+	8.9	}	_{-	7.2	}$	&	1494	$\pm$	802	&	7	\\
PG0953+414	&	0.2341	&	45.186	$\pm$	0.013	&	150.1	$^{+	21.6	}	_{-	22.6	}$	&	3002	$\pm$	398	&	7	\\
NGC3227	&	0.0039	&	42.236	$\pm$	0.106	&	3.8	$^{+	0.8	}	_{-	0.8	}$	&	3578	$\pm$	83	&	7{ $^\ddagger$}	\\
NGC3516	&	0.0088	&	42.787	$\pm$	0.205	&	11.7	$^{+	1.0	}	_{-	1.5	}$	&	5175	$\pm$	96	&	7{ $^\ddagger$}	\\
SBS1116+583A	&	0.0279	&	42.138	$\pm$	0.231	&	2.3	$^{+	0.6	}	_{-	0.5	}$	&	3604	$\pm$	1123	&	7	\\
Arp151	&	0.0211	&	42.548	$\pm$	0.101	&	4.0	$^{+	0.5	}	_{-	0.7	}$	&	2357	$\pm$	142	&	7	\\
NGC3783	&	0.0097	&	42.558	$\pm$	0.180	&	10.2	$^{+	3.3	}	_{-	2.3	}$	&	3093	$\pm$	529	&	7{ $^\ddagger$}	\\
Mrk1310	&	0.0196	&	42.293	$\pm$	0.145	&	3.7	$^{+	0.6	}	_{-	0.6	}$	&	1602	$\pm$	250	&	7	\\
NGC4051	&	0.0023	&	41.898	$\pm$	0.152	&	2.1	$^{+	0.9	}	_{-	0.7	}$	&	1034	$\pm$	41	&	7{ $^\ddagger$}	\\
NGC4151	&	0.0033	&	42.091	$\pm$	0.207	&	6.6	$^{+	1.1	}	_{-	0.8	}$	&	4711	$\pm$	750	&	7{ $^\ddagger$}	\\
Mrk202	&	0.0210	&	42.260	$\pm$	0.144	&	3.0	$^{+	1.7	}	_{-	1.1	}$	&	1354	$\pm$	250	&	7	\\
NGC4253	&	0.0129	&	42.570	$\pm$	0.122	&	6.2	$^{+	1.6	}	_{-	1.2	}$	&	834	$\pm$	1260	&	7	\\
PG1229+204	&	0.0630	&	43.697	$\pm$	0.047	&	37.8	$^{+	27.6	}	_{-	15.3	}$	&	3415	$\pm$	320	&	7	\\
NGC4593	&	0.0090	&	42.621	$\pm$	0.370	&	4.0	$^{+	0.8	}	_{-	0.7	}$	&	4268	$\pm$	551	&	7	\\
NGC4748	&	0.0146	&	42.556	$\pm$	0.120	&	5.5	$^{+	1.6	}	_{-	2.2	}$	&	1212	$\pm$	173	&	7	\\
PG1307+085	&	0.1550	&	44.849	$\pm$	0.015	&	105.6	$^{+	36.0	}	_{-	46.6	}$	&	5058	$\pm$	524	&	7	\\
Mrk279	&	0.0305	&	43.705	$\pm$	0.074	&	16.7	$^{+	3.9	}	_{-	3.9	}$	&	3385	$\pm$	349	&	7	\\
PG1411+442	&	0.0896	&	44.563	$\pm$	0.020	&	124.3	$^{+	61.0	}	_{-	61.7	}$	&	2398	$\pm$	353	&	7	\\
PG1426+015	&	0.0866	&	44.629	$\pm$	0.024	&	95.0	$^{+	29.9	}	_{-	37.1	}$	&	6323	$\pm$	1295	&	7	\\
Mrk817	&	0.0315	&	43.743	$\pm$	0.089	&	19.9	$^{+	9.9	}	_{-	6.7	}$	&	4122	$\pm$	1197	&	7	\\
Mrk290	&	0.0296	&	43.168	$\pm$	0.057	&	8.7	$^{+	1.2	}	_{-	1.0	}$	&	4270	$\pm$	157	&	7	\\
PG1613+658	&	0.1290	&	44.774	$\pm$	0.022	&	40.1	$^{+	15.0	}	_{-	15.2	}$	&	7897	$\pm$	1792	&	7	\\
PG1617+175	&	0.1124	&	44.391	$\pm$	0.017	&	71.5	$^{+	29.6	}	_{-	33.7	}$	&	4718	$\pm$	991	&	7	\\
PG1700+518	&	0.2920	&	45.586	$\pm$	0.007	&	251.8	$^{+	45.9	}	_{-	38.8	}$	&	1846	$\pm$	682	&	7	\\
3C390.3	&	0.0561	&	44.434	$\pm$	0.576	&	44.5	$^{+	27.7	}	_{-	17.0	}$	&	10415	$\pm$	1971	&	7	\\
NGC6814	&	0.0052	&	42.120	$\pm$	0.285	&	6.6	$^{+	0.9	}	_{-	0.9	}$	&	3277	$\pm$	297	&	7	\\
Mrk509	&	0.0344	&	44.193	$\pm$	0.045	&	79.6	$^{+	6.1	}	_{-	5.4	}$	&	2715	$\pm$	101	&	7	\\
PG2130+099	&	0.0630	&	44.203	$\pm$	0.028	&	9.6	$^{+	1.2	}	_{-	1.2	}$	&	2097	$\pm$	102	&	7	\\
NGC7469	&	0.0163	&	43.506	$\pm$	0.108	&	10.8	$^{+	3.4	}	_{-	1.3	}$	&	1066	$\pm$	84	&	7	\\
PG1211+143	&	0.0809	&	44.728	$\pm$	0.081	&	93.8	$^{+	25.6	}	_{-	42.1	}$	&	2012	$\pm$	37	&	8	\\
PG0844+349	&	0.0640	&	44.218	$\pm$	0.071	&	32.3	$^{+	13.7	}	_{-	13.4	}$	&	2436	$\pm$	329	&	8	\\
NGC5273	&	0.0036	&	41.535	$\pm$	0.160	&	2.2	$^{+	1.2	}	_{-	1.6	}$	&	4615	$\pm$	330	&	9{ $^\ddagger$}	\\
Mrk1511	&	0.0339	&	43.162	$\pm$	0.062	&	5.7	$^{+	0.9	}	_{-	0.8	}$	&	4171	$\pm$	137	&	10	\\
KA1858-4850	&	0.0780	&	43.428	$\pm$	0.047	&	13.5	$^{+	2.0	}	_{-	2.3	}$	&	1511	$\pm$	68	&	11	\\
MCG6-30-15	&	0.0078	&	41.643	$\pm$	0.108	&	5.7	$^{+	1.8	}	_{-	1.7	}$	&	1422	$\pm$	416	&	12	\\
UGC06728	&	0.0065	&	41.864	$\pm$	0.081	&	1.4	$^{+	0.7	}	_{-	0.8	}$	&	1145	$\pm$	58	&	13	\\
MCG+08-11-011	&	0.0205	&	43.330	$\pm$	0.111	&	15.7	$^{+	0.5	}	_{-	0.5	}$	&	1159	$\pm$	8	&	14	\\
NGC2617	&	0.0142	&	42.667	$\pm$	0.159	&	4.3	$^{+	1.1	}	_{-	1.4	}$	&	5303	$\pm$	49	&	14{ $^\ddagger$}	\\
3C382	&	0.0579	&	43.835	$\pm$	0.102	&	40.5	$^{+	8.0	}	_{-	3.7	}$	&	3619	$\pm$	282	&	14	\\
Mrk374	&	0.0426	&	43.774	$\pm$	0.042	&	14.8	$^{+	5.8	}	_{-	3.3	}$	&	3250	$\pm$	19	&	14	\\
\hline \noalign{\vskip 0.1cm} 																					
\multicolumn{7}{c}{\citet{lu2016}}\\																					
\hline \noalign{\vskip 0.1cm} 																					
NGC5548	&	0.0172	&	43.210	$\pm$	0.120	&	7.2	$^{+	1.3	}	_{-	0.4	}$	&	9912	$\pm$	362	&	15{ $^\ddagger$}	\\

\hline \noalign{\vskip 0.1cm} 																					
\multicolumn{7}{c}{\citet{zhang2018}} \\																					
\hline \noalign{\vskip 0.1cm} 																					
3C 273	&	0.1583	&	45.965	$\pm$	0.016	&	146.8	$^{+	8.3	}	_{-	12.1	}$	&	3314	$\pm$	59	&	16	\\

\hline \noalign{\vskip 0.1cm}       
\end{longtable}
\end{center}
\footnotesize{{\sc Notes.} Columns are as follows: (1) Object name. Discarded object of the analysis are marked by $\dagger$ symbol. { Double-peak H$\alpha$ profiles are marked by $\ddagger$ symbol.} (2) Redshift. (3) Luminosity at 5100 \AA.  (4) Delay time at rest-frame. (5) FWHM of \hb\ emission line. (6) References: 1: \citet{du2014}, 2: \citet{wangSEAMBH2014}, 3:  \citet{du2015}, 4: \citet{du2016}, 5: \citet{du2018}, 6: \citet{grier2017}, 7: \citet{bentz2013}, 8: \citet{bentz2009}, 9: \citet{bentz2014}, 10: \citet{barth2013}, 11: \citet{pei2014}, 12: \citet{bentz2016a}, 13: \citet{bentz2016b}, 14: \citet{fausnaugh2017}, 15: \citet{lu2016}, 16: \citet{zhang2018}. 
}

\newpage

\begin{center}
\scriptsize
\small
\begin{longtable}{ccccccc}
\caption{Observational properties for the full sample} \label{tab:measurements} \\ 
 \hline\hline\noalign{\vskip 0.1cm}          
\multirow{2}{*}{Object} & \multirow{1}{*}{log \mbh} &  \multirow{2}{*}{log \mdot} & \multirow{2}{*}{\LLEdd} & \multirow{2}{*}{log \DRhb} & \multirow{2}{*}{\fblrc} & \multirow{2}{*}{\fvar} \\
 & [$M_\odot$] & & & \\
(1) & (2) & (3) & (4) & (5) & (6) & (7)\\

\hline \noalign{\vskip 0.1cm} 																																										
\multicolumn{7}{c}{SEAMBH sample} \\ 																																										
\hline \noalign{\vskip 0.1cm} 																																										
Mrk335	&	7.08	$^{+	0.16	}	_{-	0.13	}$	&	0.97	$^{+	0.33	}	_{-	0.27	}$	&	0.33	$^{+	0.15	}	_{-	0.13	}$	&	-0.25	$^{+	0.15	}	_{-	0.12	}$	&	2.48	$\pm$	0.24	&	0.030	$^	1	$	\\
Mrk142	&	6.50	$^{+	0.50	}	_{-	0.23	}$	&	1.88	$^{+	0.99	}	_{-	0.47	}$	&	0.85	$^{+	1.00	}	_{-	0.50	}$	&	-0.50	$^{+	0.50	}	_{-	0.23	}$	&	3.43	$\pm$	0.15	&	0.066	$^	1	$	\\
IRASF12397	&	6.79	$^{+	0.37	}	_{-	0.28	}$	&	2.25	$^{+	0.74	}	_{-	0.57	}$	&	1.89	$^{+	1.65	}	_{-	1.30	}$	&	-0.66	$^{+	0.25	}	_{-	0.09	}$	&	2.96	$\pm$	1.07	&	0.041	$^	1	$	\\
Mrk486	&	7.24	$^{+	0.14	}	_{-	0.06	}$	&	0.54	$^{+	0.29	}	_{-	0.14	}$	&	0.19	$^{+	0.08	}	_{-	0.05	}$	&	0.01	$^{+	0.14	}	_{-	0.07	}$	&	2.71	$\pm$	0.11	&	0.034	$^	1	$	\\
Mrk382	&	6.50	$^{+	0.24	}	_{-	0.21	}$	&	1.18	$^{+	0.50	}	_{-	0.44	}$	&	0.29	$^{+	0.18	}	_{-	0.16	}$	&	-0.18	$^{+	0.18	}	_{-	0.13	}$	&	3.77	$\pm$	0.89	&	0.041	$^	1	$	\\
IRAS04416	&	6.78	$^{+	0.45	}	_{-	0.05	}$	&	2.63	$^{+	0.91	}	_{-	0.11	}$	&	3.34	$^{+	3.57	}	_{-	0.82	}$	&	-0.65	$^{+	0.46	}	_{-	0.06	}$	&	3.60	$\pm$	0.12	&	0.020	$^	1	$	\\
MCG06	&	6.92	$^{+	0.16	}	_{-	0.10	}$	&	-0.34	$^{+	0.36	}	_{-	0.26	}$	&	0.04	$^{+	0.02	}	_{-	0.02	}$	&	0.56	$^{+	0.17	}	_{-	0.12	}$	&	4.20	$\pm$	0.29	&	0.092	$^	1	$	\\
Mrk493	&	6.14	$^{+	0.05	}	_{-	0.10	}$	&	1.88	$^{+	0.15	}	_{-	0.23	}$	&	0.65	$^{+	0.19	}	_{-	0.23	}$	&	0.01	$^{+	0.07	}	_{-	0.11	}$	&	7.90	$\pm$	0.14	&	0.031	$^	1	$	\\
Mrk1044	&	6.46	$^{+	0.14	}	_{-	0.11	}$	&	1.22	$^{+	0.31	}	_{-	0.27	}$	&	0.30	$^{+	0.13	}	_{-	0.12	}$	&	-0.02	$^{+	0.15	}	_{-	0.13	}$	&	4.86	$\pm$	0.11	&	0.037	$^	1	$	\\
J080101	&	6.78	$^{+	0.51	}	_{-	0.14	}$	&	2.33	$^{+	1.02	}	_{-	0.29	}$	&	2.12	$^{+	2.51	}	_{-	0.82	}$	&	-0.75	$^{+	0.51	}	_{-	0.15	}$	&	2.73	$\pm$	0.03	&	0.039				\\
J081456	&	7.44	$^{+	0.14	}	_{-	0.29	}$	&	0.59	$^{+	0.29	}	_{-	0.59	}$	&	0.24	$^{+	0.09	}	_{-	0.17	}$	&	-0.14	$^{+	0.14	}	_{-	0.30	}$	&	2.10	$\pm$	0.06	&	0.031				\\
J093922	&	6.53	$^{+	0.08	}	_{-	0.23	}$	&	2.53	$^{+	0.17	}	_{-	0.46	}$	&	2.37	$^{+	0.67	}	_{-	1.36	}$	&	-0.49	$^{+	0.09	}	_{-	0.23	}$	&	4.71	$\pm$	0.07	&	0.036				\\
J080131	&	6.57	$^{+	0.29	}	_{-	0.14	}$	&	2.27	$^{+	0.57	}	_{-	0.29	}$	&	1.64	$^{+	1.14	}	_{-	0.64	}$	&	-0.44	$^{+	0.29	}	_{-	0.15	}$	&	4.37	$\pm$	0.05	&	0.047				\\
J085946	&	7.30	$^{+	0.24	}	_{-	0.33	}$	&	1.50	$^{+	0.48	}	_{-	0.66	}$	&	0.88	$^{+	0.52	}	_{-	0.69	}$	&	-0.14	$^{+	0.14	}	_{-	0.30	}$	&	3.13	$\pm$	0.03	&	0.048				\\
J102339	&	7.17	$^{+	0.35	}	_{-	0.07	}$	&	1.29	$^{+	0.69	}	_{-	0.15	}$	&	0.58	$^{+	0.48	}	_{-	0.15	}$	&	-0.18	$^{+	0.35	}	_{-	0.08	}$	&	3.09	$\pm$	0.06	&	0.032				\\
J074352	&	7.93	$^{+	0.05	}	_{-	0.04	}$	&	1.68	$^{+	0.11	}	_{-	0.09	}$	&	1.88	$^{+	0.45	}	_{-	0.43	}$	&	-0.61	$^{+	0.08	}	_{-	0.07	}$	&	1.53	$\pm$	0.02	&	0.060				\\
J075051	&	7.68	$^{+	0.12	}	_{-	0.06	}$	&	2.14	$^{+	0.24	}	_{-	0.13	}$	&	3.11	$^{+	1.08	}	_{-	0.78	}$	&	-0.41	$^{+	0.13	}	_{-	0.08	}$	&	2.77	$\pm$	0.02	&	0.032				\\
J075101	&	7.22	$^{+	0.11	}	_{-	0.08	}$	&	1.40	$^{+	0.22	}	_{-	0.18	}$	&	0.71	$^{+	0.23	}	_{-	0.21	}$	&	-0.17	$^{+	0.11	}	_{-	0.09	}$	&	3.21	$\pm$	0.08	&	0.065				\\
J075949	&	7.44	$^{+	0.33	}	_{-	0.19	}$	&	0.90	$^{+	0.66	}	_{-	0.39	}$	&	0.39	$^{+	0.31	}	_{-	0.19	}$	&	0.01	$^{+	0.33	}	_{-	0.19	}$	&	2.99	$\pm$	0.03	&	0.094				\\
J081441	&	7.20	$^{+	0.18	}	_{-	0.13	}$	&	1.02	$^{+	0.36	}	_{-	0.27	}$	&	0.39	$^{+	0.18	}	_{-	0.14	}$	&	-0.10	$^{+	0.18	}	_{-	0.13	}$	&	2.99	$\pm$	0.03	&	0.051				\\
J083553	&	6.88	$^{+	0.19	}	_{-	0.19	}$	&	2.40	$^{+	0.38	}	_{-	0.38	}$	&	2.53	$^{+	1.22	}	_{-	1.22	}$	&	-0.67	$^{+	0.19	}	_{-	0.19	}$	&	3.04	$\pm$	0.03	&	0.052				\\
J084533	&	6.78	$^{+	0.14	}	_{-	0.11	}$	&	2.72	$^{+	0.29	}	_{-	0.23	}$	&	3.82	$^{+	1.49	}	_{-	1.27	}$	&	-0.55	$^{+	0.15	}	_{-	0.12	}$	&	4.34	$\pm$	0.05	&	0.040				\\
J093302	&	7.08	$^{+	0.09	}	_{-	0.10	}$	&	1.79	$^{+	0.26	}	_{-	0.28	}$	&	1.17	$^{+	0.48	}	_{-	0.50	}$	&	-0.41	$^{+	0.12	}	_{-	0.12	}$	&	2.96	$\pm$	0.05	&	0.036				\\
J100402	&	7.44	$^{+	0.59	}	_{-	0.06	}$	&	2.89	$^{+	1.17	}	_{-	0.11	}$	&	8.29	$^{+	11.32	}	_{-	2.00	}$	&	-0.83	$^{+	0.59	}	_{-	0.08	}$	&	2.49	$\pm$	0.001	&	0.048				\\
J101000	&	7.46	$^{+	0.37	}	_{-	0.12	}$	&	1.71	$^{+	0.74	}	_{-	0.24	}$	&	1.37	$^{+	1.19	}	_{-	0.47	}$	&	-0.49	$^{+	0.37	}	_{-	0.13	}$	&	2.21	$\pm$	0.001	&	0.065				\\
\hline \noalign{\vskip 0.1cm} 																																										
\multicolumn{7}{c}{SDSS sample} \\ 																																										
\hline \noalign{\vskip 0.1cm} 																																										
J140812	&	7.59	$^{+	0.12	}	_{-	0.14	}$	&	-0.96	$^{+	0.24	}	_{-	0.29	}$	&	0.03	$^{+	0.01	}	_{-	0.01	}$	&	-0.05	$^{+	0.06	}	_{-	0.10	}$	&	1.06	$\pm$	0.16	&	0.043				\\
J141923	&	7.30	$^{+	0.03	}	_{-	0.06	}$	&	-0.43	$^{+	0.06	}	_{-	0.11	}$	&	0.05	$^{+	0.01	}	_{-	0.01	}$	&	0.01	$^{+	0.05	}	_{-	0.07	}$	&	1.66	$\pm$	0.01	&	0.076				\\
J140759	&	7.63	$^{+	0.35	}	_{-	0.18	}$	&	-0.41	$^{+	0.70	}	_{-	0.35	}$	&	0.06	$^{+	0.05	}	_{-	0.03	}$	&	-0.09	$^{+	0.35	}	_{-	0.18	}$	&	1.29	$\pm$	0.01	&	0.038				\\
J141729	&	7.96	$^{+	0.45	}	_{-	0.17	}$	&	-1.49	$^{+	0.90	}	_{-	0.34	}$	&	0.01	$^{+	0.02	}	_{-	0.01	}$	&	-0.41	$^{+	0.45	}	_{-	0.17	}$	&	0.44	$\pm$	0.01	&	0.033				\\
J141645.15	&	7.73	$^{+	0.13	}	_{-	0.12	}$	&	-1.15	$^{+	0.26	}	_{-	0.24	}$	&	0.02	$^{+	0.01	}	_{-	0.01	}$	&	-0.41	$^{+	0.14	}	_{-	0.13	}$	&	0.57	$\pm$	0.01	&	0.068				\\
J142135	&	6.86	$^{+	0.10	}	_{-	0.10	}$	&	0.98	$^{+	0.20	}	_{-	0.20	}$	&	0.28	$^{+	0.09	}	_{-	0.09	}$	&	-0.66	$^{+	0.11	}	_{-	0.11	}$	&	1.57	$\pm$	0.04	&	0.038				\\
J141625	&	7.56	$^{+	0.09	}	_{-	0.13	}$	&	0.31	$^{+	0.19	}	_{-	0.27	}$	&	0.17	$^{+	0.05	}	_{-	0.06	}$	&	-0.33	$^{+	0.10	}	_{-	0.14	}$	&	1.35	$\pm$	0.01	&	0.058				\\
J142103	&	8.12	$^{+	0.02	}	_{-	0.02	}$	&	-1.30	$^{+	0.05	}	_{-	0.06	}$	&	0.02	$^{+	0.005	}	_{-	0.005	}$	&	0.54	$^{+	0.04	}	_{-	0.04	}$	&	1.63	$\pm$	0.03	&	--				\\
J142038	&	8.04	$^{+	0.08	}	_{-	0.10	}$	&	-1.40	$^{+	0.16	}	_{-	0.20	}$	&	0.02	$^{+	0.01	}	_{-	0.01	}$	&	0.16	$^{+	0.09	}	_{-	0.10	}$	&	0.96	$\pm$	0.01	&	0.065				\\
J142043	&	7.36	$^{+	0.04	}	_{-	0.05	}$	&	-0.12	$^{+	0.07	}	_{-	0.10	}$	&	0.08	$^{+	0.02	}	_{-	0.02	}$	&	-0.44	$^{+	0.05	}	_{-	0.06	}$	&	1.03	$\pm$	0.03	&	0.036				\\
J141041	&	8.04	$^{+	0.08	}	_{-	0.05	}$	&	-0.85	$^{+	0.17	}	_{-	0.10	}$	&	0.04	$^{+	0.01	}	_{-	0.01	}$	&	-0.09	$^{+	0.09	}	_{-	0.06	}$	&	0.89	$\pm$	0.01	&	0.074				\\
J141318	&	7.66	$^{+	0.03	}	_{-	0.07	}$	&	0.08	$^{+	0.05	}	_{-	0.13	}$	&	0.13	$^{+	0.03	}	_{-	0.03	}$	&	-0.19	$^{+	0.04	}	_{-	0.07	}$	&	1.39	$\pm$	0.02	&	0.060				\\
J141955	&	7.99	$^{+	0.24	}	_{-	0.19	}$	&	-1.39	$^{+	0.48	}	_{-	0.39	}$	&	0.02	$^{+	0.01	}	_{-	0.01	}$	&	-0.18	$^{+	0.23	}	_{-	0.18	}$	&	0.63	$\pm$	0.06	&	0.066				\\
J141645.58	&	6.90	$^{+	0.13	}	_{-	0.07	}$	&	1.21	$^{+	0.26	}	_{-	0.15	}$	&	0.42	$^{+	0.15	}	_{-	0.11	}$	&	-0.43	$^{+	0.13	}	_{-	0.08	}$	&	2.37	$\pm$	0.06	&	0.053				\\
J141324	&	8.27	$^{+	0.19	}	_{-	0.10	}$	&	-1.12	$^{+	0.37	}	_{-	0.20	}$	&	0.03	$^{+	0.02	}	_{-	0.01	}$	&	-0.09	$^{+	0.19	}	_{-	0.10	}$	&	0.71	$\pm$	0.02	&	0.046				\\
J141214	&	7.47	$^{+	0.13	}	_{-	0.16	}$	&	1.15	$^{+	0.26	}	_{-	0.33	}$	&	0.58	$^{+	0.21	}	_{-	0.25	}$	&	-0.41	$^{+	0.09	}	_{-	0.13	}$	&	1.88	$\pm$	0.25	&	0.034				\\
J140518	&	7.98	$^{+	0.15	}	_{-	0.09	}$	&	0.04	$^{+	0.31	}	_{-	0.17	}$	&	0.16	$^{+	0.06	}	_{-	0.04	}$	&	-0.09	$^{+	0.16	}	_{-	0.09	}$	&	1.40	$\pm$	0.01	&	0.072				\\
J141018	&	7.77	$^{+	0.10	}	_{-	0.13	}$	&	-0.68	$^{+	0.20	}	_{-	0.27	}$	&	0.04	$^{+	0.01	}	_{-	0.02	}$	&	-0.10	$^{+	0.08	}	_{-	0.13	}$	&	1.06	$\pm$	0.09	&	0.044				\\
J141123	&	7.63	$^{+	0.05	}	_{-	0.03	}$	&	0.42	$^{+	0.10	}	_{-	0.06	}$	&	0.22	$^{+	0.05	}	_{-	0.05	}$	&	-0.48	$^{+	0.06	}	_{-	0.04	}$	&	1.13	$\pm$	0.01	&	0.071				\\
J142039	&	7.87	$^{+	0.03	}	_{-	0.07	}$	&	-0.03	$^{+	0.05	}	_{-	0.13	}$	&	0.13	$^{+	0.03	}	_{-	0.03	}$	&	-0.29	$^{+	0.04	}	_{-	0.07	}$	&	1.08	$\pm$	0.03	&	0.079				\\
J141724	&	7.73	$^{+	0.54	}	_{-	0.12	}$	&	0.01	$^{+	1.08	}	_{-	0.23	}$	&	0.12	$^{+	0.16	}	_{-	0.04	}$	&	-0.52	$^{+	0.54	}	_{-	0.12	}$	&	0.85	$\pm$	0.01	&	0.064				\\
J141004	&	7.95	$^{+	0.04	}	_{-	0.04	}$	&	-0.08	$^{+	0.08	}	_{-	0.07	}$	&	0.13	$^{+	0.03	}	_{-	0.03	}$	&	0.08	$^{+	0.05	}	_{-	0.05	}$	&	1.68	$\pm$	0.04	&	0.028				\\
J141706	&	6.76	$^{+	0.26	}	_{-	0.13	}$	&	2.25	$^{+	0.53	}	_{-	0.25	}$	&	1.83	$^{+	1.17	}	_{-	0.65	}$	&	-0.61	$^{+	0.26	}	_{-	0.13	}$	&	3.20	$\pm$	0.03	&	0.052				\\
J142010	&	7.96	$^{+	0.21	}	_{-	0.17	}$	&	-0.30	$^{+	0.42	}	_{-	0.34	}$	&	0.09	$^{+	0.05	}	_{-	0.04	}$	&	-0.47	$^{+	0.20	}	_{-	0.16	}$	&	0.72	$\pm$	0.08	&	0.133				\\
J141712	&	7.08	$^{+	0.17	}	_{-	0.18	}$	&	0.14	$^{+	0.34	}	_{-	0.36	}$	&	0.09	$^{+	0.04	}	_{-	0.04	}$	&	-0.01	$^{+	0.08	}	_{-	0.10	}$	&	2.31	$\pm$	0.49	&	0.189				\\
J141115	&	8.06	$^{+	0.10	}	_{-	0.02	}$	&	-0.15	$^{+	0.20	}	_{-	0.04	}$	&	0.12	$^{+	0.04	}	_{-	0.03	}$	&	0.003	$^{+	0.10	}	_{-	0.04	}$	&	1.39	$\pm$	0.02	&	0.064				\\
J141112	&	7.49	$^{+	0.05	}	_{-	0.04	}$	&	0.70	$^{+	0.11	}	_{-	0.09	}$	&	0.30	$^{+	0.07	}	_{-	0.07	}$	&	-0.28	$^{+	0.06	}	_{-	0.05	}$	&	1.79	$\pm$	0.03	&	0.037				\\
J141417	&	8.11	$^{+	0.14	}	_{-	0.18	}$	&	-1.63	$^{+	0.28	}	_{-	0.36	}$	&	0.01	$^{+	0.01	}	_{-	0.01	}$	&	-0.01	$^{+	0.10	}	_{-	0.15	}$	&	0.66	$\pm$	0.09	&	0.149				\\
J141031	&	7.93	$^{+	0.03	}	_{-	0.13	}$	&	-0.34	$^{+	0.06	}	_{-	0.26	}$	&	0.08	$^{+	0.02	}	_{-	0.03	}$	&	0.02	$^{+	0.03	}	_{-	0.13	}$	&	1.36	$\pm$	0.05	&	0.053				\\
J141941	&	7.68	$^{+	0.06	}	_{-	0.12	}$	&	0.92	$^{+	0.12	}	_{-	0.24	}$	&	0.48	$^{+	0.12	}	_{-	0.17	}$	&	-0.32	$^{+	0.07	}	_{-	0.12	}$	&	1.75	$\pm$	0.03	&	0.055				\\
J141135	&	7.34	$^{+	0.21	}	_{-	0.18	}$	&	0.87	$^{+	0.43	}	_{-	0.37	}$	&	0.35	$^{+	0.18	}	_{-	0.16	}$	&	-0.30	$^{+	0.21	}	_{-	0.19	}$	&	2.00	$\pm$	0.06	&	0.096				\\
J140904	&	8.39	$^{+	0.33	}	_{-	0.19	}$	&	-1.07	$^{+	0.67	}	_{-	0.39	}$	&	0.04	$^{+	0.03	}	_{-	0.02	}$	&	-0.54	$^{+	0.32	}	_{-	0.18	}$	&	0.38	$\pm$	0.05	&	0.099				\\
J142052	&	7.49	$^{+	0.05	}	_{-	0.04	}$	&	2.10	$^{+	0.10	}	_{-	0.07	}$	&	2.54	$^{+	0.58	}	_{-	0.56	}$	&	-1.02	$^{+	0.07	}	_{-	0.06	}$	&	1.30	$\pm$	0.01	&	0.022				\\
J141147	&	6.84	$^{+	0.10	}	_{-	0.10	}$	&	1.86	$^{+	0.21	}	_{-	0.20	}$	&	1.06	$^{+	0.33	}	_{-	0.32	}$	&	-0.73	$^{+	0.11	}	_{-	0.10	}$	&	2.18	$\pm$	0.07	&	0.092				\\
J141532	&	7.13	$^{+	0.16	}	_{-	0.15	}$	&	1.43	$^{+	0.33	}	_{-	0.29	}$	&	0.70	$^{+	0.30	}	_{-	0.27	}$	&	-0.18	$^{+	0.17	}	_{-	0.15	}$	&	3.36	$\pm$	0.09	&	0.274				\\
J142023	&	7.52	$^{+	0.17	}	_{-	0.20	}$	&	0.79	$^{+	0.33	}	_{-	0.40	}$	&	0.35	$^{+	0.15	}	_{-	0.18	}$	&	-0.72	$^{+	0.17	}	_{-	0.20	}$	&	1.03	$\pm$	0.04	&	0.107				\\
J142049	&	8.29	$^{+	0.09	}	_{-	0.09	}$	&	-0.43	$^{+	0.18	}	_{-	0.18	}$	&	0.10	$^{+	0.03	}	_{-	0.03	}$	&	-0.10	$^{+	0.10	}	_{-	0.10	}$	&	0.97	$\pm$	0.02	&	0.097				\\
J142112	&	7.74	$^{+	0.13	}	_{-	0.11	}$	&	0.49	$^{+	0.25	}	_{-	0.22	}$	&	0.26	$^{+	0.09	}	_{-	0.08	}$	&	-0.54	$^{+	0.12	}	_{-	0.10	}$	&	1.03	$\pm$	0.08	&	0.075				\\
J141606	&	8.52	$^{+	0.16	}	_{-	0.21	}$	&	-0.36	$^{+	0.32	}	_{-	0.42	}$	&	0.13	$^{+	0.05	}	_{-	0.07	}$	&	-0.45	$^{+	0.16	}	_{-	0.21	}$	&	0.57	$\pm$	0.02	&	0.071				\\
J141859	&	8.00	$^{+	0.12	}	_{-	0.15	}$	&	0.85	$^{+	0.24	}	_{-	0.30	}$	&	0.56	$^{+	0.19	}	_{-	0.22	}$	&	-0.70	$^{+	0.13	}	_{-	0.16	}$	&	0.90	$\pm$	0.01	&	0.056				\\
J141952	&	8.59	$^{+	0.08	}	_{-	0.08	}$	&	-1.31	$^{+	0.16	}	_{-	0.15	}$	&	0.03	$^{+	0.01	}	_{-	0.01	}$	&	-0.14	$^{+	0.08	}	_{-	0.07	}$	&	0.54	$\pm$	0.03	&	0.166				\\
J142417	&	7.32	$^{+	0.09	}	_{-	0.10	}$	&	0.98	$^{+	0.20	}	_{-	0.22	}$	&	0.40	$^{+	0.13	}	_{-	0.13	}$	&	-0.01	$^{+	0.07	}	_{-	0.08	}$	&	3.12	$\pm$	0.31	&	0.275				\\
J141856$^\dagger$	&	7.48	$^{+	0.17	}	_{-	0.05	}$	&	2.60	$^{+	0.33	}	_{-	0.11	}$	&	5.50	$^{+	2.37	}	_{-	1.31	}$	&	-1.06	$^{+	0.17	}	_{-	0.08	}$	&	1.56	$\pm$	0.03	&	0.144				\\
J141314$^\dagger$	&	7.23	$^{+	0.12	}	_{-	0.12	}$	&	1.81	$^{+	0.25	}	_{-	0.25	}$	&	1.34	$^{+	0.48	}	_{-	0.47	}$	&	-0.16	$^{+	0.06	}	_{-	0.06	}$	&	3.93	$\pm$	0.60	&	0.216				\\
\hline \noalign{\vskip 0.1cm} 																																										
\multicolumn{7}{c}{Bentz collection} \\ 																																										
\hline \noalign{\vskip 0.1cm} 																																										
PG0026+129	&	7.81	$^{+	0.27	}	_{-	0.27	}$	&	1.33	$^{+	0.44	}	_{-	0.55	}$	&	1.00	$^{+	0.65	}	_{-	0.66	}$	&	0.001	$^{+	0.11	}	_{-	0.12	}$	&	3.12	$\pm$	1.05	&	0.173	$^	2	$	\\
PG0052+251	&	8.48	$^{+	0.14	}	_{-	0.14	}$	&	-0.27	$^{+	0.43	}	_{-	0.28	}$	&	0.14	$^{+	0.06	}	_{-	0.06	}$	&	-0.004	$^{+	0.13	}	_{-	0.12	}$	&	1.11	$\pm$	0.12	&	0.199	$^	2	$	\\
Fairall9	&	8.21	$^{+	0.12	}	_{-	0.14	}$	&	-0.96	$^{+	0.50	}	_{-	0.29	}$	&	0.04	$^{+	0.01	}	_{-	0.02	}$	&	-0.28	$^{+	0.09	}	_{-	0.11	}$	&	0.61	$\pm$	0.07	&	0.328	$^	2	$	\\
Mrk590	&	7.39	$^{+	0.30	}	_{-	0.29	}$	&	-0.05	$^{+	0.43	}	_{-	0.66	}$	&	0.09	$^{+	0.08	}	_{-	0.07	}$	&	0.15	$^{+	0.16	}	_{-	0.15	}$	&	2.32	$\pm$	0.86	&	0.108	$^	2	$	\\
3C120	&	7.46	$^{+	0.23	}	_{-	0.21	}$	&	0.58	$^{+	0.40	}	_{-	0.45	}$	&	0.24	$^{+	0.15	}	_{-	0.14	}$	&	-0.11	$^{+	0.16	}	_{-	0.13	}$	&	2.14	$\pm$	0.53	&	0.178	$^	2	$	\\
Ark120	&	8.36	$^{+	0.11	}	_{-	0.10	}$	&	-1.42	$^{+	0.43	}	_{-	0.43	}$	&	0.02	$^{+	0.01	}	_{-	0.01	}$	&	0.14	$^{+	0.17	}	_{-	0.16	}$	&	0.82	$\pm$	0.06	&	0.072	$^	2	$	\\
Mrk79	&	7.86	$^{+	0.32	}	_{-	0.31	}$	&	-0.71	$^{+	0.43	}	_{-	0.63	}$	&	0.05	$^{+	0.04	}	_{-	0.03	}$	&	-0.16	$^{+	0.16	}	_{-	0.14	}$	&	0.93	$\pm$	0.35	&	0.091	$^	2	$	\\
PG0804+761	&	8.07	$^{+	0.37	}	_{-	0.37	}$	&	0.72	$^{+	0.43	}	_{-	0.74	}$	&	0.48	$^{+	0.42	}	_{-	0.42	}$	&	0.16	$^{+	0.07	}	_{-	0.07	}$	&	2.60	$\pm$	1.28	&	0.176	$^	2	$	\\
Mrk110	&	7.05	$^{+	0.49	}	_{-	0.48	}$	&	0.88	$^{+	0.43	}	_{-	0.98	}$	&	0.28	$^{+	0.33	}	_{-	0.32	}$	&	0.06	$^{+	0.17	}	_{-	0.14	}$	&	3.68	$\pm$	2.31	&	0.184	$^	2	$	\\
PG0953+414	&	8.42	$^{+	0.13	}	_{-	0.13	}$	&	0.42	$^{+	0.44	}	_{-	0.27	}$	&	0.40	$^{+	0.14	}	_{-	0.15	}$	&	0.02	$^{+	0.08	}	_{-	0.08	}$	&	1.63	$\pm$	0.25	&	0.136	$^	2	$	\\
NGC3227	&	6.98	$^{+	0.09	}	_{-	0.09	}$	&	-1.11	$^{+	0.43	}	_{-	0.25	}$	&	0.01	$^{+	0.005	}	_{-	0.005	}$	&	-0.01	$^{+	0.13	}	_{-	0.13	}$	&	1.32	$\pm$	0.04	&	0.094	$^	2	$	\\
NGC3516	&	7.79	$^{+	0.04	}	_{-	0.06	}$	&	-1.90	$^{+	0.45	}	_{-	0.33	}$	&	0.01	$^{+	0.004	}	_{-	0.004	}$	&	0.19	$^{+	0.13	}	_{-	0.13	}$	&	0.86	$\pm$	0.02	&	0.289	$^	2	$	\\
SBS1116+583A	&	6.77	$^{+	0.29	}	_{-	0.29	}$	&	-0.84	$^{+	0.43	}	_{-	0.67	}$	&	0.02	$^{+	0.01	}	_{-	0.01	}$	&	-0.17	$^{+	0.18	}	_{-	0.17	}$	&	1.31	$\pm$	0.48	&	0.043	$^	3	$	\\
Arp151	&	6.64	$^{+	0.08	}	_{-	0.09	}$	&	0.03	$^{+	0.48	}	_{-	0.24	}$	&	0.06	$^{+	0.02	}	_{-	0.02	}$	&	-0.15	$^{+	0.10	}	_{-	0.11	}$	&	2.16	$\pm$	0.15	&	0.120	$^	3	$	\\
NGC3783	&	7.28	$^{+	0.20	}	_{-	0.18	}$	&	-1.24	$^{+	0.40	}	_{-	0.45	}$	&	0.01	$^{+	0.01	}	_{-	0.01	}$	&	0.25	$^{+	0.18	}	_{-	0.15	}$	&	1.57	$\pm$	0.31	&	0.192	$^	2	$	\\
Mrk1310	&	6.27	$^{+	0.15	}	_{-	0.15	}$	&	0.39	$^{+	0.43	}	_{-	0.37	}$	&	0.07	$^{+	0.04	}	_{-	0.04	}$	&	-0.05	$^{+	0.12	}	_{-	0.12	}$	&	3.39	$\pm$	0.62	&	0.051	$^	3	$	\\
NGC4051	&	5.64	$^{+	0.19	}	_{-	0.15	}$	&	1.05	$^{+	0.37	}	_{-	0.38	}$	&	0.12	$^{+	0.07	}	_{-	0.07	}$	&	-0.08	$^{+	0.22	}	_{-	0.18	}$	&	5.66	$\pm$	0.26	&	0.059	$^	2	$	\\
NGC4151	&	7.46	$^{+	0.16	}	_{-	0.15	}$	&	-2.29	$^{+	0.42	}	_{-	0.43	}$	&	0.003	$^{+	0.002	}	_{-	0.002	}$	&	0.31	$^{+	0.15	}	_{-	0.14	}$	&	0.96	$\pm$	0.18	&	0.058	$^	2	$	\\
Mrk202	&	6.03	$^{+	0.29	}	_{-	0.23	}$	&	0.82	$^{+	0.35	}	_{-	0.50	}$	&	0.12	$^{+	0.09	}	_{-	0.08	}$	&	-0.12	$^{+	0.27	}	_{-	0.19	}$	&	4.13	$\pm$	0.89	&	0.027	$^	3	$	\\
NGC4253	&	5.93	$^{+	1.32	}	_{-	1.31	}$	&	1.49	$^{+	0.43	}	_{-	2.64	}$	&	0.30	$^{+	0.92	}	_{-	0.92	}$	&	0.03	$^{+	0.14	}	_{-	0.12	}$	&	7.28	$\pm$	12.87	&	0.053	$^	3	$	\\
PG1229+204	&	7.94	$^{+	0.33	}	_{-	0.19	}$	&	-0.84	$^{+	0.26	}	_{-	0.39	}$	&	0.04	$^{+	0.03	}	_{-	0.02	}$	&	0.21	$^{+	0.32	}	_{-	0.18	}$	&	1.40	$\pm$	0.15	&	0.107	$^	2	$	\\
NGC4593	&	7.15	$^{+	0.14	}	_{-	0.14	}$	&	-0.89	$^{+	0.43	}	_{-	0.62	}$	&	0.02	$^{+	0.02	}	_{-	0.02	}$	&	-0.19	$^{+	0.22	}	_{-	0.22	}$	&	1.08	$\pm$	0.16	&	0.114	$^	2	$	\\
NGC4748	&	6.20	$^{+	0.18	}	_{-	0.21	}$	&	0.93	$^{+	0.51	}	_{-	0.47	}$	&	0.16	$^{+	0.08	}	_{-	0.09	}$	&	-0.02	$^{+	0.15	}	_{-	0.20	}$	&	4.70	$\pm$	0.79	&	0.045	$^	3	$	\\
PG1307+085	&	8.72	$^{+	0.17	}	_{-	0.21	}$	&	-0.68	$^{+	0.53	}	_{-	0.42	}$	&	0.09	$^{+	0.04	}	_{-	0.05	}$	&	0.04	$^{+	0.15	}	_{-	0.20	}$	&	0.88	$\pm$	0.11	&	0.113	$^	2	$	\\
Mrk279	&	7.57	$^{+	0.14	}	_{-	0.14	}$	&	-0.10	$^{+	0.43	}	_{-	0.29	}$	&	0.09	$^{+	0.04	}	_{-	0.04	}$	&	-0.15	$^{+	0.11	}	_{-	0.11	}$	&	1.41	$\pm$	0.17	&	0.082	$^	2	$	\\
PG1411+442	&	8.15	$^{+	0.25	}	_{-	0.25	}$	&	0.04	$^{+	0.44	}	_{-	0.50	}$	&	0.18	$^{+	0.11	}	_{-	0.11	}$	&	0.27	$^{+	0.22	}	_{-	0.22	}$	&	2.12	$\pm$	0.36	&	0.105	$^	2	$	\\
PG1426+015	&	8.87	$^{+	0.22	}	_{-	0.25	}$	&	-1.31	$^{+	0.48	}	_{-	0.49	}$	&	0.04	$^{+	0.02	}	_{-	0.02	}$	&	0.12	$^{+	0.14	}	_{-	0.17	}$	&	0.68	$\pm$	0.16	&	0.173	$^	2	$	\\
Mrk817	&	7.82	$^{+	0.33	}	_{-	0.29	}$	&	-0.54	$^{+	0.38	}	_{-	0.60	}$	&	0.06	$^{+	0.05	}	_{-	0.04	}$	&	-0.09	$^{+	0.22	}	_{-	0.16	}$	&	1.12	$\pm$	0.38	&	0.050	$^	4	$	\\
Mrk290	&	7.49	$^{+	0.07	}	_{-	0.06	}$	&	-0.74	$^{+	0.40	}	_{-	0.15	}$	&	0.03	$^{+	0.01	}	_{-	0.01	}$	&	-0.14	$^{+	0.08	}	_{-	0.07	}$	&	1.08	$\pm$	0.05	&	0.180	$^	4	$	\\
PG1613+658	&	8.69	$^{+	0.26	}	_{-	0.26	}$	&	-0.73	$^{+	0.44	}	_{-	0.51	}$	&	0.08	$^{+	0.05	}	_{-	0.05	}$	&	-0.34	$^{+	0.17	}	_{-	0.17	}$	&	0.52	$\pm$	0.14	&	0.123	$^	2	$	\\
PG1617+175	&	8.49	$^{+	0.26	}	_{-	0.27	}$	&	-0.91	$^{+	0.46	}	_{-	0.55	}$	&	0.05	$^{+	0.03	}	_{-	0.04	}$	&	0.12	$^{+	0.18	}	_{-	0.21	}$	&	0.96	$\pm$	0.24	&	0.191	$^	2	$	\\
PG1700+518	&	8.23	$^{+	0.33	}	_{-	0.33	}$	&	1.42	$^{+	0.43	}	_{-	0.66	}$	&	1.58	$^{+	1.24	}	_{-	1.23	}$	&	0.03	$^{+	0.10	}	_{-	0.09	}$	&	2.87	$\pm$	1.24	&	0.060	$^	2	$	\\
3C390.3	&	8.98	$^{+	0.32	}	_{-	0.23	}$	&	-1.81	$^{+	0.40	}	_{-	0.98	}$	&	0.02	$^{+	0.03	}	_{-	0.03	}$	&	-0.11	$^{+	0.41	}	_{-	0.35	}$	&	0.38	$\pm$	0.08	&	0.343	$^	2	$	\\
NGC6814	&	7.14	$^{+	0.10	}	_{-	0.10	}$	&	-1.62	$^{+	0.43	}	_{-	0.47	}$	&	0.01	$^{+	0.005	}	_{-	0.005	}$	&	0.29	$^{+	0.18	}	_{-	0.18	}$	&	1.47	$\pm$	0.16	&	0.068	$^	3	$	\\
Mrk509	&	8.06	$^{+	0.05	}	_{-	0.04	}$	&	-0.34	$^{+	0.42	}	_{-	0.11	}$	&	0.09	$^{+	0.02	}	_{-	0.02	}$	&	0.27	$^{+	0.05	}	_{-	0.05	}$	&	1.83	$\pm$	0.08	&	0.181	$^	2	$	\\
PG2130+099	&	6.92	$^{+	0.07	}	_{-	0.07	}$	&	1.96	$^{+	0.43	}	_{-	0.14	}$	&	1.33	$^{+	0.35	}	_{-	0.35	}$	&	-0.65	$^{+	0.06	}	_{-	0.06	}$	&	2.48	$\pm$	0.14	&	0.086	$^	2	$	\\
NGC7469	&	6.38	$^{+	0.15	}	_{-	0.09	}$	&	1.99	$^{+	0.30	}	_{-	0.24	}$	&	0.92	$^{+	0.44	}	_{-	0.35	}$	&	-0.23	$^{+	0.15	}	_{-	0.09	}$	&	5.46	$\pm$	0.50	&	0.150	$^	2	$	\\
PG1211+143	&	7.87	$^{+	0.12	}	_{-	0.20	}$	&	0.84	$^{+	0.66	}	_{-	0.41	}$	&	0.50	$^{+	0.19	}	_{-	0.26	}$	&	0.06	$^{+	0.13	}	_{-	0.20	}$	&	2.60	$\pm$	0.06	&	0.134	$^	2	$	\\
PG0844+349	&	7.57	$^{+	0.22	}	_{-	0.21	}$	&	0.67	$^{+	0.43	}	_{-	0.44	}$	&	0.30	$^{+	0.17	}	_{-	0.17	}$	&	-0.13	$^{+	0.19	}	_{-	0.19	}$	&	2.08	$\pm$	0.33	&	0.105	$^	2	$	\\
NGC5273	&	6.96	$^{+	0.24	}	_{-	0.32	}$	&	-2.13	$^{+	0.55	}	_{-	0.69	}$	&	0.003	$^{+	0.002	}	_{-	0.002	}$	&	0.13	$^{+	0.27	}	_{-	0.34	}$	&	0.98	$\pm$	0.08	&	0.059	$^	5	$	\\
Mrk1511	&	7.29	$^{+	0.07	}	_{-	0.07	}$	&	-0.34	$^{+	0.41	}	_{-	0.17	}$	&	0.05	$^{+	0.02	}	_{-	0.02	}$	&	-0.32	$^{+	0.09	}	_{-	0.08	}$	&	1.11	$\pm$	0.04	&	0.150	$^	6	$	\\
KA1858-4850	&	6.78	$^{+	0.08	}	_{-	0.08	}$	&	1.07	$^{+	0.48	}	_{-	0.18	}$	&	0.31	$^{+	0.09	}	_{-	0.09	}$	&	-0.09	$^{+	0.08	}	_{-	0.09	}$	&	3.63	$\pm$	0.19	&	0.084	$^	7	$	\\
MCG6-30-15	&	6.35	$^{+	0.29	}	_{-	0.28	}$	&	-0.75	$^{+	0.43	}	_{-	0.59	}$	&	0.01	$^{+	0.01	}	_{-	0.01	}$	&	0.49	$^{+	0.17	}	_{-	0.16	}$	&	3.90	$\pm$	1.34	&	0.132	$^	8	$	\\
UGC06728	&	5.56	$^{+	0.22	}	_{-	0.25	}$	&	1.17	$^{+	0.49	}	_{-	0.52	}$	&	0.14	$^{+	0.08	}	_{-	0.09	}$	&	-0.24	$^{+	0.24	}	_{-	0.26	}$	&	5.03	$\pm$	0.30	&	0.090	$^	9	$	\\
MCG+08-11-011	&	6.62	$^{+	0.02	}	_{-	0.02	}$	&	1.25	$^{+	0.43	}	_{-	0.17	}$	&	0.36	$^{+	0.12	}	_{-	0.12	}$	&	0.03	$^{+	0.07	}	_{-	0.07	}$	&	4.95	$\pm$	0.04	&	0.100	$^	{10}	$	\\
NGC2617	&	7.37	$^{+	0.11	}	_{-	0.14	}$	&	-1.26	$^{+	0.48	}	_{-	0.36	}$	&	0.01	$^{+	0.01	}	_{-	0.01	}$	&	-0.18	$^{+	0.15	}	_{-	0.17	}$	&	0.84	$\pm$	0.01	&	0.090	$^	{10}	$	\\
3C382	&	8.02	$^{+	0.11	}	_{-	0.04	}$	&	-0.79	$^{+	0.28	}	_{-	0.17	}$	&	0.05	$^{+	0.02	}	_{-	0.01	}$	&	0.17	$^{+	0.11	}	_{-	0.07	}$	&	1.31	$\pm$	0.12	&	0.090	$^	{10}	$	\\
Mrk374	&	7.49	$^{+	0.17	}	_{-	0.10	}$	&	0.18	$^{+	0.26	}	_{-	0.20	}$	&	0.13	$^{+	0.06	}	_{-	0.04	}$	&	-0.24	$^{+	0.17	}	_{-	0.10	}$	&	1.48	$\pm$	0.01	&	0.030	$^	{10}	$	\\
\hline \noalign{\vskip 0.1cm} 																																										
\multicolumn{7}{c}{\citet{lu2016}} \\ 																																										
\hline \noalign{\vskip 0.1cm} 																																										
NGC5548	&	8.14	$^{+	0.08	}	_{-	0.04	}$	&	-1.98	$^{+	0.25	}	_{-	0.20	}$	&	0.008	$^{+	0.003	}	_{-	0.003	}$	&	-0.25	$^{+	0.11	}	_{-	0.08	}$	&	0.40	$\pm$	0.02	&	0.230	$^	{11}	$	\\
\hline \noalign{\vskip 0.1cm} 																																										
\multicolumn{7}{c}{\citet{zhang2018}} \\ 																																										
\hline \noalign{\vskip 0.1cm} 																																										
3C 273	&	8.50	$^{+	0.03	}	_{-	0.04	}$	&	1.44	$^{+	0.06	}	_{-	0.08	}$	&	2.01	$^{+	0.43	}	_{-	0.45	}$	&	-0.41	$^{+	0.08	}	_{-	0.08	}$	&	1.45	$\pm$	0.03	&	0.052	$^	2	$	\\

\hline \noalign{\vskip 0.1cm}       
\end{longtable}
\end{center}
\footnotesize{{\sc Notes.} Columns are as follows: (1) Object name. Discarded object of the analysis are marked $\dagger$ symbol. (2) Black hole mass in units of solar masses considering. (3) Dimensionless accretion rate, see Equation~(\ref{equ:mdot}). (4) Eddington ratio. (5) Deviations of BLR size from $R_{H\beta}-L_{5100}$ relation. (6) Virial factor anti-correlated with the FWHM of \hb, see Equation~(\ref{equ:fblrc}). \fvar\ value. In some cases is specified the origin of the estimation: (1) \citet{hu2015}, (2) \citet{peterson2004}, (3) \citet{bentz2009}, (4) \citet{denney2010}, (5) \citet{bentz2014}, (6) \citet{barth2013}, (7) \citet{pei2014}, (8) \citet{bentz2016a}, (9) \citet{bentz2016b}, (10) \citet{fausnaugh2017}, (11) \citet{lu2016}. \fvar\ was estimated for the sources without reference. Columns 2,3,4 and 5 have been computed considering \fblr=1. }

\newpage

\begin{center}
\scriptsize
\small
\begin{longtable}{ccccc}
\caption{Observational properties corrected by the effect of dimensionless accretion rate} \label{tab:mea_corr} \\ 
 \hline\hline\noalign{\vskip 0.1cm}          
\multirow{2}{*}{Object} & \multirow{2}{*}{log \mdotc} &   \multirow{2}{*}{$\Delta R\mathrm{_{H\beta,{ \dot{\mathscr{M}}\mathrm{^{c}}}}}$} & \multirow{1}{*}{$\tau\mathrm{_{corr}}$} & \multirow{1}{*}{ $D\mathrm{_{L,corr}}$}   \\
 & & & [days] & [Mpc] \\
(1) & (2) & (3) & (4) & (5) \\ 
\hline \noalign{\vskip 0.1cm} 													\hline \noalign{\vskip 0.1cm} 																													
\multicolumn{5}{c}{SEAMBH sample} \\ 																													
\hline \noalign{\vskip 0.1cm} 																													
Mrk335	&	0.19	$^{+	0.34	}	_{-	0.28	}$	&	-0.28	$^{+	0.10	}	_{-	0.08	}$	&	26.7	$^{+	8.8	}	_{-	6.5	}$	&	136.1	$\pm$	38.9	\\
Mrk142	&	0.81	$^{+	0.99	}	_{-	0.47	}$	&	-0.46	$^{+	0.28	}	_{-	0.14	}$	&	18.4	$^{+	21.0	}	_{-	9.8	}$	&	199.7	$\pm$	167.0	\\
IRASF12397	&	1.31	$^{+	0.80	}	_{-	0.65	}$	&	-0.60	$^{+	0.23	}	_{-	0.19	}$	&	38.6	$^{+	21.9	}	_{-	7.2	}$	&	192.6	$\pm$	72.5	\\
Mrk486	&	-0.32	$^{+	0.29	}	_{-	0.14	}$	&	-0.14	$^{+	0.08	}	_{-	0.04	}$	&	32.5	$^{+	10.3	}	_{-	3.7	}$	&	271.9	$\pm$	58.5	\\
Mrk382	&	0.03	$^{+	0.54	}	_{-	0.49	}$	&	-0.24	$^{+	0.15	}	_{-	0.14	}$	&	12.9	$^{+	5.0	}	_{-	3.4	}$	&	180.8	$\pm$	59.1	\\
IRAS04416	&	1.52	$^{+	0.91	}	_{-	0.12	}$	&	-0.66	$^{+	0.26	}	_{-	0.05	}$	&	60.5	$^{+	63.2	}	_{-	6.4	}$	&	490.6	$\pm$	282.2	\\
MCG06	&	-1.59	$^{+	0.36	}	_{-	0.26	}$	&	0.22	$^{+	0.11	}	_{-	0.08	}$	&	14.4	$^{+	5.0	}	_{-	2.9	}$	&	327.5	$\pm$	90.1	\\
Mrk493	&	0.09	$^{+	0.15	}	_{-	0.23	}$	&	-0.25	$^{+	0.04	}	_{-	0.07	}$	&	20.8	$^{+	2.1	}	_{-	4.7	}$	&	268.0	$\pm$	43.9	\\
Mrk1044	&	-0.15	$^{+	0.31	}	_{-	0.27	}$	&	-0.19	$^{+	0.09	}	_{-	0.08	}$	&	16.1	$^{+	5.1	}	_{-	4.1	}$	&	114.9	$\pm$	32.8	\\
J080101	&	1.46	$^{+	1.02	}	_{-	0.29	}$	&	-0.64	$^{+	0.29	}	_{-	0.09	}$	&	36.4	$^{+	42.5	}	_{-	11.8	}$	&	602.2	$\pm$	449.9	\\
J081456	&	-0.05	$^{+	0.29	}	_{-	0.59	}$	&	-0.21	$^{+	0.08	}	_{-	0.17	}$	&	39.7	$^{+	12.6	}	_{-	26.8	}$	&	767.7	$\pm$	380.7	\\
J093922	&	1.18	$^{+	0.17	}	_{-	0.46	}$	&	-0.56	$^{+	0.05	}	_{-	0.13	}$	&	43.6	$^{+	7.7	}	_{-	23.1	}$	&	1241.0	$\pm$	438.1	\\
J080131	&	0.99	$^{+	0.57	}	_{-	0.29	}$	&	-0.51	$^{+	0.16	}	_{-	0.08	}$	&	37.0	$^{+	24.3	}	_{-	12.0	}$	&	1159.0	$\pm$	567.9	\\
J085946	&	0.51	$^{+	0.48	}	_{-	0.66	}$	&	-0.37	$^{+	0.14	}	_{-	0.19	}$	&	82.0	$^{+	45.2	}	_{-	62.0	}$	&	2141.8	$\pm$	1400.3	\\
J102339	&	0.31	$^{+	0.69	}	_{-	0.15	}$	&	-0.32	$^{+	0.20	}	_{-	0.05	}$	&	51.6	$^{+	41.1	}	_{-	8.1	}$	&	1019.5	$\pm$	485.2	\\
J074352	&	1.31	$^{+	0.11	}	_{-	0.09	}$	&	-0.60	$^{+	0.04	}	_{-	0.04	}$	&	174.2	$^{+	20.6	}	_{-	16.7	}$	&	1563.7	$\pm$	167.5	\\
J075051	&	1.25	$^{+	0.24	}	_{-	0.13	}$	&	-0.58	$^{+	0.07	}	_{-	0.05	}$	&	254.6	$^{+	71.5	}	_{-	37.8	}$	&	4086.4	$\pm$	877.5	\\
J075101	&	0.39	$^{+	0.22	}	_{-	0.18	}$	&	-0.34	$^{+	0.06	}	_{-	0.05	}$	&	66.2	$^{+	15.8	}	_{-	12.6	}$	&	968.3	$\pm$	208.2	\\
J075949	&	-0.05	$^{+	0.66	}	_{-	0.39	}$	&	-0.21	$^{+	0.19	}	_{-	0.11	}$	&	71.8	$^{+	54.3	}	_{-	31.2	}$	&	1803.4	$\pm$	1072.6	\\
J081441	&	0.07	$^{+	0.36	}	_{-	0.26	}$	&	-0.25	$^{+	0.10	}	_{-	0.08	}$	&	44.7	$^{+	18.4	}	_{-	13.3	}$	&	1264.3	$\pm$	447.5	\\
J083553	&	1.43	$^{+	0.38	}	_{-	0.38	}$	&	-0.63	$^{+	0.11	}	_{-	0.11	}$	&	53.4	$^{+	23.3	}	_{-	23.3	}$	&	1107.6	$\pm$	482.4	\\
J084533	&	1.44	$^{+	0.29	}	_{-	0.23	}$	&	-0.64	$^{+	0.09	}	_{-	0.07	}$	&	78.5	$^{+	26.1	}	_{-	20.5	}$	&	2307.1	$\pm$	683.8	\\
J093302	&	0.85	$^{+	0.26	}	_{-	0.28	}$	&	-0.47	$^{+	0.08	}	_{-	0.08	}$	&	55.9	$^{+	11.2	}	_{-	12.7	}$	&	1144.7	$\pm$	244.0	\\
J100402	&	2.10	$^{+	1.17	}	_{-	0.11	}$	&	-0.82	$^{+	0.33	}	_{-	0.05	}$	&	214.2	$^{+	289.5	}	_{-	28.0	}$	&	2182.5	$\pm$	1616.9	\\
J101000	&	1.02	$^{+	0.74	}	_{-	0.24	}$	&	-0.52	$^{+	0.21	}	_{-	0.07	}$	&	91.0	$^{+	77.2	}	_{-	25.0	}$	&	1677.6	$\pm$	941.8	\\
\hline \noalign{\vskip 0.1cm} 																													
\multicolumn{5}{c}{SDSS sample} \\ 																													
\hline \noalign{\vskip 0.1cm} 																													
J140812	&	-1.00	$^{+	0.27	}	_{-	0.32	}$	&	0.06	$^{+	0.08	}	_{-	0.09	}$	&	9.2	$^{+	0.9	}	_{-	1.9	}$	&	450.4	$\pm$	68.6	\\
J141923	&	-0.87	$^{+	0.06	}	_{-	0.11	}$	&	0.02	$^{+	0.03	}	_{-	0.04	}$	&	11.3	$^{+	0.7	}	_{-	1.4	}$	&	767.1	$\pm$	71.5	\\
J140759	&	-0.63	$^{+	0.70	}	_{-	0.35	}$	&	-0.05	$^{+	0.20	}	_{-	0.10	}$	&	18.3	$^{+	14.7	}	_{-	7.4	}$	&	843.3	$\pm$	509.6	\\
J141729	&	-0.78	$^{+	0.90	}	_{-	0.34	}$	&	-0.01	$^{+	0.26	}	_{-	0.10	}$	&	5.6	$^{+	5.8	}	_{-	2.1	}$	&	513.3	$\pm$	364.0	\\
J141645.15	&	-0.66	$^{+	0.26	}	_{-	0.25	}$	&	-0.04	$^{+	0.08	}	_{-	0.07	}$	&	5.5	$^{+	1.7	}	_{-	1.5	}$	&	571.3	$\pm$	165.7	\\
J142135	&	0.58	$^{+	0.21	}	_{-	0.21	}$	&	-0.39	$^{+	0.06	}	_{-	0.06	}$	&	9.6	$^{+	2.2	}	_{-	2.2	}$	&	756.9	$\pm$	174.7	\\
J141625	&	0.05	$^{+	0.19	}	_{-	0.27	}$	&	-0.24	$^{+	0.06	}	_{-	0.08	}$	&	26.3	$^{+	5.6	}	_{-	8.0	}$	&	1252.6	$\pm$	323.5	\\
J142103	&	-1.72	$^{+	0.06	}	_{-	0.06	}$	&	0.26	$^{+	0.04	}	_{-	0.04	}$	&	41.4	$^{+	1.8	}	_{-	1.8	}$	&	2868.9	$\pm$	124.0	\\
J142038	&	-1.36	$^{+	0.16	}	_{-	0.20	}$	&	0.16	$^{+	0.05	}	_{-	0.06	}$	&	17.5	$^{+	3.3	}	_{-	4.0	}$	&	1503.1	$\pm$	310.2	\\
J142043	&	-0.15	$^{+	0.08	}	_{-	0.10	}$	&	-0.19	$^{+	0.03	}	_{-	0.03	}$	&	9.1	$^{+	0.6	}	_{-	0.9	}$	&	1096.8	$\pm$	93.0	\\
J141041	&	-0.74	$^{+	0.17	}	_{-	0.10	}$	&	-0.02	$^{+	0.05	}	_{-	0.03	}$	&	22.8	$^{+	4.4	}	_{-	2.5	}$	&	1825.9	$\pm$	275.1	\\
J141318	&	-0.21	$^{+	0.05	}	_{-	0.13	}$	&	-0.17	$^{+	0.02	}	_{-	0.04	}$	&	29.4	$^{+	1.6	}	_{-	4.4	}$	&	2080.1	$\pm$	213.2	\\
J141955	&	-0.98	$^{+	0.49	}	_{-	0.40	}$	&	0.05	$^{+	0.14	}	_{-	0.11	}$	&	9.5	$^{+	5.0	}	_{-	3.9	}$	&	1498.1	$\pm$	700.0	\\
J141645.58	&	0.46	$^{+	0.26	}	_{-	0.15	}$	&	-0.36	$^{+	0.08	}	_{-	0.05	}$	&	19.5	$^{+	5.7	}	_{-	3.2	}$	&	2352.9	$\pm$	539.8	\\
J141324	&	-0.83	$^{+	0.37	}	_{-	0.20	}$	&	0.01	$^{+	0.11	}	_{-	0.06	}$	&	25.1	$^{+	10.7	}	_{-	5.7	}$	&	2317.8	$\pm$	759.0	\\
J141214	&	0.60	$^{+	0.29	}	_{-	0.35	}$	&	-0.40	$^{+	0.08	}	_{-	0.10	}$	&	53.5	$^{+	10.5	}	_{-	16.0	}$	&	2950.6	$\pm$	730.8	\\
J140518	&	-0.26	$^{+	0.31	}	_{-	0.17	}$	&	-0.16	$^{+	0.09	}	_{-	0.05	}$	&	59.5	$^{+	21.2	}	_{-	11.9	}$	&	3619.4	$\pm$	1004.9	\\
J141018	&	-0.73	$^{+	0.21	}	_{-	0.28	}$	&	-0.02	$^{+	0.06	}	_{-	0.08	}$	&	17.0	$^{+	3.0	}	_{-	4.7	}$	&	2464.1	$\pm$	562.8	\\
J141123	&	0.31	$^{+	0.10	}	_{-	0.06	}$	&	-0.32	$^{+	0.03	}	_{-	0.02	}$	&	27.0	$^{+	2.9	}	_{-	1.7	}$	&	2101.4	$\pm$	177.8	\\
J142039	&	-0.10	$^{+	0.06	}	_{-	0.13	}$	&	-0.20	$^{+	0.02	}	_{-	0.04	}$	&	32.8	$^{+	1.4	}	_{-	4.8	}$	&	2535.0	$\pm$	238.8	\\
J141724	&	0.15	$^{+	1.07	}	_{-	0.23	}$	&	-0.27	$^{+	0.30	}	_{-	0.07	}$	&	18.9	$^{+	23.4	}	_{-	5.0	}$	&	1763.9	$\pm$	1327.3	\\
J141004	&	-0.53	$^{+	0.08	}	_{-	0.08	}$	&	-0.08	$^{+	0.03	}	_{-	0.03	}$	&	64.2	$^{+	5.0	}	_{-	4.8	}$	&	5113.1	$\pm$	391.8	\\
J141706	&	1.24	$^{+	0.53	}	_{-	0.25	}$	&	-0.58	$^{+	0.15	}	_{-	0.08	}$	&	39.4	$^{+	23.9	}	_{-	11.4	}$	&	3318.1	$\pm$	1483.7	\\
J142010	&	-0.01	$^{+	0.43	}	_{-	0.35	}$	&	-0.22	$^{+	0.12	}	_{-	0.10	}$	&	21.5	$^{+	9.6	}	_{-	7.5	}$	&	2096.5	$\pm$	835.3	\\
J141712	&	-0.59	$^{+	0.39	}	_{-	0.41	}$	&	-0.06	$^{+	0.11	}	_{-	0.12	}$	&	14.4	$^{+	2.1	}	_{-	3.0	}$	&	3916.2	$\pm$	689.3	\\
J141115	&	-0.44	$^{+	0.20	}	_{-	0.05	}$	&	-0.10	$^{+	0.06	}	_{-	0.02	}$	&	62.4	$^{+	14.1	}	_{-	2.5	}$	&	4955.6	$\pm$	661.1	\\
J141112	&	0.20	$^{+	0.11	}	_{-	0.09	}$	&	-0.28	$^{+	0.04	}	_{-	0.03	}$	&	39.3	$^{+	4.8	}	_{-	3.8	}$	&	4003.1	$\pm$	441.5	\\
J141417	&	-1.27	$^{+	0.30	}	_{-	0.38	}$	&	0.13	$^{+	0.09	}	_{-	0.11	}$	&	11.5	$^{+	2.4	}	_{-	3.8	}$	&	2803.4	$\pm$	745.8	\\
J141031	&	-0.61	$^{+	0.07	}	_{-	0.26	}$	&	-0.06	$^{+	0.03	}	_{-	0.08	}$	&	40.6	$^{+	1.2	}	_{-	11.7	}$	&	4857.2	$\pm$	773.3	\\
J141941	&	0.44	$^{+	0.12	}	_{-	0.24	}$	&	-0.35	$^{+	0.04	}	_{-	0.07	}$	&	68.4	$^{+	8.8	}	_{-	18.7	}$	&	4947.3	$\pm$	992.8	\\
J141135	&	0.27	$^{+	0.43	}	_{-	0.37	}$	&	-0.30	$^{+	0.12	}	_{-	0.11	}$	&	35.5	$^{+	17.4	}	_{-	14.9	}$	&	4509.7	$\pm$	2049.9	\\
J140904	&	-0.23	$^{+	0.68	}	_{-	0.40	}$	&	-0.16	$^{+	0.19	}	_{-	0.12	}$	&	16.9	$^{+	12.5	}	_{-	6.7	}$	&	1923.8	$\pm$	1094.6	\\
J142052	&	1.87	$^{+	0.10	}	_{-	0.07	}$	&	-0.76	$^{+	0.05	}	_{-	0.04	}$	&	68.2	$^{+	7.5	}	_{-	5.7	}$	&	2810.9	$\pm$	271.8	\\
J141147	&	1.18	$^{+	0.21	}	_{-	0.20	}$	&	-0.56	$^{+	0.07	}	_{-	0.06	}$	&	23.4	$^{+	5.5	}	_{-	5.1	}$	&	3188.0	$\pm$	722.4	\\
J141532	&	0.38	$^{+	0.33	}	_{-	0.29	}$	&	-0.34	$^{+	0.09	}	_{-	0.08	}$	&	57.4	$^{+	21.4	}	_{-	19.0	}$	&	7326.8	$\pm$	2585.2	\\
J142023	&	0.76	$^{+	0.33	}	_{-	0.40	}$	&	-0.44	$^{+	0.10	}	_{-	0.12	}$	&	23.7	$^{+	8.9	}	_{-	10.9	}$	&	2827.5	$\pm$	1181.0	\\
J142049	&	-0.40	$^{+	0.18	}	_{-	0.18	}$	&	-0.11	$^{+	0.06	}	_{-	0.06	}$	&	59.9	$^{+	12.4	}	_{-	12.4	}$	&	5688.3	$\pm$	1174.8	\\
J142112	&	0.46	$^{+	0.26	}	_{-	0.23	}$	&	-0.36	$^{+	0.08	}	_{-	0.07	}$	&	32.4	$^{+	8.5	}	_{-	6.9	}$	&	4128.8	$\pm$	974.1	\\
J141606	&	0.12	$^{+	0.32	}	_{-	0.42	}$	&	-0.26	$^{+	0.09	}	_{-	0.12	}$	&	58.7	$^{+	21.3	}	_{-	28.4	}$	&	4298.4	$\pm$	1820.2	\\
J141859	&	0.95	$^{+	0.24	}	_{-	0.30	}$	&	-0.50	$^{+	0.07	}	_{-	0.09	}$	&	64.0	$^{+	17.6	}	_{-	22.0	}$	&	4366.7	$\pm$	1348.6	\\
J141952	&	-0.77	$^{+	0.17	}	_{-	0.16	}$	&	-0.01	$^{+	0.05	}	_{-	0.05	}$	&	33.6	$^{+	5.7	}	_{-	5.2	}$	&	4911.1	$\pm$	798.6	\\
J142417	&	-0.01	$^{+	0.22	}	_{-	0.23	}$	&	-0.22	$^{+	0.06	}	_{-	0.07	}$	&	60.9	$^{+	7.6	}	_{-	9.2	}$	&	10753.6	$\pm$	1481.3	\\
\hline \noalign{\vskip 0.1cm} 																													
\multicolumn{5}{c}{Bentz collection} \\ 																													
\hline \noalign{\vskip 0.1cm} 																													
PG0026+129	&	0.34	$^{+	0.61	}	_{-	0.62	}$	&	-0.32	$^{+	0.17	}	_{-	0.18	}$	&	234.4	$^{+	50.9	}	_{-	59.8	}$	&	1761.2	$\pm$	415.8	\\
PG0052+251	&	-0.36	$^{+	0.30	}	_{-	0.30	}$	&	-0.13	$^{+	0.09	}	_{-	0.09	}$	&	120.2	$^{+	32.8	}	_{-	32.3	}$	&	1191.0	$\pm$	322.3	\\
Fairall9	&	-0.54	$^{+	0.27	}	_{-	0.30	}$	&	-0.08	$^{+	0.08	}	_{-	0.09	}$	&	20.8	$^{+	3.8	}	_{-	5.1	}$	&	151.1	$\pm$	32.6	\\
Mrk590	&	-0.78	$^{+	0.74	}	_{-	0.73	}$	&	-0.01	$^{+	0.21	}	_{-	0.21	}$	&	26.0	$^{+	6.6	}	_{-	5.4	}$	&	180.4	$\pm$	41.6	\\
3C120	&	-0.09	$^{+	0.53	}	_{-	0.50	}$	&	-0.20	$^{+	0.15	}	_{-	0.14	}$	&	41.9	$^{+	13.9	}	_{-	10.6	}$	&	206.4	$\pm$	60.3	\\
Ark120	&	-1.24	$^{+	0.44	}	_{-	0.44	}$	&	0.12	$^{+	0.13	}	_{-	0.13	}$	&	29.7	$^{+	6.4	}	_{-	5.9	}$	&	171.3	$\pm$	35.3	\\
Mrk79	&	-0.64	$^{+	0.72	}	_{-	0.71	}$	&	-0.05	$^{+	0.20	}	_{-	0.20	}$	&	17.3	$^{+	6.0	}	_{-	5.4	}$	&	82.4	$\pm$	27.2	\\
PG0804+761	&	-0.11	$^{+	0.85	}	_{-	0.85	}$	&	-0.20	$^{+	0.24	}	_{-	0.24	}$	&	231.7	$^{+	29.7	}	_{-	29.8	}$	&	1278.3	$\pm$	164.1	\\
Mrk110	&	-0.25	$^{+	1.13	}	_{-	1.12	}$	&	-0.16	$^{+	0.32	}	_{-	0.32	}$	&	36.7	$^{+	12.8	}	_{-	10.3	}$	&	286.4	$\pm$	90.1	\\
PG0953+414	&	0.001	$^{+	0.30	}	_{-	0.30	}$	&	-0.23	$^{+	0.09	}	_{-	0.09	}$	&	253.8	$^{+	36.5	}	_{-	38.2	}$	&	2587.7	$\pm$	381.0	\\
NGC3227	&	-1.36	$^{+	0.25	}	_{-	0.25	}$	&	0.16	$^{+	0.08	}	_{-	0.08	}$	&	2.7	$^{+	0.6	}	_{-	0.6	}$	&	11.9	$\pm$	2.5	\\
NGC3516	&	-1.77	$^{+	0.32	}	_{-	0.33	}$	&	0.27	$^{+	0.10	}	_{-	0.10	}$	&	6.2	$^{+	0.5	}	_{-	0.8	}$	&	33.3	$\pm$	3.6	\\
SBS1116+583A	&	-1.07	$^{+	0.75	}	_{-	0.74	}$	&	0.08	$^{+	0.21	}	_{-	0.21	}$	&	1.9	$^{+	0.5	}	_{-	0.4	}$	&	68.9	$\pm$	16.5	\\
Arp151	&	-0.63	$^{+	0.22	}	_{-	0.25	}$	&	-0.05	$^{+	0.07	}	_{-	0.07	}$	&	4.5	$^{+	0.6	}	_{-	0.8	}$	&	74.3	$\pm$	11.1	\\
NGC3783	&	-1.63	$^{+	0.52	}	_{-	0.48	}$	&	0.23	$^{+	0.15	}	_{-	0.14	}$	&	6.0	$^{+	1.9	}	_{-	1.3	}$	&	46.3	$\pm$	12.7	\\
Mrk1310	&	-0.67	$^{+	0.41	}	_{-	0.41	}$	&	-0.04	$^{+	0.12	}	_{-	0.12	}$	&	4.0	$^{+	0.7	}	_{-	0.7	}$	&	85.7	$\pm$	13.9	\\
NGC4051	&	-0.45	$^{+	0.44	}	_{-	0.38	}$	&	-0.10	$^{+	0.13	}	_{-	0.11	}$	&	2.6	$^{+	1.1	}	_{-	0.9	}$	&	8.7	$\pm$	3.3	\\
NGC4151	&	-2.25	$^{+	0.47	}	_{-	0.46	}$	&	0.41	$^{+	0.14	}	_{-	0.14	}$	&	2.6	$^{+	0.4	}	_{-	0.3	}$	&	10.2	$\pm$	1.5	\\
Mrk202	&	-0.42	$^{+	0.65	}	_{-	0.53	}$	&	-0.11	$^{+	0.19	}	_{-	0.15	}$	&	3.9	$^{+	2.2	}	_{-	1.4	}$	&	89.5	$\pm$	41.8	\\
NGC4253	&	-0.23	$^{+	3.05	}	_{-	3.05	}$	&	-0.16	$^{+	0.86	}	_{-	0.86	}$	&	9.0	$^{+	2.3	}	_{-	1.7	}$	&	89.8	$\pm$	20.3	\\
PG1229+204	&	-1.13	$^{+	0.66	}	_{-	0.40	}$	&	0.09	$^{+	0.19	}	_{-	0.12	}$	&	30.6	$^{+	22.4	}	_{-	12.4	}$	&	419.3	$\pm$	237.9	\\
NGC4593	&	-0.95	$^{+	0.64	}	_{-	0.63	}$	&	0.04	$^{+	0.18	}	_{-	0.18	}$	&	3.6	$^{+	0.7	}	_{-	0.6	}$	&	23.6	$\pm$	4.4	\\
NGC4748	&	-0.42	$^{+	0.42	}	_{-	0.49	}$	&	-0.11	$^{+	0.12	}	_{-	0.14	}$	&	7.1	$^{+	2.1	}	_{-	2.9	}$	&	82.8	$\pm$	28.8	\\
PG1307+085	&	-0.58	$^{+	0.36	}	_{-	0.44	}$	&	-0.07	$^{+	0.10	}	_{-	0.13	}$	&	122.7	$^{+	41.8	}	_{-	54.1	}$	&	1165.7	$\pm$	455.9	\\
Mrk279	&	-0.40	$^{+	0.31	}	_{-	0.31	}$	&	-0.11	$^{+	0.09	}	_{-	0.09	}$	&	21.8	$^{+	5.1	}	_{-	5.1	}$	&	137.2	$\pm$	32.0	\\
PG1411+442	&	-0.61	$^{+	0.52	}	_{-	0.52	}$	&	-0.06	$^{+	0.15	}	_{-	0.15	}$	&	141.3	$^{+	69.3	}	_{-	70.1	}$	&	1038.8	$\pm$	512.7	\\
PG1426+015	&	-0.97	$^{+	0.50	}	_{-	0.53	}$	&	0.05	$^{+	0.14	}	_{-	0.15	}$	&	85.1	$^{+	26.8	}	_{-	33.2	}$	&	559.4	$\pm$	197.3	\\
Mrk817	&	-0.64	$^{+	0.74	}	_{-	0.67	}$	&	-0.05	$^{+	0.21	}	_{-	0.19	}$	&	22.2	$^{+	11.0	}	_{-	7.5	}$	&	143.1	$\pm$	59.7	\\
Mrk290	&	-0.81	$^{+	0.16	}	_{-	0.15	}$	&	0.001	$^{+	0.05	}	_{-	0.05	}$	&	8.7	$^{+	1.2	}	_{-	1.0	}$	&	101.7	$\pm$	13.0	\\
PG1613+658	&	-0.17	$^{+	0.56	}	_{-	0.56	}$	&	-0.18	$^{+	0.16	}	_{-	0.16	}$	&	60.8	$^{+	22.7	}	_{-	23.0	}$	&	515.3	$\pm$	194.1	\\
PG1617+175	&	-0.87	$^{+	0.56	}	_{-	0.59	}$	&	0.02	$^{+	0.16	}	_{-	0.17	}$	&	68.3	$^{+	28.3	}	_{-	32.2	}$	&	773.7	$\pm$	342.5	\\
PG1700+518	&	0.50	$^{+	0.76	}	_{-	0.76	}$	&	-0.37	$^{+	0.22	}	_{-	0.21	}$	&	590.8	$^{+	107.8	}	_{-	91.2	}$	&	4894.0	$\pm$	824.1	\\
3C390.3	&	-0.97	$^{+	1.09	}	_{-	1.00	}$	&	0.05	$^{+	0.31	}	_{-	0.28	}$	&	40.0	$^{+	24.9	}	_{-	15.3	}$	&	207.4	$\pm$	104.3	\\
NGC6814	&	-1.95	$^{+	0.48	}	_{-	0.48	}$	&	0.32	$^{+	0.14	}	_{-	0.14	}$	&	3.1	$^{+	0.4	}	_{-	0.4	}$	&	20.0	$\pm$	2.7	\\
Mrk509	&	-0.87	$^{+	0.12	}	_{-	0.12	}$	&	0.02	$^{+	0.04	}	_{-	0.04	}$	&	76.5	$^{+	5.9	}	_{-	5.2	}$	&	312.7	$\pm$	22.6	\\
PG2130+099	&	1.17	$^{+	0.15	}	_{-	0.15	}$	&	-0.56	$^{+	0.05	}	_{-	0.05	}$	&	34.9	$^{+	4.4	}	_{-	4.4	}$	&	266.8	$\pm$	33.4	\\
NGC7469	&	0.51	$^{+	0.35	}	_{-	0.25	}$	&	-0.37	$^{+	0.10	}	_{-	0.07	}$	&	25.5	$^{+	8.0	}	_{-	3.1	}$	&	106.8	$\pm$	23.2	\\
PG1211+143	&	0.01	$^{+	0.27	}	_{-	0.41	}$	&	-0.23	$^{+	0.08	}	_{-	0.12	}$	&	159.7	$^{+	43.6	}	_{-	71.7	}$	&	868.9	$\pm$	313.6	\\
PG0844+349	&	0.03	$^{+	0.47	}	_{-	0.46	}$	&	-0.24	$^{+	0.13	}	_{-	0.13	}$	&	55.8	$^{+	23.7	}	_{-	23.2	}$	&	426.6	$\pm$	179.0	\\
NGC5273	&	-2.12	$^{+	0.55	}	_{-	0.69	}$	&	0.37	$^{+	0.16	}	_{-	0.20	}$	&	0.9	$^{+	0.5	}	_{-	0.7	}$	&	9.4	$\pm$	6.0	\\
Mrk1511	&	-0.43	$^{+	0.18	}	_{-	0.17	}$	&	-0.11	$^{+	0.05	}	_{-	0.05	}$	&	7.3	$^{+	1.1	}	_{-	1.0	}$	&	97.5	$\pm$	14.7	\\
KA1858-4850	&	-0.05	$^{+	0.17	}	_{-	0.19	}$	&	-0.21	$^{+	0.05	}	_{-	0.06	}$	&	22.1	$^{+	3.3	}	_{-	3.8	}$	&	516.2	$\pm$	82.6	\\
MCG6-30-15	&	-1.94	$^{+	0.67	}	_{-	0.66	}$	&	0.32	$^{+	0.19	}	_{-	0.19	}$	&	2.7	$^{+	0.9	}	_{-	0.8	}$	&	48.5	$\pm$	14.7	\\
UGC06728	&	-0.23	$^{+	0.46	}	_{-	0.52	}$	&	-0.16	$^{+	0.13	}	_{-	0.15	}$	&	2.0	$^{+	1.0	}	_{-	1.2	}$	&	21.1	$\pm$	11.3	\\
MCG+08-11-011	&	-0.14	$^{+	0.17	}	_{-	0.17	}$	&	-0.19	$^{+	0.05	}	_{-	0.05	}$	&	24.3	$^{+	0.8	}	_{-	0.8	}$	&	164.0	$\pm$	5.3	\\
NGC2617	&	-1.10	$^{+	0.33	}	_{-	0.36	}$	&	0.08	$^{+	0.10	}	_{-	0.11	}$	&	3.5	$^{+	0.9	}	_{-	1.1	}$	&	34.0	$\pm$	9.7	\\
3C382	&	-1.02	$^{+	0.28	}	_{-	0.17	}$	&	0.06	$^{+	0.08	}	_{-	0.06	}$	&	35.1	$^{+	6.9	}	_{-	3.2	}$	&	376.6	$\pm$	54.6	\\
Mrk374	&	-0.16	$^{+	0.35	}	_{-	0.20	}$	&	-0.18	$^{+	0.10	}	_{-	0.06	}$	&	22.5	$^{+	8.8	}	_{-	5.0	}$	&	189.7	$\pm$	58.3	\\
\hline \noalign{\vskip 0.1cm} 																													
\multicolumn{5}{c}{\citet{lu2016}} \\ 																													
\hline \noalign{\vskip 0.1cm} 																													
NGC5548	&	-1.19	$^{+	0.25	}	_{-	0.20	}$	&	0.11	$^{+	0.08	}	_{-	0.06	}$	&	5.6	$^{+	1.0	}	_{-	0.3	}$	&	34.7	$\pm$	4.1	\\
\hline \noalign{\vskip 0.1cm} 																													
\multicolumn{5}{c}{\citet{zhang2018}} \\ 																													
\hline \noalign{\vskip 0.1cm} 																													
3C 273	&	1.12	$^{+	0.07	}	_{-	0.08	}$	&	-0.54	$^{+	0.03	}	_{-	0.03	}$	&	514.1	$^{+	29.1	}	_{-	42.4	}$	&	1380.8	$\pm$	96.0	\\

\hline \noalign{\vskip 0.1cm}       
\end{longtable}
\end{center}
\footnotesize{{\sc Notes.} Columns are as follows: (1) Object name. (2) Dimensionless accretion rate. (3) Deviation of the expected \RL\ estimated from the Equation~(\ref{equ:linearfit}). (4) Time delay corrected by the dimensionless accretion rate in unit of days. (6) Luminosity distance in units of Mpc.}

\newpage

\begin{table*}[]
\caption{Orthogonal linear fit parameters}
\begin{center}
\label{tab:linearfit} 
    \begin{tabular}{cccccc}
     \hline\hline\noalign{\vskip 0.1cm}
     & X & $\alpha$ & $\beta$ & $P$ & $rms$ \\ 
     (1) & (2) & (3) & (4) & (5) & (6)  \\ 
      \hline \noalign{\vskip 0.1cm}
      \multirow{2}{*}{\fblr} & \multirow{1}{*}{\mdot} & \multirow{1}{*}{-0.143$\pm$0.018} & \multirow{1}{*}{-0.136$\pm$0.023} & \multirow{1}{*}{0.572}  & \multirow{1}{*}{0.243} \\
 & \LLEdd & -0.271$\pm$0.030 & -0.396$\pm$0.032 & 0.605 & 0.626 \\
       \hline \noalign{\vskip 0.1cm} 
    \multirow{2}{*}{\fblrc} & \multirow{1}{*}{\mdotc} & \multirow{1}{*}{-0.283$\pm$0.017} & \multirow{1}{*}{-0.228$\pm$0.016} & \multirow{1}{*}{0.822}  & \multirow{1}{*}{0.172} \\
 & \LLEddc & -0.394$\pm$0.030 & -0.589$\pm$0.036 & 0.744 & 0.199 \\
      \hline \noalign{\vskip 0.1cm}
    \end{tabular}
\end{center}
\end{table*}
    \footnotesize{ {\sc Notes.}  Columns are as follows: (1) Virial factor. (2) Accretion parameter: dimensionless accretion rate or Eddington ratio. (3) Slope of Equation~(\ref{equ:linearfit}). (4) Ordinate of Equation~(\ref{equ:linearfit}). (5) Pearson coefficient. (6) $rms$ value. Rows 1 and 2 correspond to the estimations for virial factor equal to 1, \fblr. Rows 3 and 4 correspond to the estimations for a virial factor anti-correlated with the FWHM, \fblrc. }

\end{document}